\documentclass[aps,prc,twocolumn,groupedaddress,showpacs,amsmath,amssymb,floatfix,superscriptaddress]{revtex4}
\usepackage{graphicx}
\usepackage{times}

\newcommand{\nuebar}{$\overline{\nu}_{e}$}

\begin{document}



\newcommand{\DefaultFigureAngle}{270}


\title{Production of radioactive isotopes through cosmic muon spallation in KamLAND}

\newcommand{\tohoku}{\affiliation{Research Center for Neutrino
    Science, Tohoku University, Sendai 980-8578, Japan}}
\newcommand{\alabama}{\affiliation{Department of Physics and
    Astronomy, University of Alabama, Tuscaloosa, Alabama 35487, USA}}
\newcommand{\lbl}{\affiliation{Physics Department, University of
    California, Berkeley and \\ Lawrence Berkeley National Laboratory, 
Berkeley, California 94720, USA}}
\newcommand{\caltech}{\affiliation{W.~K.~Kellogg Radiation Laboratory,
    California Institute of Technology, Pasadena, California 91125, USA}}
\newcommand{\colostate}{\affiliation{Department of Physics, Colorado
    State University, Fort Collins, Colorado 80523, USA}}
\newcommand{\drexel}{\affiliation{Physics Department, Drexel
    University, Philadelphia, Pennsylvania 19104, USA}}
\newcommand{\hawaii}{\affiliation{Department of Physics and Astronomy,
    University of Hawaii at Manoa, Honolulu, Hawaii 96822, USA}}
\newcommand{\kansas}{\affiliation{Department of Physics,
    Kansas State University, Manhattan, Kansas 66506, USA}}
\newcommand{\lsu}{\affiliation{Department of Physics and Astronomy,
    Louisiana State University, Baton Rouge, Louisiana 70803, USA}}
\newcommand{\stanford}{\affiliation{Physics Department, Stanford
    University, Stanford, California 94305, USA}}
\newcommand{\ut}{\affiliation{Department of Physics and
    Astronomy, University of Tennessee, Knoxville, Tennessee 37996, USA}}
\newcommand{\tunl}{\affiliation{Triangle Universities Nuclear
    Laboratory, Durham, North Carolina 27708, USA and \\
Physics Departments at Duke University, North Carolina Central University,
and the University of North Carolina at Chapel Hill}}
\newcommand{\wisc}{\affiliation{Department of Physics, University
    of Wisconsin, Madison, Wisconsin 53706, USA}}  
\newcommand{\cnrs}{\affiliation{CEN Bordeaux-Gradignan, IN2P3-CNRS and
    University Bordeaux I, F-33175 Gradignan Cedex, France}}
\newcommand{\ipmu}{\affiliation{Institute for the Physics and Mathematics of the 
    Universe, Tokyo University, Kashiwa 277-8568, Japan}}
\newcommand{\nikhef}{\affiliation{Nikhef, Science Park 105, 1098 XG Amsterdam, The Netherlands}}

    
\newcommand{\atlanlnow}{\altaffiliation{Present address: Subatomic Physics Group, Los Alamos National Laboratory, Los Alamos, NM 87545, USA}}
    
\newcommand{\atksunow}{\altaffiliation{Present address: Department of Physics, Kansas State University, Manhattan, Kansas 66506, USA}}

\newcommand{\atokayamanow}{\altaffiliation{Present address: Center of Quantum Universe, Okayama University, Okayama 700-8530, Japan}}

\newcommand{\atregisnow}{\altaffiliation{Present address: Department of Physics and Computational Science, Regis University, Denver, Colorado 80221, USA}}

\newcommand{\atfnalnow}{\altaffiliation{Present address: Fermi National Accelerator Laboratory, Batavia, Illinois 60510, USA}}

\newcommand{\atsnolabnow}{\altaffiliation{Present address: SNOLAB, Lively, ON P3Y 1M3, Canada}}

\newcommand{\atllnlnow}{\altaffiliation{Present address: Lawrence Livermore National Laboratory, Livermore, California 94550, USA}}

\newcommand{\atucdnow}{\altaffiliation{Present address: Department of Physics, University of California, Davis, California 95616, USA}}

\newcommand{\atuwnow}{\altaffiliation{Present address:  CENPA, University of Washington, Seattle, Washington 98195, USA}}

\newcommand{\atumdnow}{\altaffiliation{Present address: Department of Physics, University of Maryland, College Park, Maryland 20742, USA}}

\newcommand{\atmitnow}{\altaffiliation{Present address: Department of Physics, Massachusetts Institute of Technology, Cambridge, MA 02139, USA}}

%
%
\author{S.~Abe}\tohoku
\author{S.~Enomoto}\tohoku\ipmu
\author{K.~Furuno}\tohoku
\author{Y.~Gando}\tohoku
\author{H.~Ikeda}\tohoku
\author{K.~Inoue}\tohoku\ipmu
\author{Y.~Kibe}\tohoku
\author{Y.~Kishimoto}\tohoku
\author{M.~Koga}\tohoku\ipmu
\author{Y.~Minekawa}\tohoku
\author{T.~Mitsui}\tohoku
\author{K.~Nakajima}\atokayamanow\tohoku 
\author{K.~Nakajima}\tohoku
\author{K.~Nakamura}\tohoku\ipmu
\author{M.~Nakamura}\tohoku
\author{I.~Shimizu}\tohoku
\author{Y.~Shimizu}\tohoku 
\author{J.~Shirai}\tohoku
\author{F.~Suekane}\tohoku
\author{A.~Suzuki}\tohoku
\author{Y.~Takemoto}\tohoku
\author{K.~Tamae}\tohoku
\author{A.~Terashima}\tohoku
\author{H.~Watanabe}\tohoku
\author{E.~Yonezawa}\tohoku
\author{S.~Yoshida}\tohoku
%
\author{A.~Kozlov}\ipmu
\author{H.~Murayama}\ipmu\lbl
%
\author{J.~Busenitz}\alabama
\author{T.~Classen}\atucdnow\alabama
\author{C.~Grant}\alabama
\author{G.~Keefer}\alabama
\author{D.S.~Leonard}\atumdnow\alabama
\author{D.~McKee}\atksunow\alabama
\author{A.~Piepke}\ipmu\alabama
%
\author{T.I.~Banks}\lbl
\author{T.~Bloxham}\lbl
\author{J.A.~Detwiler}\lbl
\author{S.J.~Freedman}\ipmu\lbl
\author{B.K.~Fujikawa}\ipmu\lbl
\author{F.~Gray}\atregisnow\lbl
\author{E.~Guardincerri}\lbl
\author{L.~Hsu}\atfnalnow\lbl
\author{K.~Ichimura}\lbl 
\author{R.~Kadel}\lbl
\author{C.~Lendvai}\lbl
\author{K.-B.~Luk}\lbl
\author{T.~O'Donnell}\lbl
\author{H.M.~Steiner}\lbl
\author{L.A.~Winslow}\atmitnow\lbl
%
\author{D.A.~Dwyer}\caltech
\author{C.~Jillings}\atsnolabnow\caltech
\author{C.~Mauger}\atlanlnow\caltech
\author{R.D.~McKeown}\caltech
\author{P.~Vogel}\caltech
\author{C.~Zhang}\caltech
%
\author{B.E.~Berger}\colostate
%
\author{C.E.~Lane}\drexel
\author{J.~Maricic}\drexel
\author{T.~Miletic}\drexel
%
\author{M.~Batygov}\hawaii
\author{J.G.~Learned}\hawaii
\author{S.~Matsuno}\hawaii
\author{S.~Pakvasa}\hawaii
%
\author{J.~Foster}\kansas
\author{G.A.~Horton-Smith}\ipmu\kansas
\author{A.~Tang}\kansas
%
\author{S.~Dazeley}\atllnlnow\lsu
%
\author{K.E.~Downum}\stanford
\author{G.~Gratta}\stanford
\author{K.~Tolich}\atuwnow\stanford
%
\author{W.~Bugg}\ut
\author{Y.~Efremenko}\ipmu\ut
\author{Y.~Kamyshkov}\ut
\author{O.~Perevozchikov}\ut
%
\author{H.J.~Karwowski}\tunl
\author{D.M.~Markoff}\tunl
\author{W.~Tornow}\tunl
%
\author{K.M.~Heeger}\ipmu\wisc 
%
\author{F.~Piquemal}\cnrs
\author{J.-S.~Ricol}\cnrs
%
\author{M.P.~Decowski}\ipmu\nikhef

\collaboration{The KamLAND Collaboration}\noaffiliation

\date{\today}

\begin{abstract}

Radioactive isotopes produced through cosmic muon spallation are a background for rare-event detection in
$\nu$\ detectors,
double-$\beta$-decay experiments,
and dark-matter searches.
Understanding the nature of cosmogenic backgrounds is particularly important for future experiments aiming to determine the $pep$\ and CNO solar neutrino fluxes,
for which the background is dominated by the spallation production of $^{11}$C.
Data from the Kamioka liquid scintillator antineutrino detector (KamLAND) provides valuable information for better understanding these backgrounds,
especially in liquid scintillators,
and for checking estimates from current simulations based upon \textsc{music}, \textsc{fluka}, and   \textsc{geant4}.
Using the time correlation between detected muons and neutron captures,
the neutron production yield in the KamLAND liquid scintillator is measured to be
\mbox{$Y_{n}=(2.8\pm0.3)\times 10^{-4}$}\ \mbox{$\mu^{-1}$g$^{-1}$cm$^{2}$}.
For other isotopes,
the production yield is determined from the observed time correlation related to known isotope lifetimes.
We find some yields are inconsistent with extrapolations based on an accelerator muon beam experiment.

\end{abstract}

\pacs{96.50.S-,96.60.-j,24.10.Lx,25.30.Mr}

\maketitle

\section{Introduction}
\label{section:Introduction}

Cosmic-ray muons and their spallation products are potential sources of background for neutrino detectors,
double-$\beta$-decay experiments,
and dark-matter searches,
even when the detectors are deployed underground.
Characterizing cosmic-ray-muon-induced backgrounds,
particularly the secondary neutrons and radioactive isotopes produced by muon-initiated spallation processes,
is essential for interpreting these experiments.

Liquid-scintillator detectors such as the KamLAND (Kamioka liquid-scintillator antineutrino detector),
Borexino~\cite{Bellini2008,Arpesella2008a,Arpesella2008b},
CANDLES~IV (calcium fluoride for studies of neutrino and dark matters by low-energy spectrometer)~\cite{Hirano2008,Kishimoto2007},
SNO+ (Sudbury Neutrino Observatory)~\cite{Zuber2007,Chen2006},
LENS (low energy neutrino spectroscopy)~\cite{Grieb2007},
and LENA (low-energy neutrino astronomy)~\cite{Undagoitia2008} are designed to detect low-energy phenomena.
In an organic liquid scintillator (LS), 
energetic muons and subsequent showers interact mostly with $^{12}$C,
the most abundant nucleus heavier than $^{1}$H in the LS,
generating neutrons and isotopes by electromagnetic or hadronic processes.
The muon-initiated spallation of carbon targets is a matter of primary interest.

Isotope production by muon-initiated spallation has been studied by an earlier experiment~\cite{Hagner2000} using the CERN Super Proton Synchrotron (SPS) muon beam.
The energy dependence was studied with 100 and 190 GeV incident muons.
The production yield at other energies is estimated from these data by extrapolation,
assuming a power-law dependence on the muon energy.
Direct measurements of the production yield by underground detectors such as
the Mont Blanc liquid scintillation detector (LSD)~\cite{Aglietta1989},
the INFN large-volume detector (LVD)~\cite{Aglietta2003}, and Borexino~\cite{Back2006}
were compared with calculations exploiting simulations based on	
\textsc{music} ~\cite{Antonioli1997}, \textsc{fluka}~\cite{Fasso2003,Ferrari2005}, and  \textsc{geant4}~\cite{Agostinelli2003,Allison2006}.
Particular attention was paid to neutron production since isotope production measurements are difficult with the small scintillator masses used in these detectors.
KamLAND, owing to its larger mass---$\sim$1 kiloton of LS---does not suffer from this difficulty and is well placed to study a variety of isotopes of interest.

This paper presents the neutron and isotope production rates in KamLAND from muon-initiated spallation based upon data collected from 5 March 2002 to 12 May 2007.
The results are compared with simulations and other experiments.
These comparisons provide important information for validating Monte Carlo simulations.

\section{Detector Description and Performance}
\label{section:DetectorDescriptionAndPerformance}

\begin{figure}
\includegraphics[angle=270,width=\columnwidth]{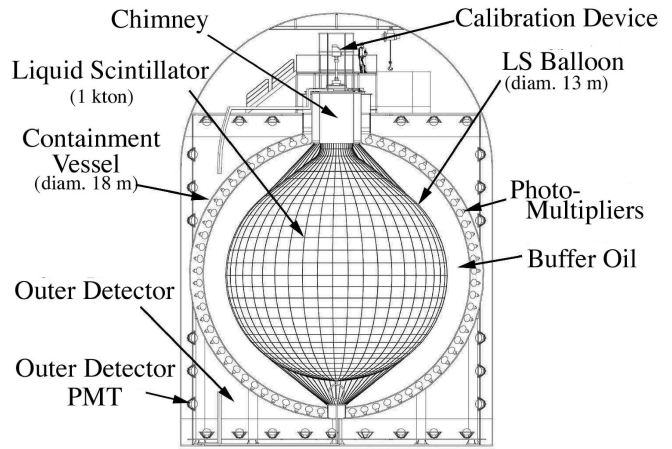}
\caption[]{Schematic diagram of the KamLAND detector at the Kamioka underground laboratory.}
\label{figure:detector}
\end{figure}

\begin{table}
\caption{\label{table:LiquidScintillator}
Elemental composition of the KamLAND liquid scintillator.
The hydrogen-to-carbon ratio was verified by elemental analysis ($\pm 2\%$).
The liquid scintillator contains traces of nitrogen and oxygen from its fluor,
PPO (C$_{15}$H$_{11}$NO),
and dissolved gases.
Measurements in similar liquid hydrocarbons~\cite{Battino1984,Hesse1995} indicate that the amount of dissolved nitrogen saturates at about 200-500 parts per million (ppm).
The range of nitrogen composition values given in this table reflects the extreme cases of zero dissolved nitrogen gas to full saturation. 
The dissolved oxygen content of liquid scintillator taken from the center of KamLAND was measured to be less than 3 ppm,
which is insignificant compared to the oxygen contribution from PPO.
}
\begin{ruledtabular}
\begin{tabular}{lcc}
 & & Number of targets \\
\raisebox{1.75ex}{Element} & \raisebox{1.75ex}{Stoichiometry} & (per kiloton) \\
\hline
Hydrogen & 1.97 & $8.47\times10^{31}$ \\
Carbon & $\equiv 1$ & $4.30\times10^{31}$ \\
Nitrogen & $1\times10^{-4}$\ to $6\times10^{-4}$ & $5\times10^{27}$\ to $3\times10^{28}$ \\
Oxygen & $1\times10^{-4}$ & $5\times10^{27}$ \\
\end{tabular}
\end{ruledtabular}
\end{table}

KamLAND is located under the peak of Ikenoyama (Ike Mountain, $36.42^{\circ}$N, $137.31^{\circ}$E),
and the vertical rock overburden is approximately 2700 meters water equivalent (mwe).
A schematic diagram of KamLAND is shown in Fig.~\ref{figure:detector}.
KamLAND consists of an active detector region of approximately 1 kiloton (kton) of ultrapure LS contained in a 13-m-diameter spherical balloon made of
135-$\mu$m-thick transparent nylon-EVOH (ethylene vinyl alcohol copolymer) composite film
and supported by a network of Kevlar ropes.
In addition to providing containment for the LS,
the balloon protects the LS against the diffusion of ambient radon from the surrounding components.
The total volume of LS in the balloon is \mbox{$1171\pm25$\ m$^{3}$}\ as determined with flow meters during the initial filling.
The LS consists of 80\% dodecane and
20\% pseudocumene (1,2,4-trimethylbenzene) by volume,
and \mbox{$1.36\pm0.03$}\ g/liter of the fluor PPO (2,5-diphenyloxazole). 
The density of the LS is 0.780\ g/cm$^3$ at $11.5^{\circ}$C.
The calculated elemental composition of the LS is given in Table~\ref{table:LiquidScintillator}.

A buffer comprising 57\% isoparaffin and 43\% dodecane oils by volume fills the region between the balloon
and the surrounding 18-m-diameter spherical stainless-steel outer vessel to shield the LS from external radiation.
The specific gravity of the buffer oil (BO) is adjusted to be 0.04\% lower than that of the LS.
An array of photomultiplier tubes (PMTs)---1325 specially developed fast PMTs masked to 17-in.-diameter and
554 older 20-in.-diameter PMTs reused from the Kamiokande experiment~\cite{Kume1983}---are mounted on the inner surface of the outer containment vessel,
providing 34\% photocathode coverage.
During the period from 5 March 2002 to 27 February 2003 the photocathode coverage was only 22\%,
since the 20-in.\ PMTs were not operated.
A 3-mm-thick acrylic barrier at 16.6~m in diameter helps prevent radon emanating from the PMT glass from entering the BO.
The inner detector (ID),
consisting of the LS and BO regions,
is surrounded by a 3.2-kton water-Cherenkov detector instrumented with 225 20-in.\ PMTs. 
This outer detector (OD) absorbs $\gamma$\ rays and neutrons from the surrounding rock and enables tagging of cosmic-ray muons.

The KamLAND front-end electronics (FEE) system is based on the
analog transient waveform digitizer (ATWD)~\cite{Kleinfelder2003}
which captures PMT signals in 128 10-bit digital samples at intervals of 1.5~ns.
Each ATWD captures three gain levels of a PMT signal to obtain a dynamic range from 1 photoelectron (p.e.) to 1000 p.e.
Each ATWD takes 27~$\mu$s to read out, so two are attached to each PMT channel to reduce dead time.
The FEE system contains discriminators set at 0.15~p.e.\ ($\sim$0.3~mV)\ threshold which send a 125-ns-long logic signal to the trigger electronics.
The trigger electronics counts the number of ID and OD PMTs above the discriminator threshold with a sampling rate of 40~MHz and
initiates readout when the number of 17-in.\ ID PMTs above the discriminator threshold ($N_{17}$) exceeds the number
corresponding to $\sim$0.8~MeV deposited energy.
The trigger system also issues independent readout commands when the number of OD PMTs above threshold exceeds a preset number.

The energy can be estimated from the $N_{\mathrm{max}}$\ parameter,
defined as the maximum value of $N_{17}$\ in a 200-ns period following the trigger command.
However,
the offline analysis takes full advantage of the information stored in the digitized PMT signals by identifying individual PMT pulses in the waveform information that is read out.
The time and integrated area (called ``charge'') are computed from the individual pulses.
For each PMT,
the average charge corresponding to a single p.e.\ is determined from \mbox{single-pulse} waveforms observed in low-occupancy events.
The ID PMT timing is calibrated with light pulses from a dye laser ($\sim$1.2-ns pulse width),
injected at the center of the detector through an optical fiber.
The vertices of spatially localized low-energy ($<$30~MeV) events are estimated by comparing calculated time-of-flights
of optical photons from the hypothetical vertex to the measured arrival times at the PMTs in KamLAND.

The reconstructed energies of events are calibrated with $\gamma$\ sources:
$^{203}$Hg, $^{68}$Ge, $^{65}$Zn, and $^{60}$Co;
and with $n+\gamma$\ sources: \mbox{$^{241}$Am+$^{9}$Be} and \mbox{$^{210}$Po+$^{13}$C~\cite{McKee2008}}.
These are deployed at various positions along the vertical axis of the detector and occasionally off the vertical axis within 5.5 m from the detector center~\cite{Berger2009}.
Such calibrations cover energies between 0.28 and 6.1~MeV.
The energy calibration is aided with studies of background contaminants
$^{40}$K and $^{208}$Tl,
\mbox{$^{212}$Bi-$^{212}$Po} and \mbox{$^{214}$Bi-$^{214}$Po} sequential decays,
$^{12}$B and $^{12}$N spallation products,
and $\gamma$'s from thermal neutron captures on $^{1}$H and $^{12}$C. 

The visible energy $E_{\mathrm{vis}}$\ of an event is computed from the measured light yield.
Specifically, $E_{\mathrm{vis}}$\ is the number of detected p.e.\ after corrections for PMT variation,
dark noise,
solid angle,
shadowing by suspension ropes,
optical transparencies,
and scattering properties in the LS.
The relationship between $E_{\mathrm{vis}}$\ and the deposited energy $E_{\mathrm{dep}}$\ of
$\gamma$'s, $e^{\pm}$'s, protons, and $\alpha$'s is nonlinear and
modeled as a combination of Birks-quenched scintillation~\cite{Birks1951,Birks1964} and Cherenkov radiation.
The scale is adjusted so that $E_{\mathrm{vis}}$\ is equal to $E_{\mathrm{dep}}$\ for the \mbox{2.225-MeV}\ $\gamma$ ray from neutron capture on $^{1}$H.
The observed energy resolution is
\mbox{$\sim$7.4\%$/\sqrt{E_{\mathrm{vis}}(\mbox{MeV})}$}\ for the period without the 20-in.\ PMTs, and
\mbox{$\sim$6.5\%$/\sqrt{E_{\mathrm{vis}}(\mbox{MeV})}$}\ for the rest of the data.

The calibration sources are also used to determine systematic deviations in position reconstruction by comparison with the source's known position.
This comparison gives an average position reconstruction uncertainty of less than 3 cm for events with energies in the range 0.28 to 6.1~MeV.

\begin{figure}[t]
\begin{center}
\includegraphics[width=\columnwidth]{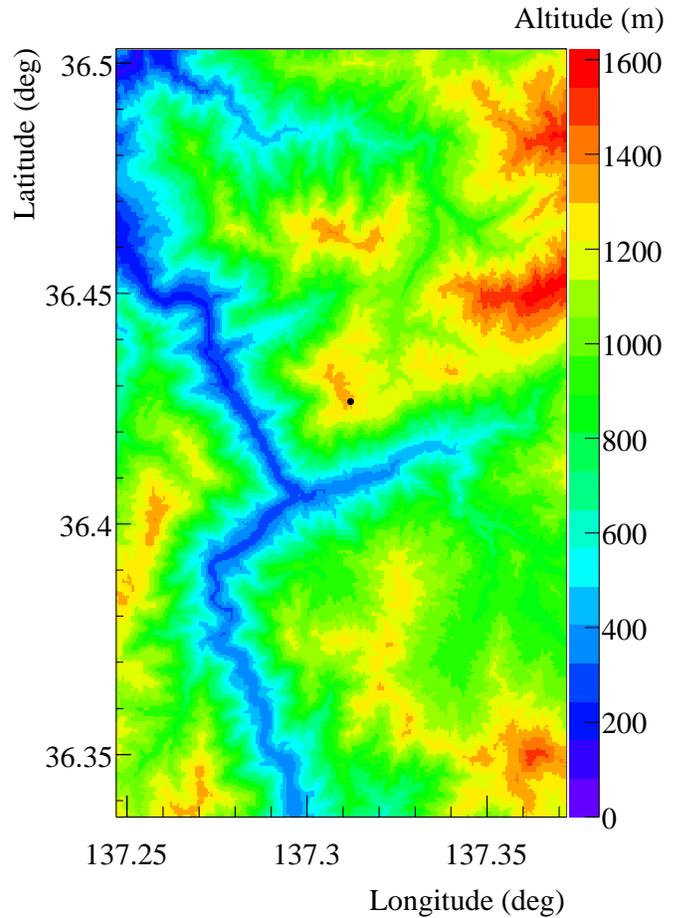}
\end{center}
\caption[shapeimage 1]{\label{figure:IkenoyamaProfile}
Ikenoyama topological profile~\cite{GeographicalSurvey}.
The black point near the center is the location of KamLAND.}
\end{figure}

\section{Cosmic-Ray Muons}
\label{section:CosmicRayMuons}

A digital map~\cite{GeographicalSurvey} of the topological profile of Ikenoyama is shown in Fig.~\ref{figure:IkenoyamaProfile}.
The vertical overburden at KamLAND is approximately 1000~m of rock and the minimum overburden corresponding to a nearby valley is approximately 900~m.

Cosmic-ray  muons are identified either by the large amount of scintillation and Cherenkov light detected by the ID PMTs,
or by the Cherenkov light detected by the OD PMTs.
Muons crossing the ID (\textit{ID muons})
are selected by requiring that one of the following conditions was satisfied:
\begin{enumerate}
\item $\mathcal{L}_{\mathrm{ID}}\geq10000$\ p.e.\ ($\sim$30\ MeV),
\item $\mathcal{L}_{\mathrm{ID}}\geq500$\ p.e.\ and $N_{\mathrm{OD}}\geq5$,
\end{enumerate}
where $\mathcal{L}_{\mathrm{ID}}$\ is the total light yield measured by the ID \mbox{17-in.} PMTs,
and $N_{\mathrm{OD}}$\ is the number of OD PMTs with signals above threshold.
Approximately 93\% of the ID muons satisfy the first selection criterion.
The ID muon track is reconstructed from arrival times of the first-arriving Cherenkov or scintillation photons at the PMTs.
Since for relativistic muons the wavefront of the scintillation light proceeds at the Cherenkov angle,
and since muons generate enough light to generate photoelectrons in every PMT,
by restricting the fit to the first-arriving photons both Cherenkov and scintillation photons can be treated identically.
The observed muon track is then established by minimizing time-of-flight deviations from hypothetical muon tracks.
The fit converges for 97\% of all ID muon events.
The majority of the events that are not reconstructed are believed to be multiple muons
or muons accompanied by large electromagnetic or hadronic showers for which the tracking model is not valid.

\begin{figure}[t]
\begin{center}
\includegraphics[angle=\DefaultFigureAngle,width=\columnwidth]{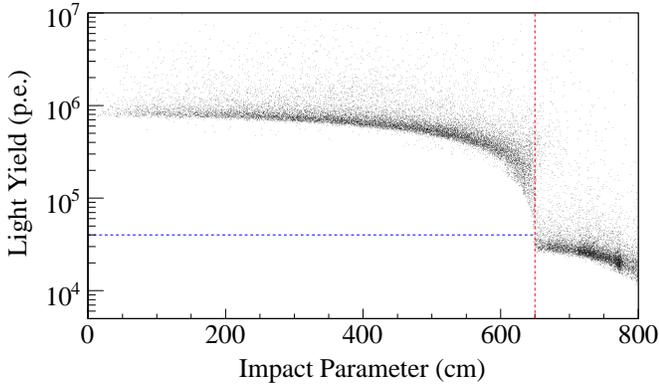}
\end{center}
\caption[muon reconstruct]{\label{figure:ChargeVersusImpactParameter}
Correlation between the light yield in the inner detector and the shortest distance from the muon track to the KamLAND detector center (impact parameter).
The vertical dashed red line represents the boundary between the LS and BO at 650~cm.
Events located to the left of this boundary and above the horizontal blue dashed line
\mbox{($\mathcal{L}_{\mathrm{ID}}=4\times10^{4}$\ p.e.)}
are designated as LS muons.
}
\end{figure}

\begin{figure}[t]
\begin{center}
\includegraphics[angle=\DefaultFigureAngle,width=\columnwidth]{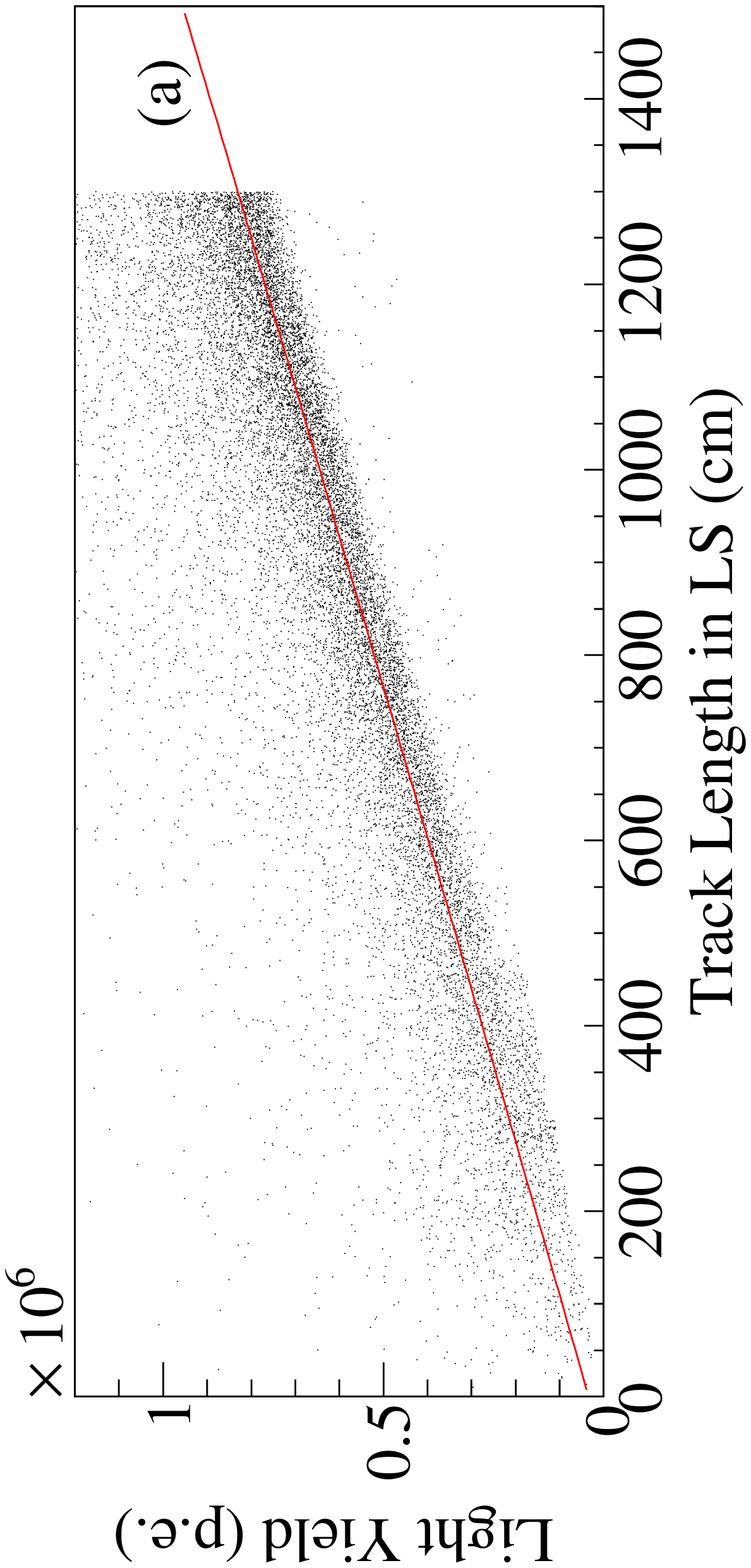}
\includegraphics[angle=\DefaultFigureAngle,width=\columnwidth]{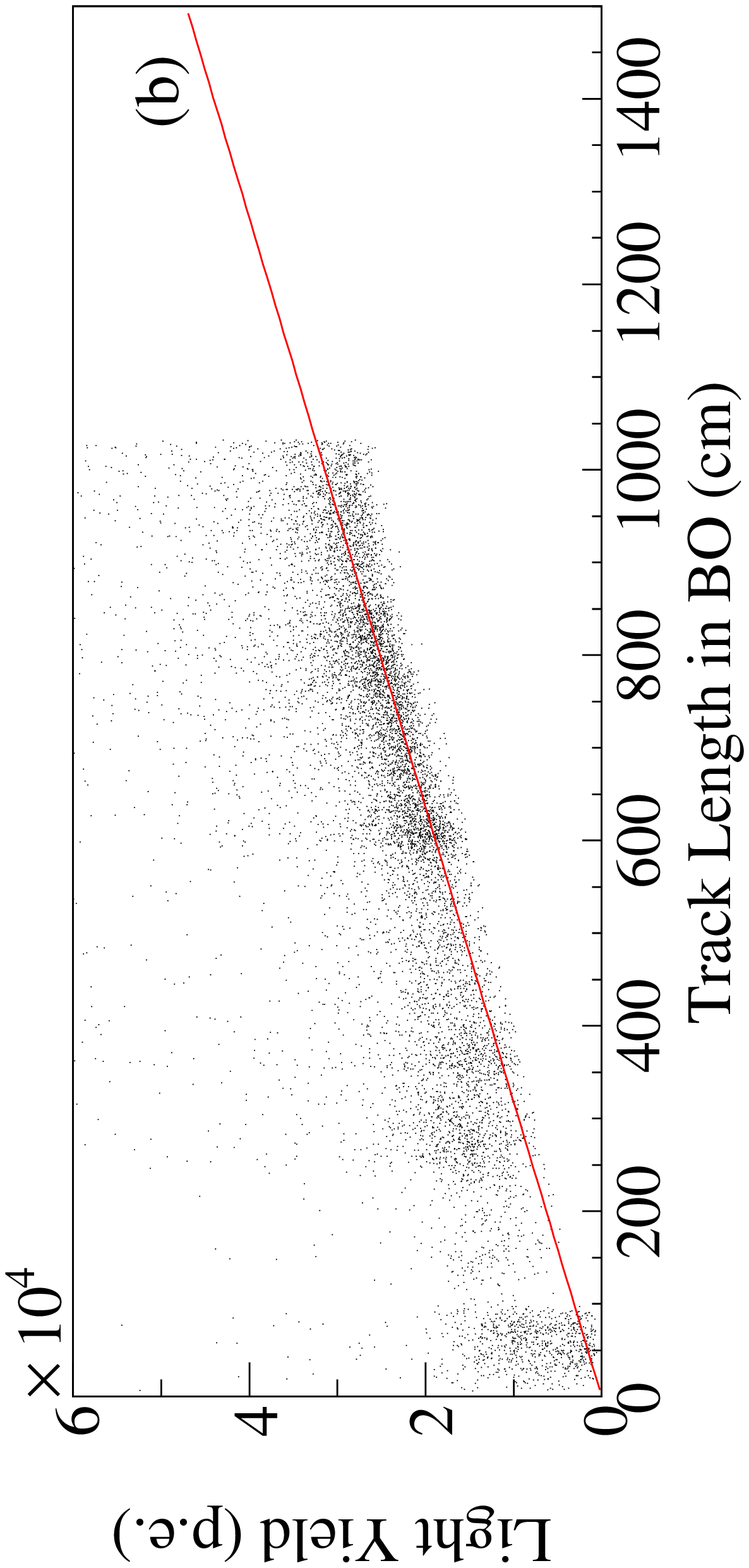}
\end{center}
\caption[muon charge]{\label{figure:ChargeVersusLength}
Correlation between the total light yield measured by the ID \mbox{17-in.} PMTs \mbox{($\mathcal{L}_{\mathrm{ID}}$)} and
the muon track length for (a) muons that pass through both the LS and BO regions and (b) muons that pass through only the BO region.
The red lines show the fitted light yield per unit length of \mbox{$\left<d\mathcal{L}_{S}/dX\right>=629\pm47$\ p.e./cm} and
\mbox{$\left<d\mathcal{L}_{\check{C}}/dX\right>=31\pm2$\ p.e./cm}.}
\end{figure}

Muons passing only through the BO produce mostly Cherenkov light,
whereas muons passing through the LS generate both Cherenkov and scintillation light.
Figure~\ref{figure:ChargeVersusImpactParameter} shows the correlation between the light yield ($\mathcal{L}_{\mathrm{ID}}$) and
the shortest distance between the reconstructed muon track and the center of KamLAND (impact parameter).
The boundary at 650~cm between the BO and LS regions is evident.
The correlations between $\mathcal{L}_{\mathrm{ID}}$\ and the reconstructed muon track length in the BO and LS regions
($L_{\mathrm{BO}}$\ and $L_{\mathrm{LS}}$, respectively)
are plotted in Figs.~\ref{figure:ChargeVersusLength}(a) and \ref{figure:ChargeVersusLength}(b).
A linear trend,
corresponding to minimum ionizing muons,
is apparent in both distributions.
The slope of each line is the light yield per unit length in the respective material.
In BO,
where the light is predominantly Cherenkov,
the light yield per unit length is found to be \mbox{$\left<d\mathcal{L}_{\check{C}}/dX\right>=31 \pm 2$\ p.e./cm};
the fit was restricted to path lengths above 700~cm,
since fits at shorter path lengths are complicated by the presence of PMTs which may obstruct some of the emitted light.
In the LS we obtain
\begin{equation}
\label{equation:ScintillationLightYield}
\frac{d\mathcal{L}_{S}}{dX}=\frac{\mathcal{L}_{\mathrm{ID}}-L_{\mathrm{BO}}\left<d\mathcal{L}_{\check{C}}/dX\right>}{L_{\mathrm{LS}}}=629\pm47\ \mbox{p.e./cm},
\end{equation}
where \mbox{$d\mathcal{L}_{S}/dX$}\ includes the Cherenkov light created in the LS.
The muons in Fig.~\ref{figure:ChargeVersusLength} generating light yields above the baseline linear trend are likely to involve secondary particles.
We define an excess light yield parameter $\Delta\mathcal{L}$,
\begin{equation}
\label{equation:ExcessLightYield}
\Delta\mathcal{L}=\mathcal{L}_{\mathrm{ID}}-L_{\mathrm{BO}}\left<\frac{d\mathcal{L}_{\check{C}}}{dX}\right>-L_{\mathrm{LS}}\left<\frac{d\mathcal{L}_{S}}{dX}\right>,
\end{equation}
for the purpose of describing \textit{showering} muons associated with secondary particles.

\begin{figure}[t]
\begin{center}
\includegraphics[angle=\DefaultFigureAngle,width=\columnwidth]{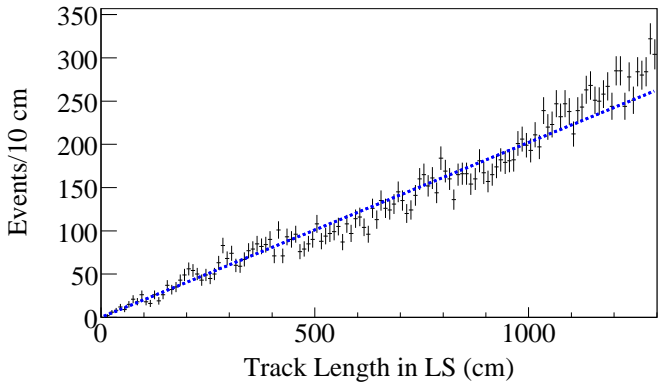}
\end{center}
\caption[muon reconstruct]{\label{figure:MuonTrackLength}
Distribution of muon track lengths in the KamLAND LS.
The dashed blue line shows the expected linear distribution for a sphere of nominal volume,
and the average track length in this case is 867~cm.
The average muon track length measured in the detector is 878 cm,
in agreement with the value \mbox{$874\pm13$\ cm} calculated using the proper LS shape and muon angular distributions.}
\end{figure}

The LS muon rate is estimated by selecting muons with
\mbox{$\mathcal{L}_{\mathrm{ID}}>4\times10^{4}$\ p.e.}\ and impact parameter \mbox{$<650$\ cm}.
The light yield cut has a negligible inefficiency for LS muons,
while the impact parameter cut eliminates LS muons that reconstruct outside the balloon because of the resolution of the fitter algorithm.
This provides the lower limit on the LS muon rate.
An upper limit is established by removing the impact parameter cut and
increasing the $\mathcal{L}_{\mathrm{ID}}$ cut to \mbox{$>10^{5}$\ p.e.}\ to eliminate muons that pass through the BO without transversing the LS.
This cut again has a small inefficiency for LS muons but does not eliminate muons which shower in the BO.
Averaging the two measurements yields \mbox{$R_{\mu}=0.198\pm0.014$\ Hz},
where the error estimate is dominated by systematic uncertainties due to temporal variations,
the difference between the two measurements,
and the presence of muon bundles,
in which multiple simultaneous muons may be tagged as single-muon events.
Using the parametrization of Ref.~\cite{Becherini2006}, the effect of muon bundles was estimated to be $<$5\%.
The muon rate corresponds to an integrated muon intensity of \mbox{$J_{\mu}=5.37\pm0.41$\ m$^{-2}$\ h$^{-1}$},
where the error includes uncertainties in the shape of the LS volume but is dominated by the uncertainty in the muon rate.
Using a cut of \mbox{$\Delta\mathcal{L}>10^{6}$\ p.e.}, which is equivalent to a \mbox{$\sim$3\ GeV} threshold,
the rate of showering muons in the LS is \mbox{$\sim$0.03\ Hz}.
It is possible that some atmospheric neutrino interactions leak into the LS muon sample.
However,
an estimate with the neutrino flux from Ref.~\cite{Honda2007} and
cross sections from Ref.~\cite{Casper2002} gives less than \mbox{$4\times10^{-5}$\ Hz} from atmospheric neutrinos.
The muon track length distribution is shown in Fig.~\ref{figure:MuonTrackLength}.
The measured average track length is \mbox{$L_{\mu}=878$\ cm},
in agreement with the calculated value of \mbox{$L_{\mu}=874\pm13$\ cm}\ where the nonspherical corrections to the balloon shape and
the muon angular distributions were taken into account.

The production yield of radioactive isotopes from muon-initiated spallation is energy dependent.
KamLAND does not measure the muon energy, so it is estimated from simulation.
There are previous estimates of the mean energy $\overline{E}_{\mu}$\ ranging from
198~GeV~\cite{Mei2006} to 285~GeV~\cite{Hagner2000,Galbiati2005b} at the Kamiokande, KamLAND, and SuperKamiokande detectors.
The authors of Ref.~\cite{Tang2006} made a detailed determination of $J_{\mu}$\ and $\overline{E}_{\mu}$\ using the
muon simulation code (\textsc{music})~\cite{Antonioli1997} to transport muons,
generated according to the \textit{modified Gaisser parameterization}~\cite{PDG2004} sea-level muon flux distribution,
through a digital profile of Ikenoyama.
Although this calculation reproduces the zenith and azimuthal angular distributions observed by KamLAND,
as shown in Fig.~10 of Ref.~\cite{Tang2006},
it overestimates $J_{\mu}$\ by 14\% relative to this work.
The calculation assumed a homogeneous rock entirely of the \textit{Inishi} type
(see Table~II of Ref.~\cite{Tang2006} for the chemical composition),
but other rock types, such as granite, limestone, and several types of metamorphic rock,
are common in Ikenoyama in unknown quantities~\cite{Kamioka1977}.
Table~\ref{table:MuonSimulationResults} shows the result of calculations of $J_{\mu}$\ and $\overline{E}_{\mu}$\ using the same simulation method of Ref.~\cite{Tang2006}
but for standard rock~\cite{Groom2001,Barrett1952} and \textit{generic skarn},
a generic mixture of rock types found in a skarn-type mine like that of Ikenoyama.
Here, generic skarn is defined to be 70\% granite and 30\% calcite by weight.
The $J_{\mu}$\ values range from \mbox{4.90\ m$^{-2}$\ h$^{-1}$},
for \mbox{$2.75$\ g/cm$^{3}$}\ specific gravity generic skarn,
to \mbox{6.71\ m$^{-2}$\ h$^{-1}$},
for \mbox{$2.65$\ g/cm$^{3}$}\ specific gravity Inishi rock,
while $\overline{E}_{\mu}$\ varies from 254~GeV to 268~GeV.
The value of $J_{\mu}$\ for \mbox{$2.70$\ g/cm$^{3}$}\ specific gravity standard rock is \mbox{5.38\ m$^{-2}$\ h$^{-1}$},
in excellent agreement with our measured value.
The value of $\overline{E}_{\mu}$\ for this rock is 259~GeV.
We take \mbox{$\overline{E}_{\mu}=260\pm8$\ GeV},
where the uncertainty is chosen to cover the full range for the various rock types.

\begin{table}[t]
\caption{\label{table:MuonSimulationResults}Calculated $J_{\mu}$\ and $\overline{E}_{\mu}$\ values for various rock types using
the simulation method from Ref.~\cite{Tang2006}.
The range in values for $J_{\mu}$\ and $\overline{E}_{\mu}$\ corresponds to varying the specific gravity of the rock from 2.65 to 2.75 \mbox{g$/$cm$^{3}$}.
The generic skarn is defined to be 70\% granite and 30\% calcite (by weight) with the following properties:
\mbox{$<$Z$>=10.22$},
\mbox{$<$A$>=20.55$\ a.u.},
and radiation length \mbox{$X_{0}=25.411$\ g$/$cm$^{2}$}.
}
\begin{ruledtabular}
\begin{tabular}{lcc}
Ikenoyama rock model & $J_{\mu}$\ (m$^{-2}$\ h$^{-1}$) & $\overline{E}_{\mu}$\ (GeV) \\
\hline
Inishi rock & 5.66-6.71 & 262-268 \\
Standard rock & 4.95-5.83 & 256-262 \\
Generic skarn & 4.90-5.82 & 254-260 \\
\hline
This measurement & $5.37\pm0.41$ & --- \\
\end{tabular}
\end{ruledtabular}
\end{table}

\section{Spallation Neutron Yield}
\label{section:SpallationNeutronYield}

Most of the neutrons produced in the KamLAND LS are captured on hydrogen or carbon atoms.
The capture cross section varies inversely with respect to velocity,
and the mean neutron capture time $\tau_{n}$\ is constant with respect to energy.
The capture time ($t$) distribution is exponential, \mbox{$P(t)\propto e^{-t/\tau_{n}}$}.
A calculation using the elemental composition of the KamLAND LS given in Table~\ref{table:LiquidScintillator}
and the thermal neutron capture cross sections from Ref.~\cite{Mughabghab1981} gives \mbox{$\tau_{n}=206\ \mu$s}.
This calculation indicates that 99.5\% of the neutrons are captured on $^{1}$H,
while the remainder are captured mostly on $^{12}$C.
The probability for capture on the other isotopes in the LS, such as $^{13}$C, is \mbox{$2\times10^{-4}$}\ or less.

\begin{figure}[t]
\begin{center}
\includegraphics[angle=\DefaultFigureAngle,width=\columnwidth]{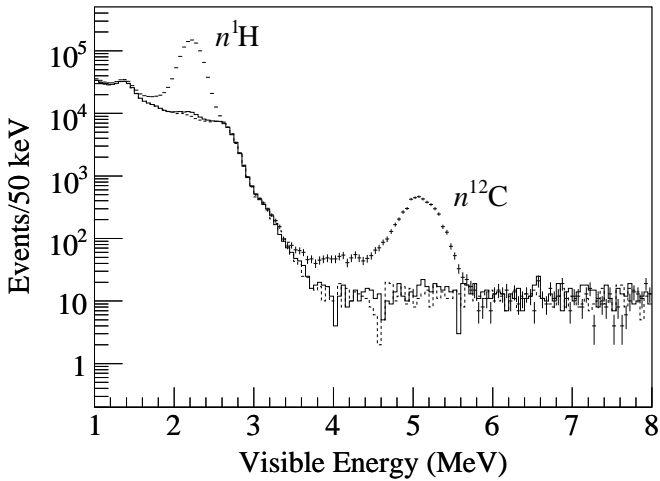}
\end{center}
\caption[spallation Neutron Simple]{\label{figure:NeutronVisibleEnergy} 
Visible energy spectra of the events following muons in the signal time window \mbox{$150\leq\Delta T<1000 \mu$s} (crosses)
compared with background in the time window \mbox{$4150\leq\Delta T<5000 \mu$s} (dashes). 
The low-energy tails in the $n$$^{1}$H and $n$$^{12}$C peaks are caused by the electronics effect discussed in the text.
In the time window \mbox{$1300\leq\Delta T<2150 \mu$s} (solid) there is a small $n$$^{1}$H peak without the low-energy tail.}
\end{figure}

\begin{figure}[t]
\begin{center}
\includegraphics[angle=\DefaultFigureAngle,width=\columnwidth]{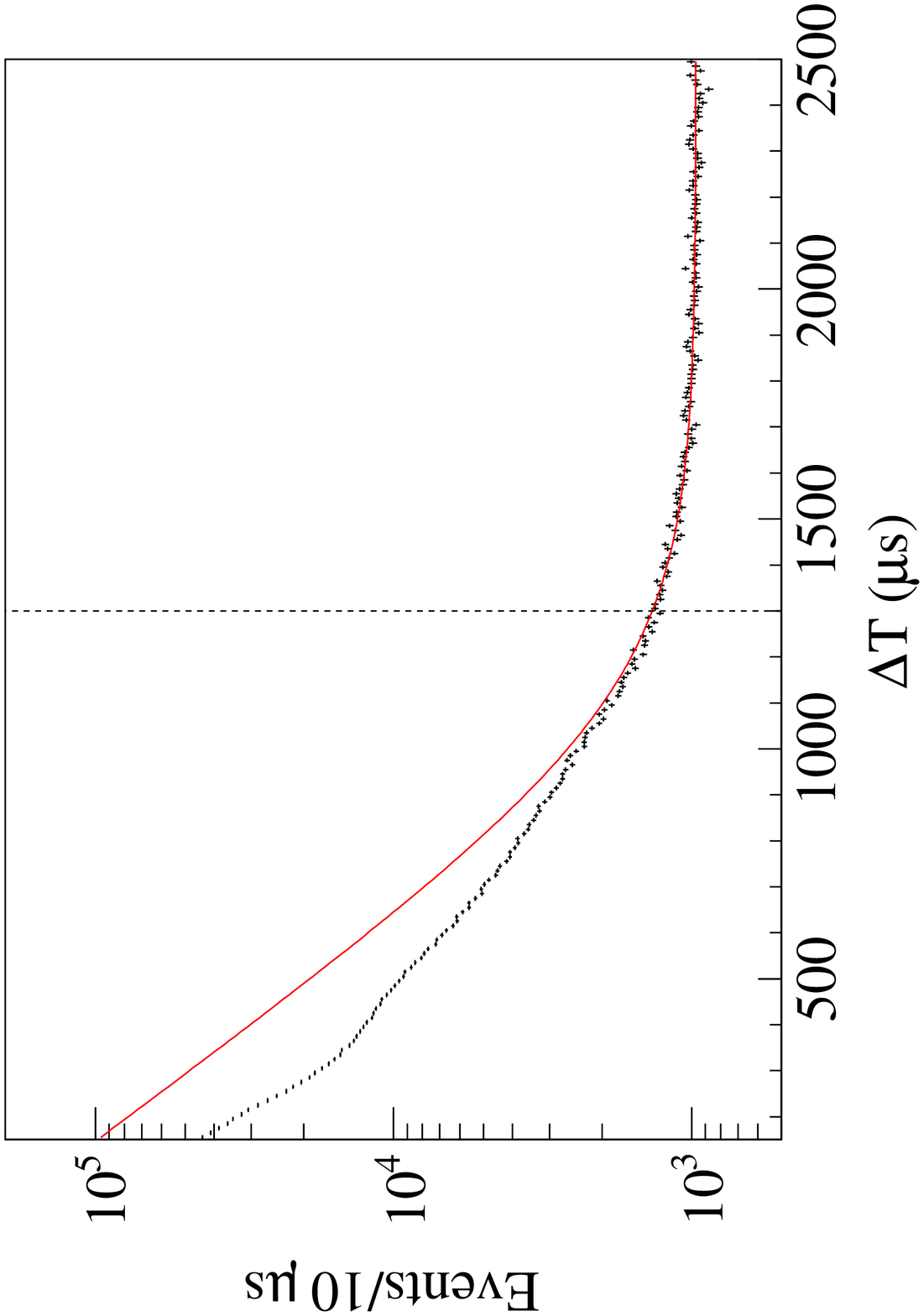}
\end{center}
\caption[Neutron Capture Time since Muon]{\label{figure:NeutronCaptureTimeSinceMuon}
Time difference between a muon and the neutron capture event, with the cuts \mbox{$1.8<E_{\mathrm{vis}}<$~2.6 MeV}.
The red line shows the fit restricted to the region \mbox{$\Delta T\geq1300 \mu$s}.
The fit results in \mbox{$N_{n}=(4.2\pm 0.3) \times 10^{6}$} with \mbox{$\chi^{2}/$d.o.f.$=98/118$}.
}
\end{figure}

Neutrons produced by muon-initiated spallation in the LS can be identified by the characteristic capture $\gamma$\ rays.
Figure~\ref{figure:NeutronVisibleEnergy} shows the $E_{\mathrm{vis}}$\ distributions in signal \mbox{($150\leq\Delta T<1000 \mu$s)}
and background \mbox{($4150\leq\Delta T<5000 \mu$s)} coincidence windows following some muons,
where \mbox{$\Delta T\equiv(t-t_{\mu})$} is the time elapsed since the muon's passage.
The $E_{\mathrm{vis}}$\ distribution clearly shows peaks from neutron captures on $^{1}$H (2.225~MeV) and $^{12}$C (4.9~MeV),
which are not evident in the background.
Figure~\ref{figure:NeutronCaptureTimeSinceMuon} shows the $\Delta T$\ distribution for events within the LS volume and with \mbox{$1.8<E_{\mathrm{vis}}<$~2.6 MeV},
which includes the single 2.225-MeV $\gamma$ ray emitted by neutron capture on $^{1}$H.
For \mbox{$\Delta T<1000 \mu$s},
there is a clear deviation from the exponential distribution 
due to the overload that large muon signals produce on individual electronics channels
and to the dead time in the system arising from the very high event multiplicity following the muon.
Both effects intervene in events that are quite different from those that KamLAND was designed to record.

The number of neutrons produced by muon-initiated spallation in the LS
is established by a binned maximum likelihood fit~\cite{Baker1984} to the data in 
Fig.~\ref{figure:NeutronCaptureTimeSinceMuon} using the function
\begin{equation}
\label{equation:NeutronCaptureTime}
r(t)=\frac{N_{n}}{\tau_{n}}e^{-(t-t_{\mu})/\tau_{n}}+r_{B},
\end{equation}
where $N_{n}$\ is the total number of neutron captures associated with the selected events,
$\tau_{n}$\ is the mean neutron capture time,
and $r_{B}$\ is the background rate,
which is assumed to be approximately constant in the region of interest \mbox{($\Delta T<2500 \mu$s)} because of the low muon rate \mbox{($\sim0.2$\ Hz)}.
To avoid the electronics-induced distortions,
the fit is restricted to the region \mbox{$\Delta T\geq1300 \mu$s}.
The parameters $N_{n}$\ and $r_{B}$\ are free,
but the mean capture time is constrained to \mbox{$\tau_{n}=207.5\pm2.8 \mu$s}\ with a Gaussian \textit{penalty} function.
This mean capture time is determined from two independent measurements of $\tau_{n}$:
a \mbox{$^{241}$Am+$^{9}$Be} calibration source,
and an analysis of a sample of neutrons generated by \textit{clean} muons.
These clean muons are identified by a multiplicity parameter $\eta_{T}$\ defined to be the number of trigger commands that follow the muon within a \mbox{10-ms} period.
Fits to the $\Delta T$\ distribution of the subset of neutrons selected with various limits on
$\eta_{T}$\ demonstrate that muons with \mbox{$\eta_{T}<10$}\ give unbiased fit residuals.
The value of $\tau_{n}$\ is \mbox{$207.5\pm0.3 \mu$s}\ from the clean muon sample
and \mbox{$205.2\pm 0.5 \mu$s} from the \mbox{$^{241}$Am+$^{9}$Be} source data. 
The observed \mbox{2.3 $\mu$s}\ discrepancy between these values is not completely understood
but is suspected to be caused by neutrons from the \mbox{$^{241}$Am+$^{9}$Be} source that are captured on the stainless-steel source capsule.
In this analysis,
we use the value \mbox{$\tau_{n}=207.5 \mu$s}\ from the nonshowering muon sample
with an uncertainty of \mbox{$\pm2.8 \mu$s}\ covering both measurements.
The fit shown in Fig.~\ref{figure:NeutronCaptureTimeSinceMuon} in the region
\mbox{$\Delta T\geq1300 \mu$s}\ results in \mbox{$N_{n}=(4.2\pm 0.3) \times 10^{6}$} and \mbox{$\chi^{2}/$d.o.f.$=98/118$}.

\begin{table}[t]
\caption{\label{table:NeutronDetectionEfficiency}
Summary of the dominant contributions to the neutron detection efficiency.
}
\begin{ruledtabular}
\begin{tabular}{lc}
Effect & Value \\
\hline
Neutron-eliminating reactions, e.g.,\ $(n,p)$ & $(96.3\pm3.7)\%$ \\
Neutron captures on $^{1}$H & $(99.5\pm0.1)\%$ \\
LS-BO boundary & $(93.3\pm 2.0)\%$ \\
Electronics dead time effects & $>98\%$ \\
\hline
Combined efficiency & $(89.4\pm3.8)\%$ \\
\end{tabular}
\end{ruledtabular}
\end{table}

The actual number of neutrons $\mathcal{N}_{n}$\ produced by muon-initiated spallation is related to the fit result $N_{n}$\ by an
efficiency $\epsilon_{n}$,
\begin{equation}
\label{equation:NumberOfNeutrons}
\mathcal{N}_{n}=\frac{N_{n}}{\epsilon_{n}},
\end{equation}
which accounts for other neutron-eliminating nuclear reactions such as \mbox{$^{12}$C$(n,p)$}\ and a net neutron loss at the LS-BO boundary.
This efficiency is calculated using the \textsc{music}-based muon simulation described in Sec.~\ref{section:CosmicRayMuons}
and the  \textsc{geant4}-based Monte Carlo of KamLAND described in Sec.~\ref{subsection:Geant4Simulation}.
The muon simulation generates muons with a three-momentum distribution appropriate to the KamLAND site inside Ikenoyama.
Muon transport is then modeled through KamLAND.
In this simulation neutrons are created and destroyed;
neutrons that survive to thermalization are tracked until they are captured.
Energy scale nonlinearities and the finite $\gamma$ ray resolution are included.
Two competing effects of the LS-BO boundary are taken into account.
The first is for neutrons produced by muons in the LS that leak out, leading to an under counting of neutrons.
The second is for neutrons produced outside of the LS that leak in, primarily from the BO, leading to an over count.
The leak-out fraction \mbox{[$(11.7\pm1.9)\%$]} is larger than the leak-in fraction \mbox{[$(5.4\pm0.4)\%$]} resulting in a net \mbox{$(6.7\pm2.0)\%$}\ correction.
The component of the efficiency related to the electronics dead time for high-multiplicity events is measured by comparing
the number of recorded waveforms with the number of trigger commands.
The correction is measured to be less than 2\% for \mbox{$\Delta T\geq1300 \mu$s}.
In counting neutrons, we use the definition by which
the \mbox{$(n,2n)$}\ reaction generates one new neutron and the \mbox{$(n,n')$}\ reaction generates no new neutrons.
This method gives an efficiency \mbox{$\epsilon_{n}=(89.4\pm3.8)\%$} (broken down in Table~\ref{table:NeutronDetectionEfficiency}).

Using Eq.~(\ref{equation:NumberOfNeutrons}) we find \mbox{$\mathcal{N}_{n}=(4.7\pm 0.4)\times 10^{6}$},
which is then used to extract the neutron production yield:
\begin{equation}
\label{equation:NeutronYieldByLength}
Y_{n}=\frac{\mathcal{N}_{n}}{R_{\mu}T_{L}\rho L_{\mu}},
\end{equation}
where \mbox{$R_{\mu}=0.198\pm0.014$\ Hz}\ is the measured rate of LS muons,
\mbox{$T_{L}=1.24 \times10^{8}$\ s}\ is the detector live time,
\mbox{$\rho=0.780\pm0.001$\ g/cm$^{3}$}\ is the density of the LS,
and \mbox{$L_{\mu}=874\pm13$\ cm}\ is the calculated mean muon track length.
The resulting neutron production yield is \mbox{$Y_{n}=(2.8\pm0.3)\times10^{-4}\ \mu^{-1}$g$^{-1}$cm$^{2}$},
with \mbox{$(64\pm5)\%$}\ of the neutrons produced by events classified as showering muons.

\section{Spallation Isotope Yield}
\label{section:SpallationIsotopeYield}

The method for determining the yields of spallation-generated isotopes is similar to the neutron analysis described in Sec.~\ref{section:SpallationNeutronYield}.
Spallation-generated isotopes are identified by their decay time relative to their creation and by their decay energy.
The decay time,
\mbox{$\Delta T\equiv t-t_{\mu}$},
is calculated for each event relative to all previous muons.
Usually,
several different isotopes decay in a given time window.
The number of each isotope produced,
$\mathcal{N}_{i}$,
is obtained from a binned maximum likelihood fit~\cite{Baker1984} to the $\Delta T$\ distribution,
using the function
\begin{equation}
\label{equation:IsotopeDeltaTime}
r(t)=\sum_{i}\frac{N_{i}}{\tau_{i}}e^{-(t-t_{\mu})/\tau_{i}}+r_{B},
\end{equation}
where
$N_{i}$\ is related to $\mathcal{N}_{i}$\ by an event selection efficiency (\mbox{$\mathcal{N}_{i}=N_{i}/\epsilon_{i}$}),
$\tau_{i}$\ is the mean lifetime,
and $r_{B}$\ is a constant rate of uncorrelated background events that are in accidental coincidence with muons.

In terms of $\mathcal{N}_{i}$, the spallation production yield for isotope $i$\ is equal to
\begin{equation}
\label{equation:IsotopeProductionYield}
Y_{i}=\frac{\mathcal{N}_{i}}{R_{\mu}T_{L}\rho L_{\mu}},
\end{equation}
where $R_{\mu}$,
$T_{L}$,
$\rho$,
and $L_{\mu}$\ are defined as in Eq.~(\ref{equation:NeutronYieldByLength}).
The spallation production rate for isotope $i$\ is equal to
\begin{equation}
\label{equation:IsotopeProductionRate}
R_{i}=\frac{\mathcal{N}_{i}}{\rho V_{T} T_{L}},
\end{equation}
where \mbox{$V_{T}=1171\pm25$\ m$^{3}$}\ is the target volume.

\subsection{$^{12}$B \textit{and} $^{12}$N}
\label{subsection:12B_12N_Yield}

\begin{figure}[t]
\begin{center}
\includegraphics[angle=\DefaultFigureAngle,width=\columnwidth]{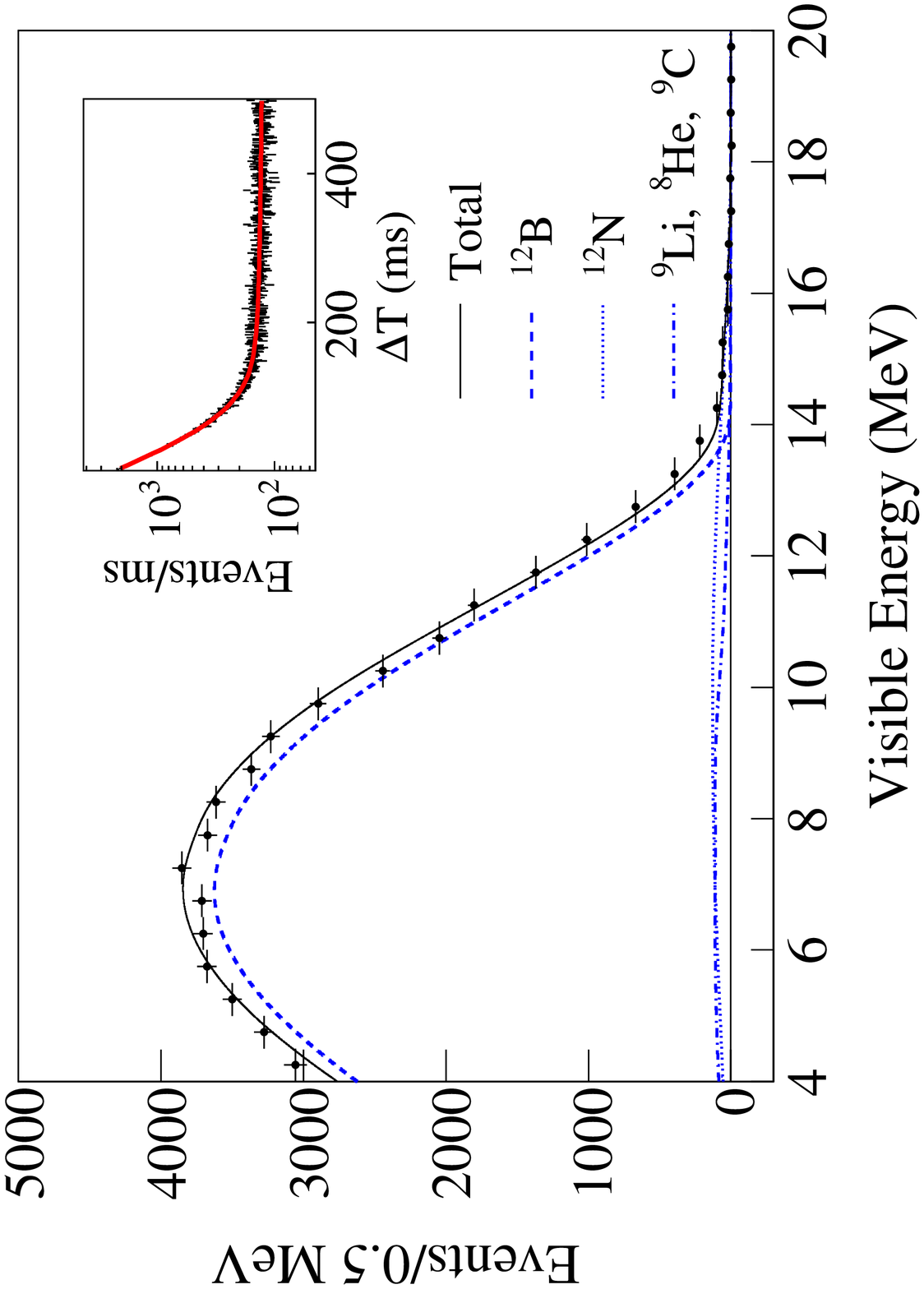}
\end{center}
\caption[spallation time correlation 20sec]{\label{figure:12B_12N_Summary}
Background-subtracted $E_{\mathrm{vis}}$\ spectrum above 4~MeV.
Signal and background events are taken from \mbox{$2\leq\Delta T<60$\ ms} and \mbox{$502\leq\Delta T<560$\ ms},
respectively.
The production rate of $^{12}$B (\mbox{$\tau=29.1$\ ms}, \mbox{$Q=13.4$\ MeV}) is estimated to be \mbox{$54.8\pm1.5$}\ \mbox{kton$^{-1}$day$^{-1}$},
from fitting Eq.~(\ref{equation:IsotopeDeltaTime}) to the $\Delta T$\ distribution shown in the inset.
This fit has a \mbox{$\chi^{2}/$d.o.f.$=509/495$}.
The production rate $^{12}$N (\mbox{$\tau=15.9$\ ms}, \mbox{$Q=17.3$\ MeV}) is estimated to be \mbox{$2.2\pm0.5$}\ \mbox{kton$^{-1}$day$^{-1}$},
from a similar fit to the $\Delta T$\ distribution for events with \mbox{$14\leq E_{\mathrm{vis}}<20$\ MeV},
where the higher $E_{\mathrm{vis}}$\ threshold is imposed to exclude $^{12}$B.
}
\end{figure}

$^{12}$B (\mbox{$\tau=29.1$\ ms}, \mbox{$Q=13.4$\ MeV})~\cite{AjzenbergSelove1990} $\beta^{-}$\ decay
and 
$^{12}$N (\mbox{$\tau=15.9$\ ms}, \mbox{$Q=17.3$\ MeV})~\cite{AjzenbergSelove1990} $\beta^{+}$\ decay
candidate events are selected via cuts on $E_{\mathrm{vis}}$\ and $\Delta T$.
The inset in Fig.~\ref{figure:12B_12N_Summary} shows the distribution of $\Delta T$\ for events with 
\mbox{$4\leq E_{\mathrm{vis}}<20$\ MeV}\ and \mbox{$2\leq\Delta T<500$\ ms}.
A binned maximum likelihood fit to the $\Delta T$\ distribution using Eq.~(\ref{equation:IsotopeDeltaTime}),
and including the long-lived isotopes $^{8}$He, $^{9}$Li, $^{9}$C, and $^{12}$N as contaminants,
yields \mbox{$N(^{12}$B$)=(5.94\pm 0.10)\times 10^{4}$}\ events with \mbox{$\chi^{2}/$d.o.f.$=509/495$}.
Longer lived isotopes give roughly constant decay rates on this time scale and fit out as a component of $r_B$.
A similar fit to the $\Delta T$\ distribution for events with \mbox{$14\leq E_{\mathrm{vis}}<20$\ MeV},
where the higher $E_{\mathrm{vis}}$\ threshold is imposed to exclude $^{12}$B,
gives \mbox{$N(^{12}$N$)=(2.8\pm0.3) \times 10^{2}$}\ events.
A comparison with the predicted $E_{\mathrm{vis}}$\ spectrum is shown in Fig.~\ref{figure:12B_12N_Summary}.
The $E_{\mathrm{vis}}$\ spectra are predicted from the allowed $^{12}$B and $^{12}$N $\beta^{\pm}$-decay spectra,
taking into account the KamLAND detector response,
normalized to the observed $N(^{12}$B$)$\ and $N(^{12}$N$)$.
The detector response model includes the energy nonlinearities and boundary effects described in Sec.~\ref{section:DetectorDescriptionAndPerformance}.

The inefficiency in identifying $^{12}$B and $^{12}$N candidates is dominated by the $E_{\mathrm{vis}}$\ cut.
The efficiencies calculated by integrating the predicted $E_{\mathrm{vis}}$\ spectra over the selection window
give \mbox{$\epsilon(^{12}$B$)=(82.9\pm0.7)\%$}\ and \mbox{$\epsilon(^{12}$N$)=(9.3\pm1.6)\%$},
where the errors come from the uncertainty in the detector response.
Using Eq.~(\ref{equation:IsotopeProductionYield}), the resultant isotope production yields are calculated to be
\mbox{$Y(^{12}$B$)=(42.9\pm3.3)\times 10^{-7}$}\ and
\mbox{$Y(^{12}$N$)=(1.8\pm0.4)\times 10^{-7}$} \mbox{$\mu^{-1}$g$^{-1}$cm$^{2}$}. 
The production rates are
\mbox{$R(^{12}$B$)=54.8\pm1.5$}\ and
\mbox{$R(^{12}$N$)=2.2\pm0.5$} \mbox{kton$^{-1}$day$^{-1}$}.

\subsection{$^{8}$Li \textit{and} $^{8}$B}
\label{subsection:8Li_8B_Yield}

\begin{figure}[t]
\begin{center}
\includegraphics[angle=\DefaultFigureAngle,width=\columnwidth]{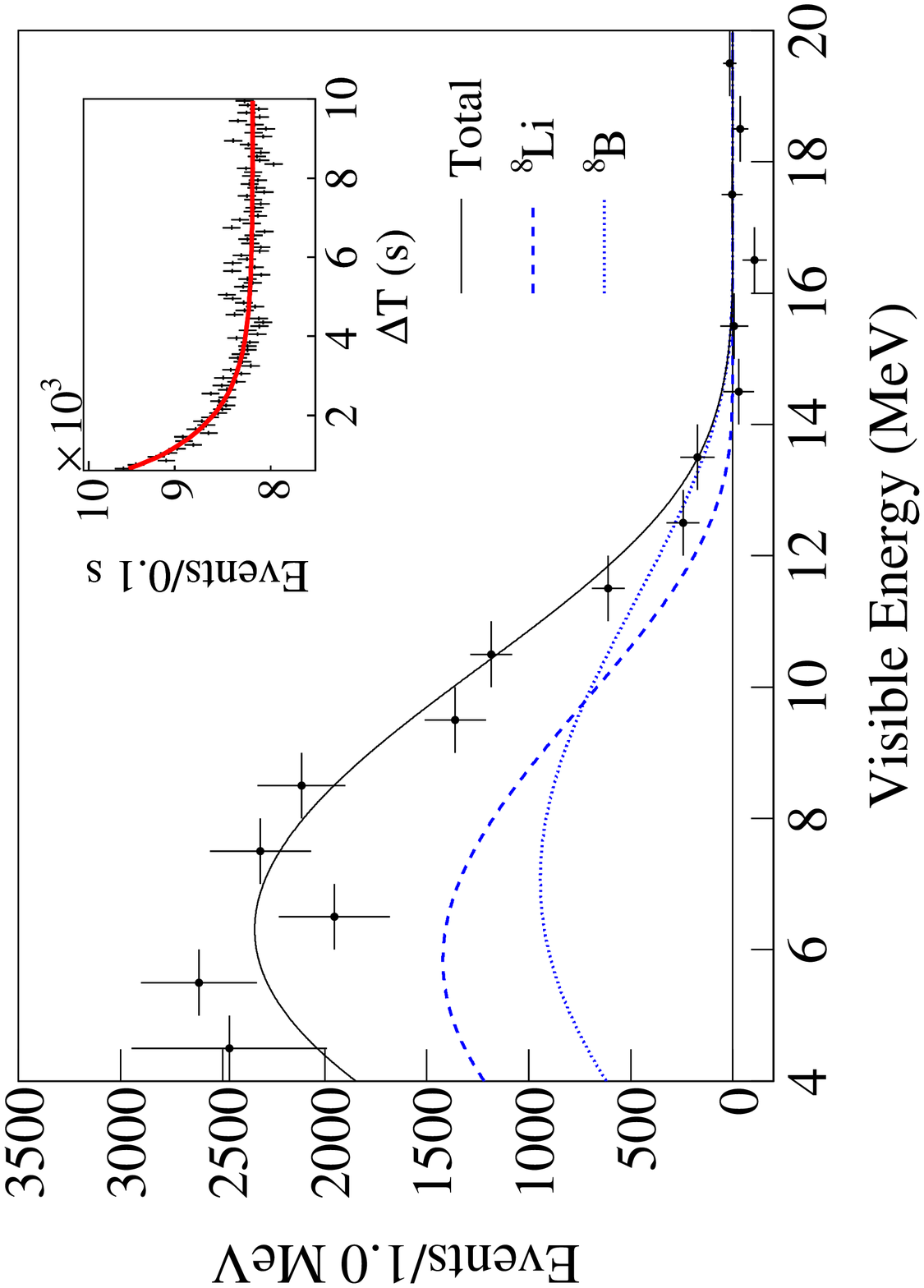}
\end{center}
\caption[spallation time correlation 20sec Without B12]{\label{figure:8B_8Li_9C_Summary}
Background-subtracted $E_{\mathrm{vis}}$\ spectrum above 4~MeV,
where signal and background events are taken from
\mbox{$0.6\leq\Delta T<4.0$\ s} and \mbox{$10.6\leq\Delta T<14.0$\ s},
respectively.
The production rates of $^{8}$Li (\mbox{$\tau=1.21$\ s}, \mbox{$Q=16.0$\ MeV}) and
$^{8}$B (\mbox{$\tau=1.11$\ s}, \mbox{$Q=18.0$\ MeV}) 
are estimated to be \mbox{$15.6\pm3.2$}\ and \mbox{$10.7\pm2.9$}\ \mbox{kton$^{-1}$day$^{-1}$},
respectively,
from simultaneously fitting the $E_{\mathrm{vis}}$\ spectrum and the $\Delta T$\ distribution shown in the inset.
The fit to the $\Delta T$\ distribution has a \mbox{$\chi^{2}/$d.o.f.$=95/91$}.
}
\end{figure}

$^{8}$Li\ (\mbox{$\tau=1.21$\ s}, \mbox{$Q=16.0$\ MeV})~\cite{Tilley2004} $\beta^{-}$\ decay and
$^{8}$B (\mbox{$\tau=1.11$\ s}, \mbox{$Q=18.0$\ MeV})~\cite{Tilley2004} $\beta^{+}$\ decay
candidate events are selected according to $E_{\mathrm{vis}}$\ and $\Delta T$.
The inset in Fig.~\ref{figure:8B_8Li_9C_Summary} shows the distribution of $\Delta T$\ for all preceding muons for events with 
\mbox{$4\leq E_{\mathrm{vis}}<20$\ MeV} and \mbox{$0.6\leq\Delta T<10$\ s},
where the lower limit on $\Delta T$\ is chosen to exclude $^{9}$C, $^8$He, $^9$Li, and other isotopes with shorter lifetimes.
Isotopes with longer lifetimes are roughly constant over the selected range of $\Delta T$.
To avoid a large accidental coincidence background from uncorrelated muons,
all showering muons and a portion of nonshowering muons whose track is within 3~m of a $^{8}$Li or $^{8}$B candidate are selected.
Figure~\ref{figure:8B_8Li_9C_Summary} shows the $E_{\mathrm{vis}}$\ distribution of $^{8}$Li and $^{8}$B candidate events
with \mbox{$0.6\leq\Delta T<4.0$\ s}\ after subtracting background estimated from the range \mbox{$10.6\leq\Delta T<14.0$\ s}.
$N(^{8}$Li$)$\ and $N(^{8}$B$)$\ are determined from a simultaneous binned maximum likelihood fit to
the $\Delta T$\ distribution and a chi-square fit to the $E_{\mathrm{vis}}$\ distribution.
The expected $E_{\mathrm{vis}}$\ spectra for KamLAND are calculated by convolving the $\beta^{\pm}$\ spectra from Refs.~\cite{Winter2006,Bhattacharya2006} 
with KamLAND's detector response.
For the fit to the $E_{\mathrm{vis}}$\ distribution,
the energy scale parameters are constrained to an allowed region determined by a prior fit to $\gamma$ ray calibration data
(described in Sec.~\ref{section:DetectorDescriptionAndPerformance}) and the $^{12}$B $E_{\mathrm{vis}}$\ distribution in Fig.~\ref{figure:12B_12N_Summary}.

\begin{figure}[t]
\begin{center}
\includegraphics[angle=\DefaultFigureAngle,width=\columnwidth]{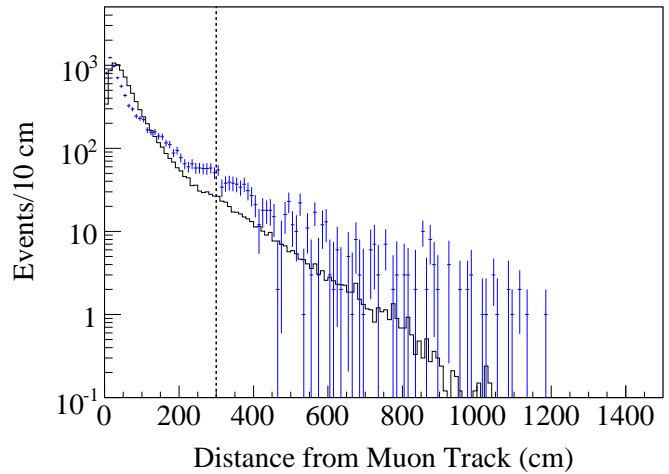}
\end{center}
\caption[track correlation of spallation products]{\label{figure:Spallation-dL-figure1-R5p5m}
Impact parameter distribution for events identified as $^{12}$B (blue points) and neutrons (black histogram) produced by muon-induced spallation.
Only nonshowering muons \mbox{($\Delta\mathcal{L}<10^{6}$\ p.e.)} were used to make this plot.
The efficiencies for the $\Delta L< 3$\ m track cut (represented by the vertical dashed line) are evaluated to be $(91.6\pm4.3)\%$ from $^{12}$B events,
and $(95.9\pm0.7)\%$ from neutron events.
}
\end{figure}

The efficiency is given by the product of $\epsilon_{E}$\ for the $E_{\mathrm{vis}}$\ selection and $\epsilon_{S}$\ for the \mbox{muon-$^{8}$Li} or
\mbox{muon-$^{8}$B} spatial correlation.
By integrating the expected spectra over \mbox{$4\leq E_{\mathrm{vis}}<20$\ MeV}, 
\mbox{$\epsilon_{E}(^{8}$Li$)$} and \mbox{$\epsilon_{E}(^{8}$B$)$}\ are estimated to be
\mbox{$(77.6\pm0.9)\%$}\ and \mbox{$(88.4\pm0.7)\%$}, respectively.
For showering muons, $\epsilon_{S}$\ is 100\% because no correlation requirement is imposed.
For nonshowering muons,
$\epsilon_{S}$\ is estimated from the $^{12}$B and $^{12}$N analysis (Sec.~\ref{subsection:12B_12N_Yield}),
and the systematic error arising from variations between isotopes is estimated with a \textsc{fluka} simulation.
Figure~\ref{figure:Spallation-dL-figure1-R5p5m} shows the impact parameter ($\Delta L$) distribution for the $^{12}$B and $^{12}$N candidate events for
nonshowering muons \mbox{($\Delta\mathcal{L}<10^{6}$\ p.e.)}.
We find that $(91.6\pm4.3)\%$\ of the candidates are within 3~m of the muon track.
This fraction is the value of $\epsilon_{S}(^{12}$B$)$\ for \mbox{$\Delta L<3$\ m}.

To obtain $\epsilon_{S}(^{8}$Li$)$\ and $\epsilon_{S}(^{8}$B$)$,
an additional correction for the difference between the \mbox{muon-$^{8}$Li} or \mbox{muon-$^{8}$B} and the \mbox{muon-$^{12}$B} spatial correlations is applied.
This correction is derived from the \textsc{fluka} simulation described in Sec.~\ref{subsection:FlukaSimulation}.
The simulation does not include the uncertainties in the muon track and the isotope decay vertex reconstruction;
it is only used to study the isotope dependance of $\epsilon_{S}$.
The range of values of $\epsilon_{S}$\ from \textsc{fluka} for different spallation isotopes is used to estimate the systematic error that should be added to
$\epsilon_{S}(^{12}$B$)$\ in order to obtain a common $\epsilon_{S}$\ for all spallation isotopes.
The resulting value,
\mbox{$\epsilon_{S}=(91.6\pm8.4)\%$},
is used for estimating the $^{8}$Li and $^{8}$B (and, later the $^{9}$C, $^{8}$He, and $^{9}$Li) yields.

\begin{figure}[t]
\begin{center}
\begin{minipage}[b]{.48\linewidth}
\vspace{1mm}
\includegraphics[angle=\DefaultFigureAngle,width=\columnwidth]{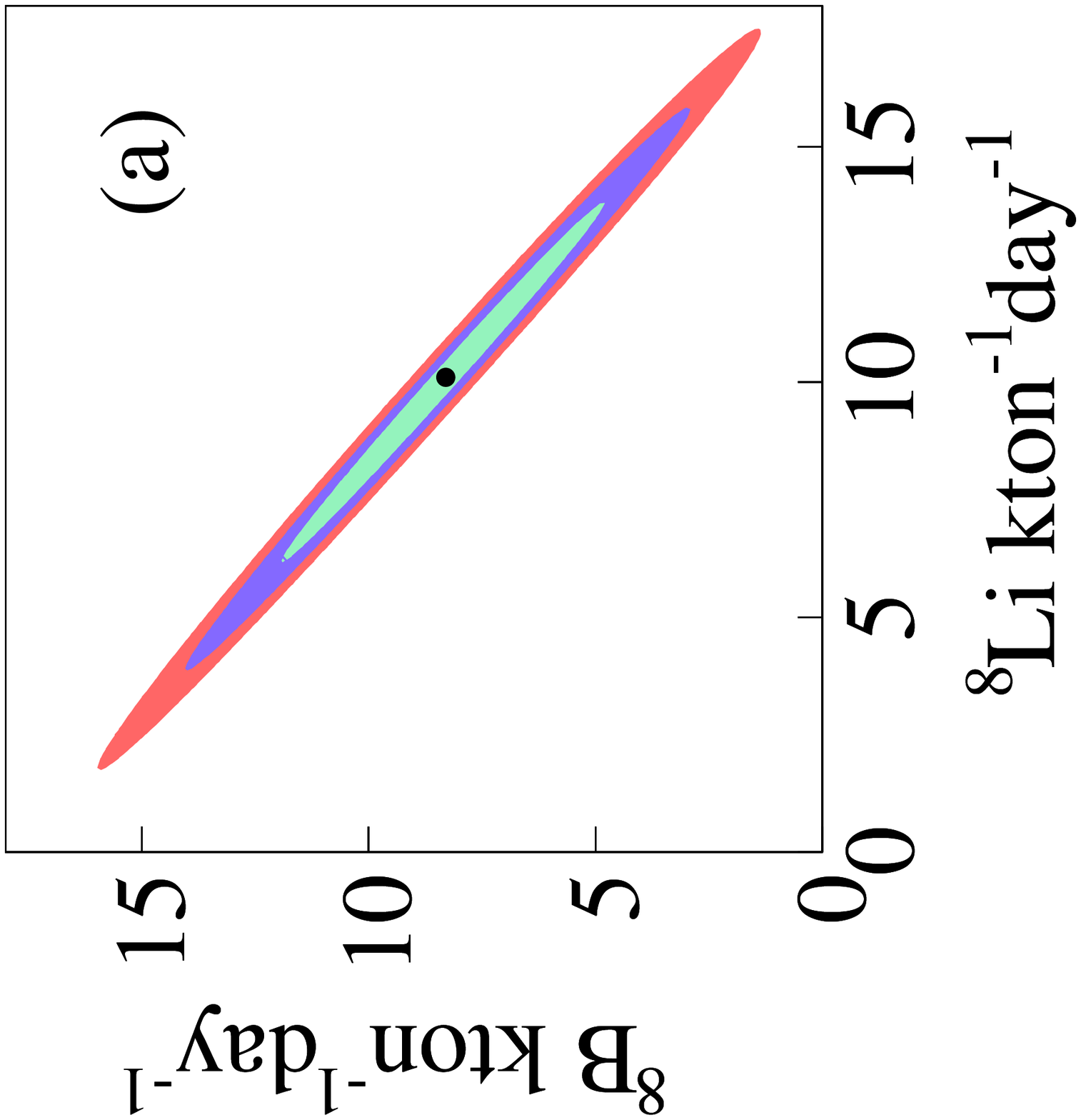}
\end{minipage}
\hspace{2mm}
\begin{minipage}[b]{.48\linewidth}
\vspace{1mm}
\includegraphics[angle=\DefaultFigureAngle,width=\columnwidth]{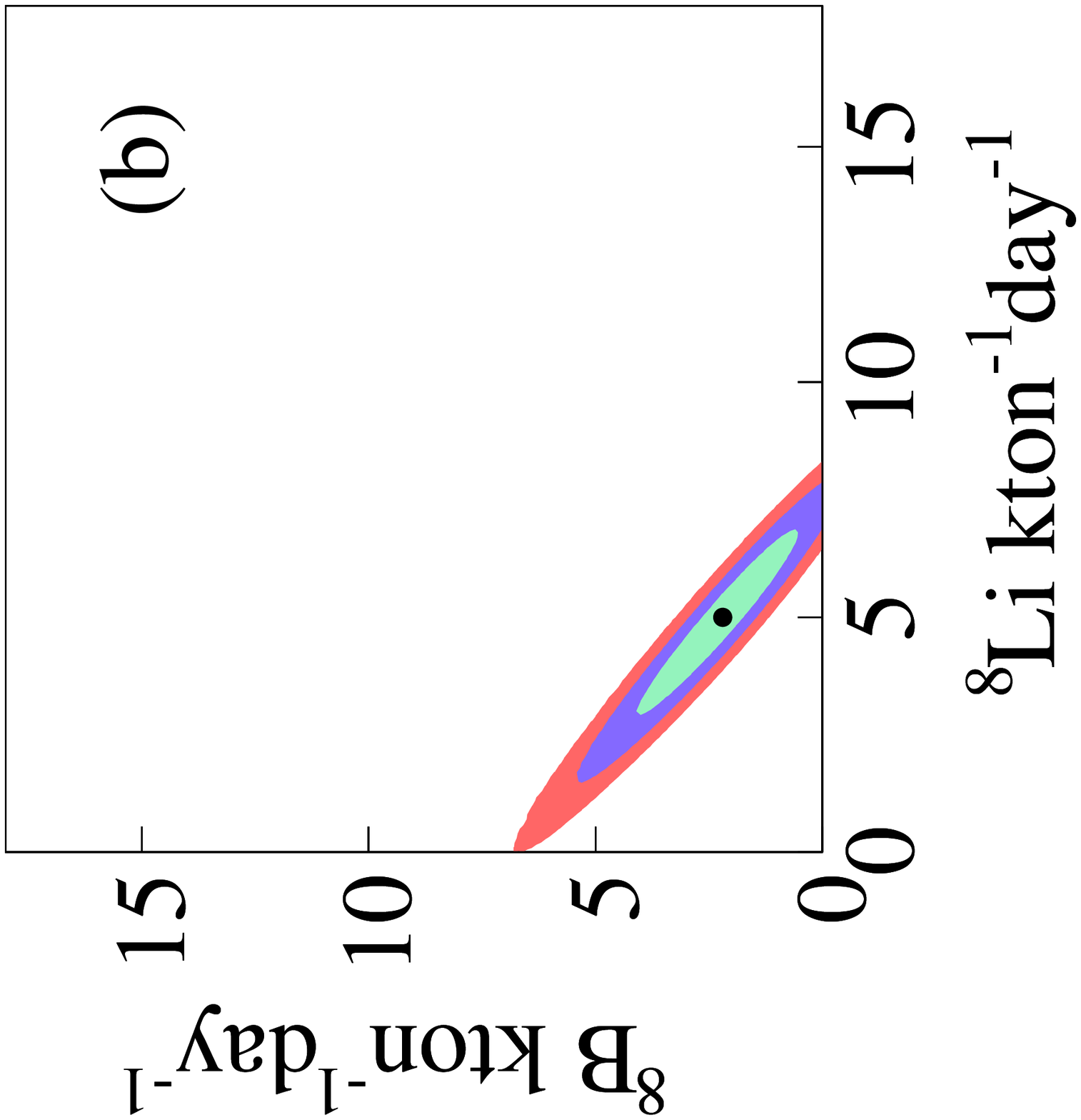}
\end{minipage}
\end{center}
\caption[chi2 B8 and Li8]{\label{figure:Chi2-from-CombinedSpectrum-B8-and-Li8-figure4}
Allowed regions for the production rates of $^{8}$Li and $^{8}$B from the combined fits of the energy spectra and
the $\Delta T$\ distributions at $1\sigma$, $2\sigma$, and $3\sigma$ C.L.\ for
(a) showering muons \mbox{($\Delta\mathcal{L}>10^{6}$\ p.e.)} and
(b) nonshowering muons \mbox{($\Delta\mathcal{L}<10^{6}$\ p.e.)} with $\Delta L< 3$\ m track cut. The black points indicate the best-fit parameters.
}
\end{figure}

Combining the above analyses, we obtain
$Y(^{8}$Li$)=(12.2\pm2.6)\times10^{-7}$\ and
$Y(^{8}$B$)=(8.4\pm2.4)\times10^{-7}$\ \mbox{$\mu^{-1}$g$^{-1}$cm$^{2}$}.
The isotope production rates are
$R(^{8}$Li$)=15.6\pm3.2$\ and
$R(^{8}$B$)=10.7\pm2.9$\ \mbox{kton$^{-1}$day$^{-1}$}.
The contour plots in Fig.~\ref{figure:Chi2-from-CombinedSpectrum-B8-and-Li8-figure4} show the correlation between $^{8}$Li and $^{8}$B.
Due to their similar lifetimes, $^{8}$Li and $^{8}$B are identified primarily by their energy spectra.

\subsection{$^{8}$He \textit{and} $^{9}$Li}
\label{subsection:8He_9Li_Yield}

$^{8}$He\ (\mbox{$\tau=171.7$\ ms}, \mbox{$Q=10.7$\ MeV})~\cite{Tilley2004} and
$^{9}$Li\ (\mbox{$\tau=257.2$\ ms}, \mbox{$Q=13.6$\ MeV})~\cite{Tilley2004} 
$\beta^{-}$-decay candidate events are selected according to the cuts \mbox{$1\leq E_{\mathrm{vis}}<13$\ MeV} and \mbox{$\Delta T<10$\ s}, 
and by the detection of a neutron following the $\beta^{-}$-decay event.
$^{8}$He\ decays to neutron-unstable excited states of $^{8}$Li\ with a \mbox{$(16\pm1)\%$}\ branching ratio~\cite{Tilley2004},
and $^{9}$Li\ decays to neutron-unstable excited states of $^{9}$Be with a \mbox{$(50.8\pm0.9)\%$}\ branching ratio~\cite{Tilley2004}.
The neutron is identified by the \mbox{2.225-MeV}\ $\gamma$ ray from radiative capture on $^{1}$H \mbox{($1.8\leq E_{\mathrm{vis}}<2.6$\ MeV)}.
The $\gamma$ ray is required to be within 200~cm and 1.0~ms of the $^{8}$He or $^{9}$Li\ $\beta^{-}$-decay candidate.
Finally,
the $^{8}$He-$^{9}$Li\ analysis is performed using a 5.5-m-radius spherical fiducial volume to reduce the number of accidental coincidences between the
$\beta^{-}$-decay candidate and external $\gamma$ ray backgrounds near the balloon.

\begin{figure}[t]
\begin{center}
\includegraphics[angle=\DefaultFigureAngle,width=\columnwidth]{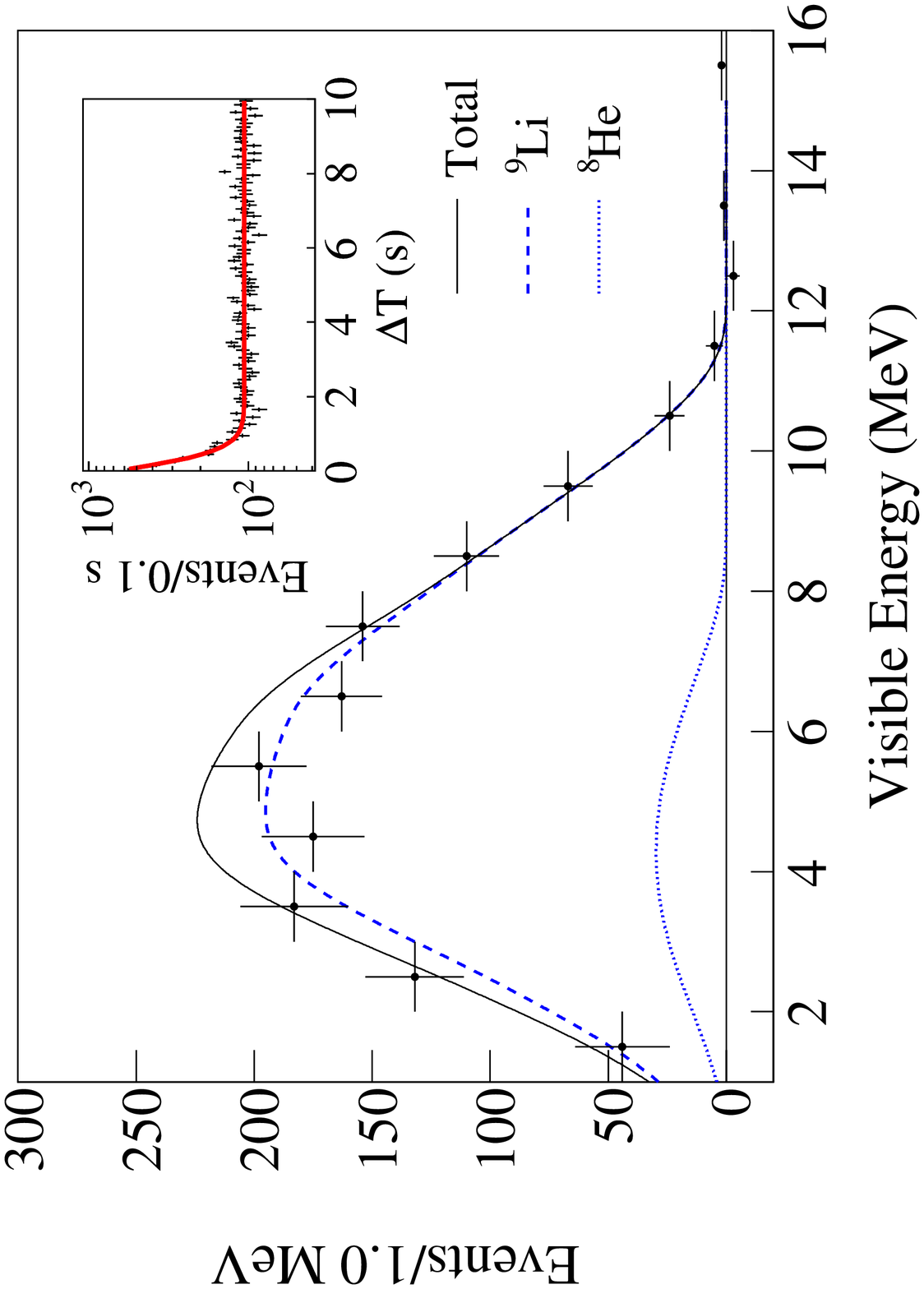}
\end{center}
\caption[spallation time correlation HeLi]{\label{figure:8He_9Li_Summary}
Background-subtracted $E_{\mathrm{vis}}$\ spectrum above 1~MeV,
where signal and background events are taken from
\mbox{$0.002\leq\Delta T<1$\ s} and \mbox{$5.002\leq\Delta T<6$\ s},
respectively.
Detection of a neutron capture following a $\beta$-decay is required to select events from the \mbox{$\beta^{-}+n$}\ decay mode.
The branching ratios for $^{9}$Li and $^{8}$He are \mbox{$(50.8\pm0.9)\%$} and \mbox{$(16\pm1)\%$},
respectively.
The production rates of
$^{8}$He (\mbox{$\tau=171.7$\ ms}, \mbox{$Q=10.7$\ MeV}) and
$^{9}$Li (\mbox{$\tau=257.2$\ ms}, \mbox{$Q=13.6$\ MeV})
are estimated to be \mbox{$1.0\pm0.5$} and \mbox{$2.8\pm0.2$}\ \mbox{kton$^{-1}$day$^{-1}$}, 
respectively,
from simultaneously fitting the $E_{\mathrm{vis}}$\ spectrum and the $\Delta T$\ distribution shown in the inset.
The fit to the $\Delta T$\ distribution has a \mbox{$\chi^{2}/$d.o.f.$=95/97$}.
}
\end{figure}

\begin{figure}[t]
\begin{center}
\begin{minipage}[b]{.48\linewidth}
\vspace{1mm}
\includegraphics[angle=\DefaultFigureAngle,width=\columnwidth]{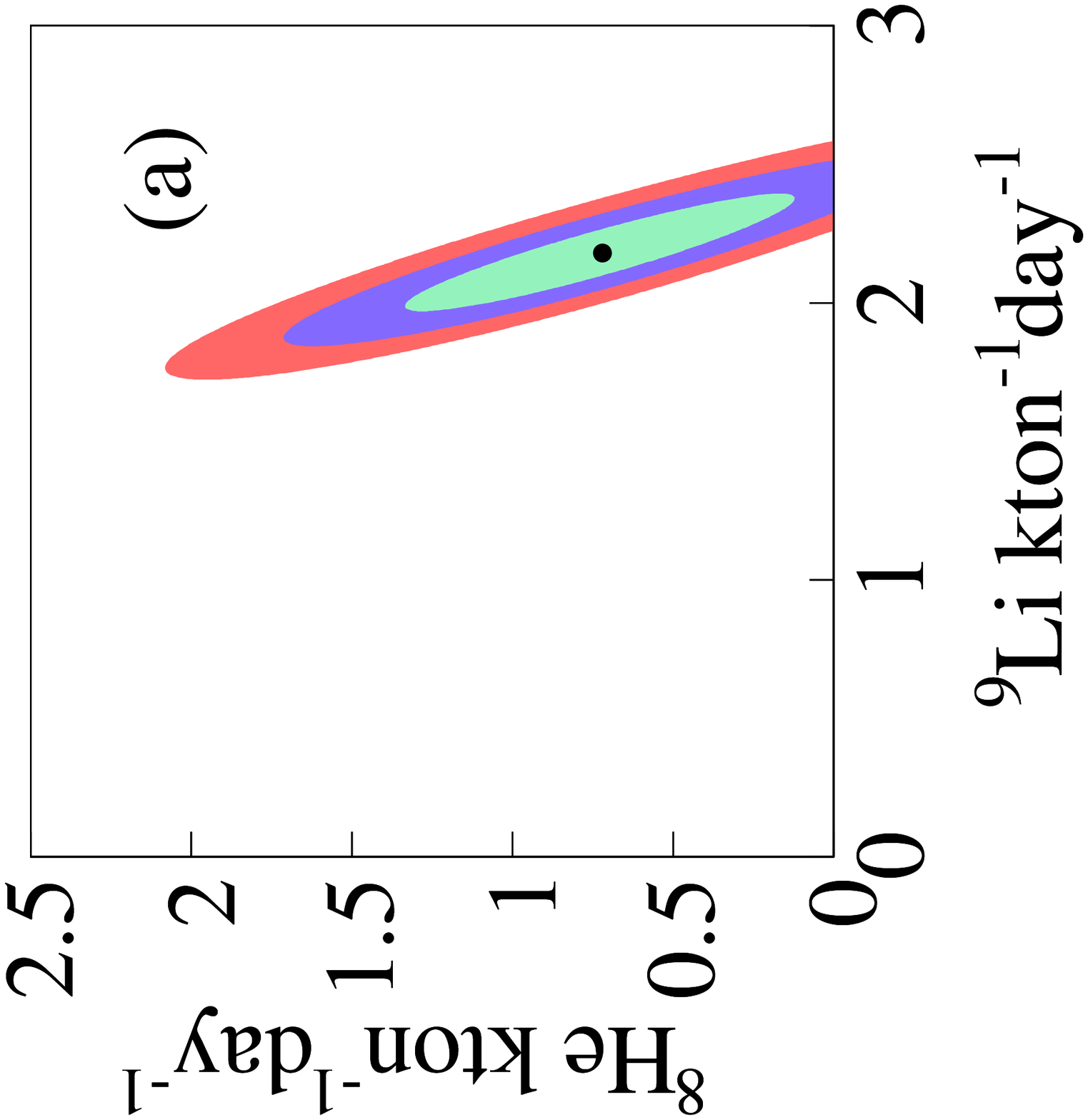}
\end{minipage}
\hspace{2mm}
\begin{minipage}[b]{.48\linewidth}
\vspace{1mm}
\includegraphics[angle=\DefaultFigureAngle,width=\columnwidth]{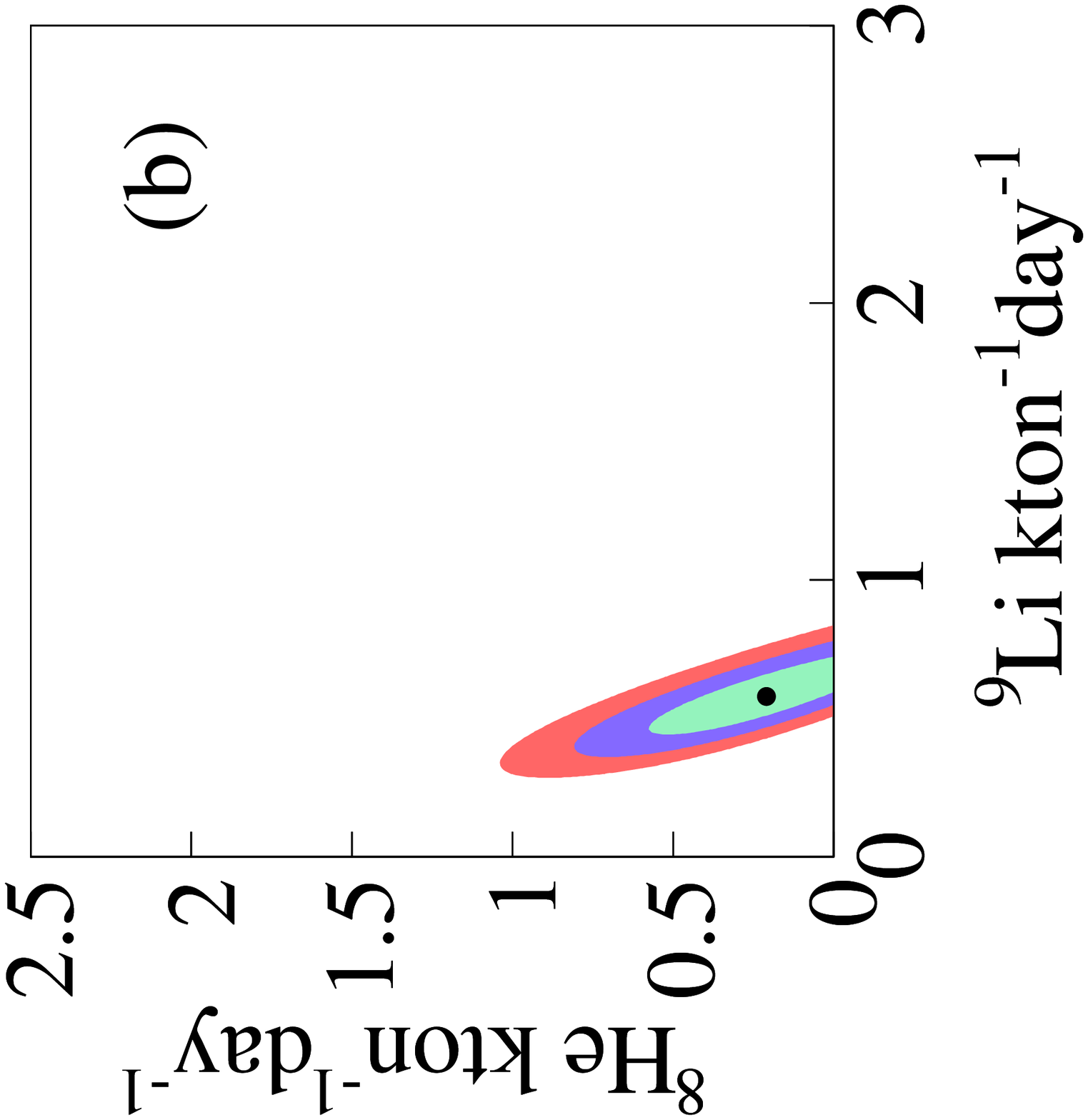}
\end{minipage}
\end{center}
\caption[chi2 He8 and Li9]{\label{figure:Chi2-from-CombinedSpectrum-He8-and-Li9-figure4}
Allowed regions for the production rates of $^{9}$Li and $^{8}$He from the combined fits of the energy spectra and
the $\Delta T$\ distributions at $1\sigma$, $2\sigma$, and $3\sigma$ C.L.\ for
(a) showering muons \mbox{($\Delta\mathcal{L}>10^{6}$\ p.e.)} and
(b) nonshowering muons \mbox{($\Delta\mathcal{L}<10^{6}$\ p.e.)} with $\Delta L< 3$\ m track cut. The black points indicate the best-fit parameters.
}
\end{figure}

The inset in Fig.~\ref{figure:8He_9Li_Summary} shows the $\Delta T$\ distribution for the events that satisfy the criteria outlined above.
Figure~\ref{figure:8He_9Li_Summary} also shows the residual $E_{\mathrm{vis}}$\ distribution corresponding to the subtraction of a background spectrum in the
\mbox{$5.002\leq\Delta T<6$\ s} window from a signal in the \mbox{$0.002\leq\Delta T<1$\ s} window.
The expected $E_{\mathrm{vis}}$\ distributions for $^{8}$He\ and $^{9}$Li\ are calculated by incorporating the KamLAND response
and adjusting for the energy deposited by the thermalizing neutron from $^{8}$He\ or $^{9}$Li decay.
$N(^{8}$He$)$\ and $N(^{9}$Li$)$\ are determined from a simultaneous binned maximum likelihood fit to
the $\Delta T$\ distribution and a chi-square fit to the $E_{\mathrm{vis}}$\ distribution.
For the fit to the $E_{\mathrm{vis}}$\ distribution,
the uncertainty in the energy scale parameters are treated in the same manner as the $^{8}$Li-$^{8}$B analysis described in Sec.~\ref{subsection:8Li_8B_Yield}.

The procedure for calculating the $^{8}$He\ and $^{9}$Li\ selection efficiency is the same as for the
$^{8}$Li and $^{8}$B efficiency analysis (Sec.~\ref{subsection:8Li_8B_Yield}),
except for the correction for the neutron detection requirement,
which is calculated with the  \textsc{geant4}-based Monte Carlo simulation described in Sec.~\ref{subsection:Geant4Simulation}.
The resultant efficiencies
\mbox{$\epsilon(^{8}$He$)=(14.9\pm1.0)\%$}\ and 
\mbox{$\epsilon(^{9}$Li$)=(46.1\pm1.1)\%$} 
include the appropriate \mbox{$\beta$-$n$} branching fractions.
Since a reduced volume is used in this analysis,
the 1.6\% fiducial volume uncertainty from Ref.~\cite{Abe2008} is included in the above efficiencies.
The resulting yields are 
\mbox{$Y(^{8}$He$)=(0.7\pm0.4)\times10^{-7}$}\ and
\mbox{$Y(^{9}$Li$)=(2.2\pm0.2)\times10^{-7}$}\ \mbox{$\mu^{-1}$g$^{-1}$cm$^{2}$}.
The production rates are
\mbox{$R(^{8}$He$)=1.0\pm0.5$}\ and
\mbox{$R(^{9}$Li$)=2.8\pm0.2$} \mbox{kton$^{-1}$day$^{-1}$}.
The contour plots in Fig.~\ref{figure:Chi2-from-CombinedSpectrum-He8-and-Li9-figure4} show the correlation between $^{9}$Li and $^{8}$He.

\subsection{$^{9}$C}
\label{subsection:9C_Yield}

\begin{figure}[t]
\begin{center}
\vspace{4mm}
\includegraphics[angle=\DefaultFigureAngle,width=\columnwidth]{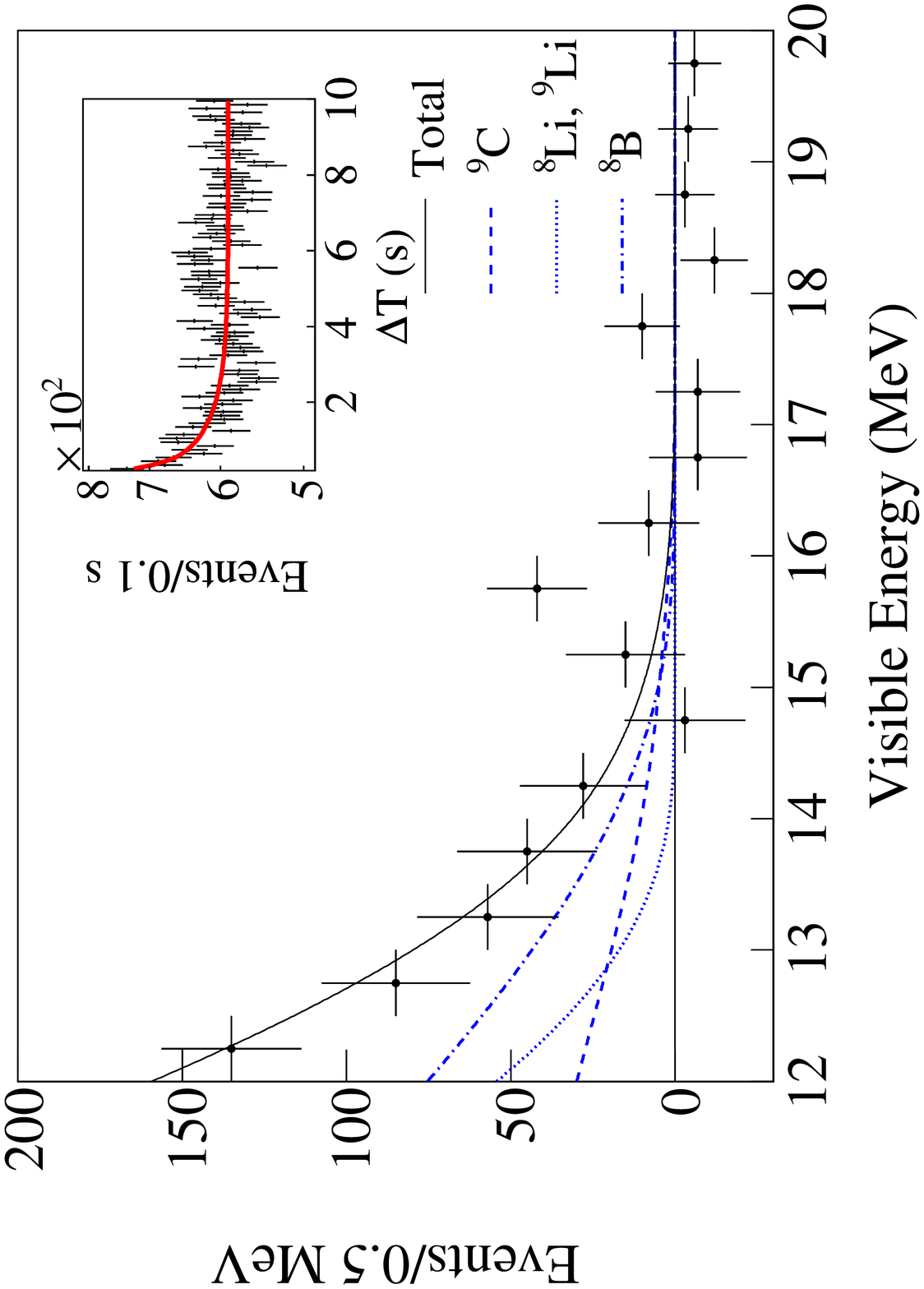}
\end{center}
\caption[spallation time correlation 20sec Without B12 (C9)]{\label{figure:9C_Summary}
Background-subtracted $E_{\mathrm{vis}}$\ spectrum above 12~MeV,
where signal and background events are taken from
\mbox{$0.2\leq\Delta T<0.6$\ s} and \mbox{$10.2\leq\Delta T<10.6$\ s},
respectively.
The production rate of $^{9}$C\ (\mbox{$\tau=182.5$\ ms}, \mbox{$Q=16.5$\ MeV}) is estimated to be \mbox{$3.8\pm1.5$}\ \mbox{kton$^{-1}$day$^{-1}$} from
a simultaneous fit to the $E_{\mathrm{vis}}$\ spectrum and the $\Delta T$\ distribution shown in the inset. 
The fit to the $\Delta T$\ distribution has a \mbox{$\chi^{2}/$d.o.f.$=97/96$}.
}
\end{figure}

The inset in Fig.~\ref{figure:9C_Summary} shows the $\Delta T$\ distribution for all events with visible energy \mbox{$12\leq E_{\mathrm{vis}}<20$\ MeV}.
The analysis region \mbox{($0.2\leq\Delta T<0.6$\ s)} contains events from $^{9}$C (\mbox{$\tau=182.5$\ ms}, \mbox{$Q=16.5$\ MeV})~\cite{Tilley2004} $\beta^{+}$\ decay.
$N(^{9}$C$)$\ is determined from a simultaneous binned maximum likelihood fit to the $\Delta T$\ distribution
and a chi-square fit to the $E_{\mathrm{vis}}$\ distribution.
The uncertainty in the energy scale parameters are treated in the same manner as the $^{8}$Li-$^{8}$B analysis described in Sec.~\ref{subsection:8Li_8B_Yield}.
In this fit,
$^{8}$Li, $^{8}$B, and $^{9}$Li are treated as possible contaminants,
the amounts are constrained to the values obtained in the previously described analyses.
This constraint includes the correlation between $^{8}$Li and $^{8}$B shown in Fig.~\ref{figure:Chi2-from-CombinedSpectrum-B8-and-Li8-figure4}.
By integrating the theoretical $^{9}$C $E_{\mathrm{vis}}$\ spectrum,
we obtain the efficiency for the \mbox{$12\leq E_{\mathrm{vis}}<20$\ MeV}\ cut of \mbox{$\epsilon(^{9}$C$)=(7.2\pm1.0)\%$}.
Combining this with the above results gives
\mbox{$Y(^{9}$C$)=(3.0\pm1.2)\times10^{-7}$}\ \mbox{$\mu^{-1}$g$^{-1}$cm$^{2}$} and
\mbox{$R(^{9}$C$)=3.8\pm1.5$}\ \mbox{kton$^{-1}$day$^{-1}$}.

\subsection{$^{11}$C}
\label{subsection:11C_Yield}

\begin{figure}[t]
\begin{center}
\includegraphics[angle=\DefaultFigureAngle,width=\columnwidth]{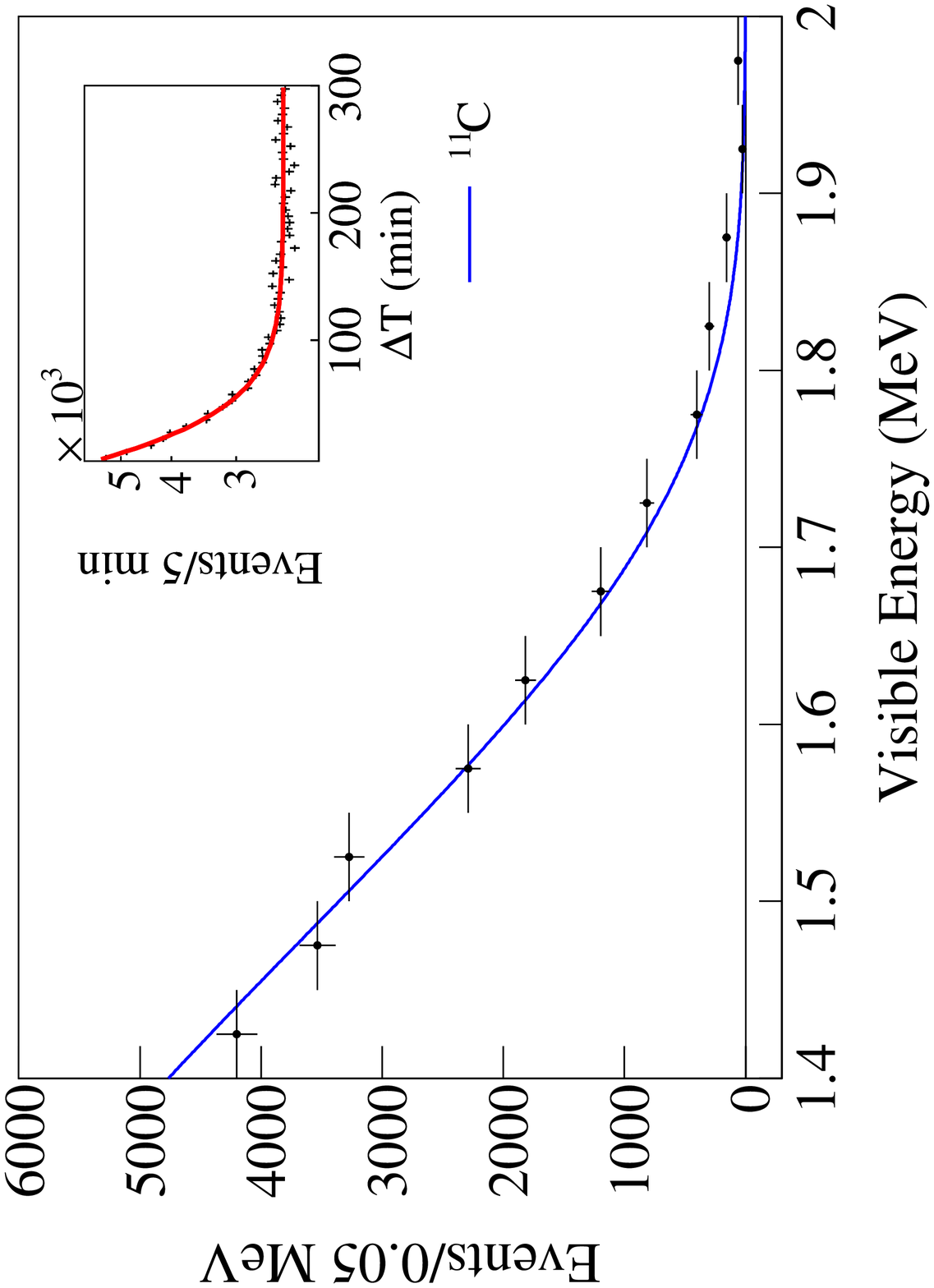}
\end{center}
\caption[spallation time correlation C11 Each Run]{\label{figure:11C_Summary}
Background-subtracted $E_{\mathrm{vis}}$\ spectrum above 1.4~MeV,
where signal and background are selected from \mbox{$5\leq\Delta T<90$\ min}\ and \mbox{$185\leq\Delta T<270$\ min},
respectively.
The production rate of $^{11}$C (\mbox{$\tau=29.4$\ min}, \mbox{$Q=1.98$\ MeV}) is estimated to be
\mbox{$1106\pm178$}\ \mbox{kton$^{-1}$day$^{-1}$}\ by fitting Eq.~(\ref{equation:IsotopeDeltaTime}) to the $\Delta T$\ distribution shown in the inset.
The fit to the $\Delta T$\ distribution has a \mbox{$\chi^{2}/$d.o.f.$=76/57$}.
}
\end{figure}

The production of $^{11}$C ($\beta^{+}$\ decay, \mbox{$\tau=29.4$\ min}, \mbox{$Q=1.98$\ MeV})~\cite{AjzenbergSelove1990}
through muon-initiated spallation is usually accompanied by a neutron,
allowing identification by the triple coincidence of the primary muon,
the spallation neutron,
and the subsequent $\beta^{+}$~\cite{Back2006,Galbiati2005a}.
The $^{11}$C $\beta^{+}$\ decays are selected in the range \mbox{$1.4\leq E_{\mathrm{vis}}<2.0$\ MeV} and
are preceded by a detected muon that is accompanied by at least one neutron capture,
identified by the \mbox{2.225-MeV}\ $\gamma$ ray
from capture on $^{1}$H.
The $\gamma$ ray is required to be in the time window \mbox{$10\leq\Delta T<2500 \mu$s} relative to the muon.
To reduce the background,
a 7-m-diameter fiducial volume is used.
To avoid inefficiencies from run boundaries and the long lifetime of $^{11}$C,
the first 5 h of the typically 24-h-long run are not used in the selection of the $^{11}$C candidates.
The number of \mbox{muon-$^{11}$C} coincidences is extracted from the $\Delta T$\ distribution for all events that meet the criteria,
shown in the inset of Fig.~\ref{figure:11C_Summary}.

The efficiency determination takes into account the visible energy range for $^{11}$C $\beta^{+}$\ decay \mbox{$(22.7\pm3.6)\%$}
and the previously discussed neutron detection efficiency (Sec.~\ref{section:SpallationNeutronYield}).
The efficiency also takes into account a correction for $^{11}$C production modes,
designated \textit{invisible modes},
which do not produce neutrons~\cite{Galbiati2005a,Back2006}.
To measure this correction,
muon-$^{11}$C event pairs were selected with and without the neutron requirement for a subset of the data
where the $^{11}$C candidate is required to be within 50~cm of the muon track;
restricting the study to a subset of the data mitigated the reduced \textit{signal-to-background} ratio associated with relaxing the neutron requirement.
The number of muon-$^{11}$C coincidences in each case was extracted from a fit of Eq.~(\ref{equation:IsotopeDeltaTime}) to the corresponding $\Delta T$\ distribution.
The visible mode efficiency,
$\epsilon_{\mathrm{vis}}$,
taken as the ratio of the number of muon-$^{11}$C pairs with one or more neutrons to the number of muon-$^{11}$C pairs without the neutron requirement,
is $(88.4\pm2.4)\%$.
Applying the correction for post-muon electronics effects,
the visible mode fraction is \mbox{$(96.3\pm2.0)\%$},
consistent with Ref.~\cite{Galbiati2005a},
which obtains \mbox{$\epsilon_{\mathrm{vis}}=95.6\%$}\ for 285-GeV muons.

Due to the relatively long $^{11}$C lifetime,
we also considered the effect of diffusion.
An analysis of $^{222}$Rn that was accidentally introduced into the center of KamLAND during the deployment
of a calibration device shows that the diffusion speed is approximately 1~mm/h.
From this study,
the effect of $^{11}$C diffusion on the efficiency is estimated to be less than $0.5$\%.
Combining this with the above results gives
\mbox{$Y(^{11}$C$)=(866\pm153)\times10^{-7}$}\ \mbox{$\mu^{-1}$g$^{-1}$cm$^{2}$}\ and
\mbox{$R(^{11}$C$)=1106\pm178$}\ \mbox{kton$^{-1}$day$^{-1}$}.

\subsection{$^{10}$C}
\label{subsection:10C_Yield}

\begin{figure}[t]
\begin{center}
\includegraphics[angle=\DefaultFigureAngle,width=\columnwidth]{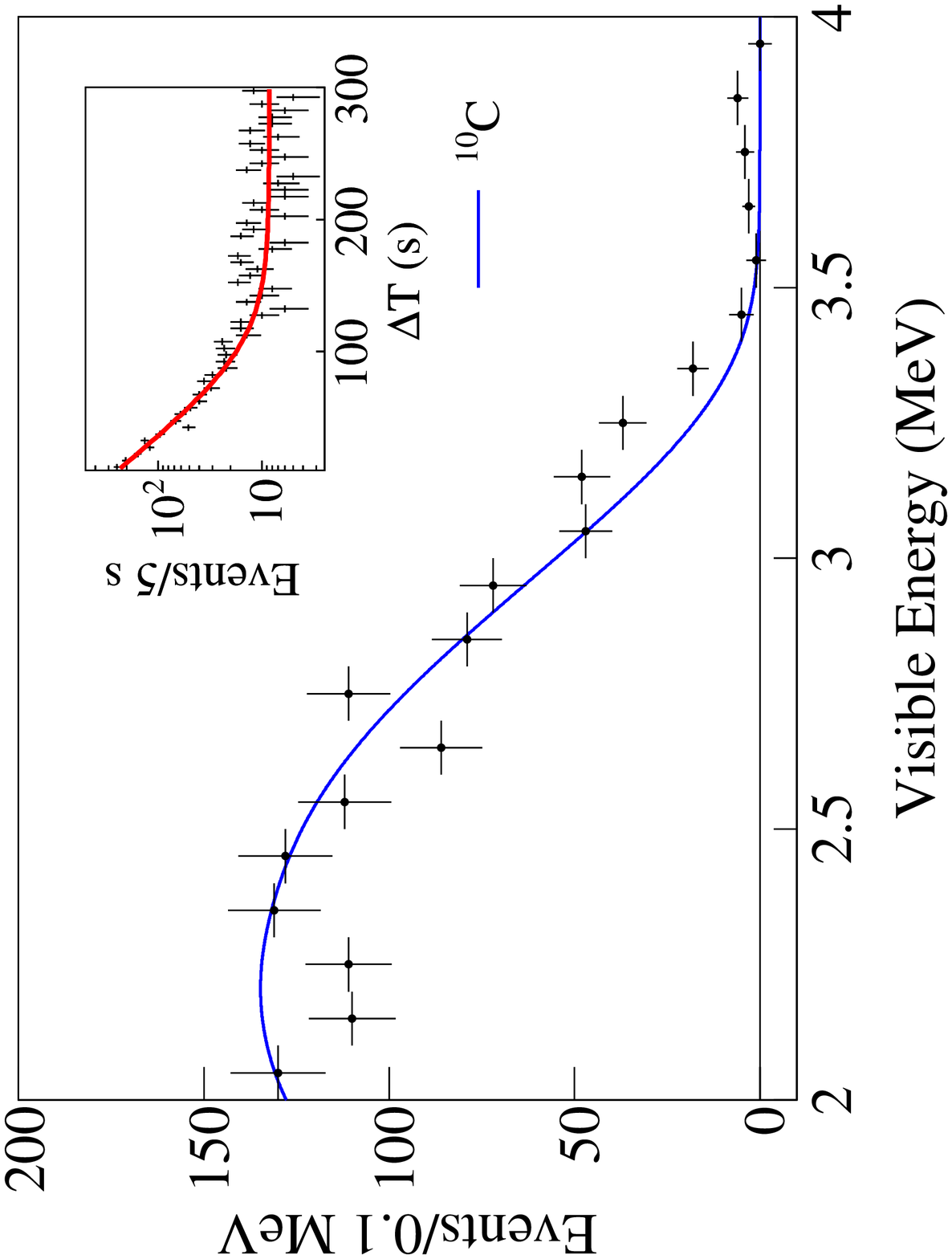}
\end{center}
\caption[spallation time correlation C10 EachRun]{\label{figure:10C_Summary}
Background-subtracted $E_{\mathrm{vis}}$\ spectrum above 2~MeV,
where signal and background are selected using \mbox{$10\leq\Delta T<90$\ s}\ and \mbox{$190\leq\Delta T<270$\ s},
respectively.
The production rate of $^{10}$C (\mbox{$\tau=27.8$\ s}, \mbox{$Q=3.65$\ MeV}) is \mbox{$21.1\pm1.8$}\ \mbox{kton$^{-1}$day$^{-1}$},
as determined by fitting Eq.~(\ref{equation:IsotopeDeltaTime}) to the $\Delta T$\ distribution shown in the inset.
The fit to the $\Delta T$\ distribution has a \mbox{$\chi^{2}/$d.o.f.$=80/56$}.
}
\end{figure}

As with $^{11}$C,
the production of $^{10}$C($\beta^{+}$\ decay, \mbox{$\tau=27.8$\ s}, \mbox{$Q=3.65$\ MeV})~\cite{Tilley2004}
through muon-initiated spallation is usually accompanied by a neutron,
so the selection criterion requiring a triple coincidence of the primary muon,
the neutron,
and the $^{10}$C candidate is used.
The neutron is identified by the \mbox{2.225-MeV}\ $n+^{1}$H capture $\gamma$ ray.
The number of $^{10}$C candidates is determined from fitting Eq.~(\ref{equation:IsotopeDeltaTime}) to the $\Delta T$\ distribution for all events identified as $^{10}$C,
shown in the inset in Fig.~\ref{figure:10C_Summary}.
$^{11}$Be is a potential background for this $^{10}$C analysis,
but the correction to the $^{10}$C yield is estimated to be less than 1\% because of the low $^{11}$Be production rate and the neutron coincidence requirement.
The efficiency for the visible energy cut \mbox{$2.0\leq E_{\mathrm{vis}}<4.0$\ MeV} is \mbox{$(73.5\pm3.2)\%$}.
The visible mode efficiency is \mbox{$\epsilon_{\mathrm{vis}}=(90.7\pm5.5)\%$}
after correcting for the electronics effects following muons. 
The final efficiency is \mbox{$(89.6\pm5.5)\%$}.
The resulting isotope yield is \mbox{$Y(^{10}$C$)=(16.5\pm1.9)\times10^{-7}$}\ \mbox{$\mu^{-1}$g$^{-1}$cm$^{2}$}\ and
the production rate is \mbox{$R(^{10}$C$)=21.1\pm1.8$}\ \mbox{kton$^{-1}$day$^{-1}$}.

\subsection{$^{11}$Be}
\label{subsection:11Be_Yield}

\begin{figure}[t]
\begin{center}
\includegraphics[angle=\DefaultFigureAngle,width=\columnwidth]{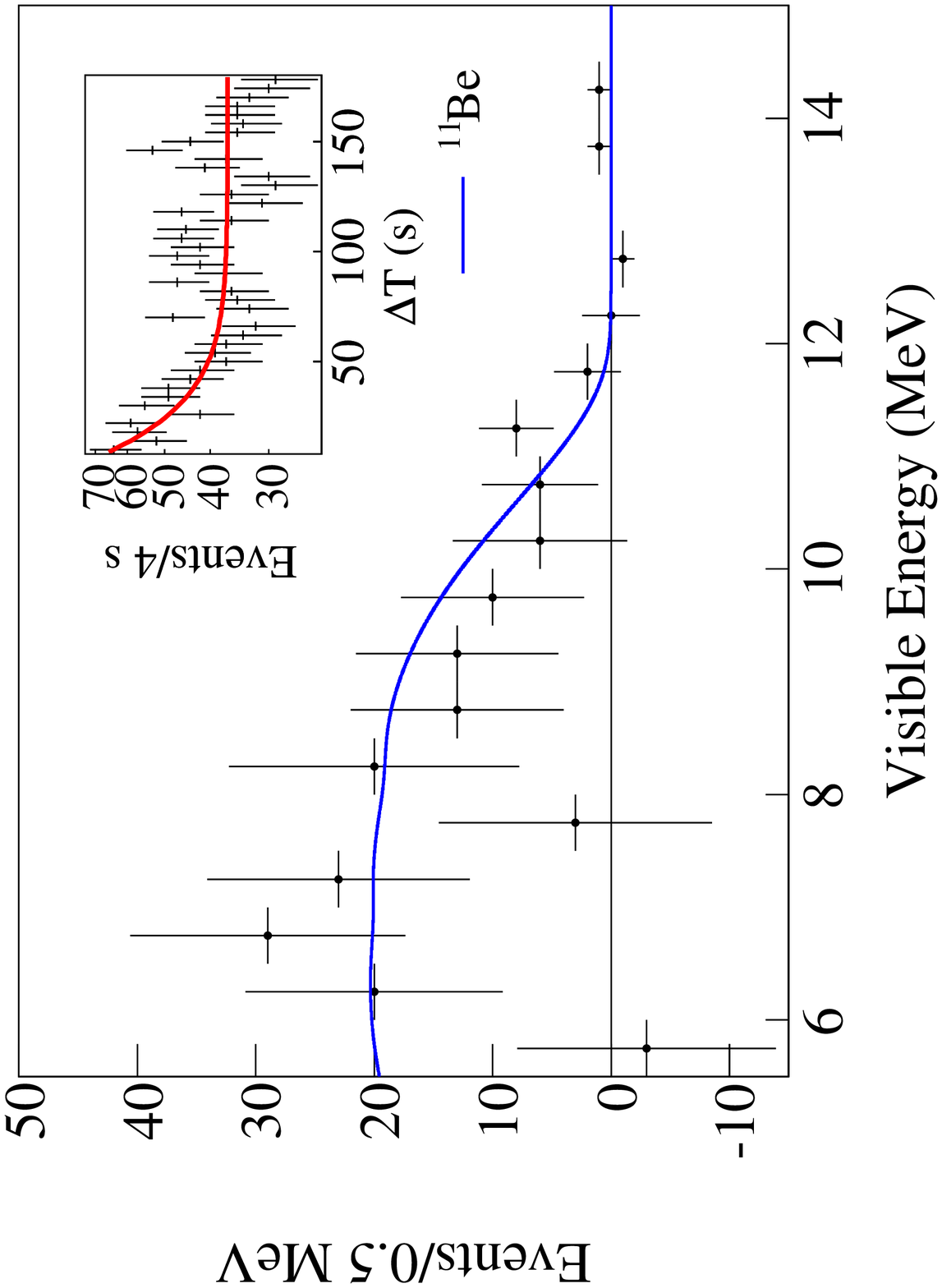}
\end{center}
\caption[example Read SingleEvent dT figure1 dQL BestFit]{\label{figure:11Be_Summary}
Background-subtracted $E_{\mathrm{vis}}$\ spectrum above 5.5~MeV for showering muons \mbox{($\Delta\mathcal{L}>10^{6}$\ p.e.)},
where signal and background are selected by \mbox{$8\leq\Delta T<60$\ s}\ and \mbox{$408\leq\Delta T<460$\ s},
respectively.
The production rate of $^{11}$Be (\mbox{$\tau=19.9$\ s}, \mbox{$Q=11.5$\  MeV}) for showering muons is estimated to be
\mbox{$1.0\pm0.2$}\ \mbox{kton$^{-1}$day$^{-1}$}\ by fitting Eq.~(\ref{equation:IsotopeDeltaTime})
to the $\Delta T$\ distribution shown in the inset \mbox{($\chi^{2}/$d.o.f.$=37/41$)}.
}
\end{figure}

The $^{11}$Be $\beta^{-}$-decay (\mbox{$\tau=19.9$\ s}, \mbox{$Q=11.5$\ MeV})~\cite{AjzenbergSelove1990}
events are selected according to \mbox{$5.5\leq E_{\mathrm{vis}}<16.0$\ MeV}.
The $E_{\mathrm{vis}}$ cut efficiency is 63.4\%. 
The inset in Fig.~\ref{figure:11Be_Summary} shows the $\Delta T$\ distributions for the events only after showering muons. 
For nonshowering muons, a tighter muon track cut \mbox{$\Delta L<1$\ m} is applied in order to reduce the background rate.
The track cut efficiency is estimated from the $^{12}$B candidates using an analysis similar to that in Sec.~\ref{subsection:8Li_8B_Yield}.
The resulting isotope yield is \mbox{$Y(^{11}$Be$)=(1.1\pm 0.2) \times 10^{-7}$}\ \mbox{$\mu^{-1}$g$^{-1}$cm$^{2}$}\ and
production rate is \mbox{$R(^{11}$Be$)=1.4\pm0.3$}\ \mbox{kton$^{-1}$day$^{-1}$}.

\section{Monte Carlo Simulations}

The  \textsc{geant4} and \textsc{fluka} simulations are used to reproduce the measurements from KamLAND.
While  \textsc{geant4} is used only to simulate neutron production,
both neutron and light isotope production are tested with \textsc{fluka}.

\subsection{\textsc{geant4}}
\label{subsection:Geant4Simulation}

 \textsc{geant4} is a widely used toolkit for performing particle tracking simulations on an event-by-event basis.
A description of the available physics processes included is given in Refs.~\cite{Agostinelli2003,Allison2006}. 
Here we compare the  \textsc{geant4} (version 9.1) prediction for neutron yield by spallation with the results obtained in Sec.~\ref{section:SpallationNeutronYield}.
We use the physics list \textsc{QGS\_BIC},
developed by the \textsc{geant4} group to support the binary cascade (\textsc{BIC}) model at lower energies
(below 10~GeV for $p$\ and $n$, and below 1.2~GeV for $\pi$).
This treatment is also appropriate for the simulation of interactions of nucleons and ions.
At higher energies, a quark-gluon string (\textsc{QGS}) model is applied for the hadronic interactions.
Neutron elastic and inelastic interactions below 20~MeV are described by a high-precision data-driven model (\textsc{NeutronHP}).
The \textsc{G4EmExtraPhysics} physics list is also used to model the photonuclear and muon-nuclear interaction processes,
which dominate the neutron production by muons in the simulation.

To estimate the neutron production yield as a function of muon energy in  \textsc{geant4},
monoenergetic muons of several energies are injected at the center of a generic hydrocarbon block of thickness 40~m.
The region more than 10~m away from the edges of the block is analyzed to avoid boundary effects.
As shown later in Fig.~\ref{figure:ProcessContribution-figure2-KamLAND},
the neutron production yields predicted by  \textsc{geant4} are systematically lower than experiment,
with the exception of point (F) from the LVD~\cite{Aglietta2003}.
These results are consistent with previous work~\cite{Araujo2005}.

\begin{figure}[t]
\begin{center}
\includegraphics[width=\columnwidth]{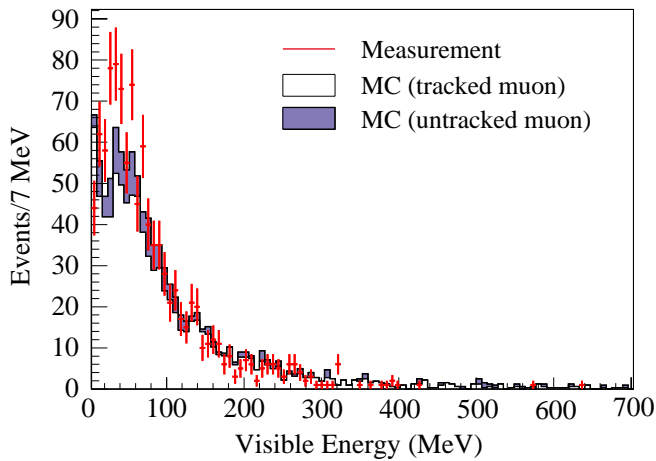}
\end{center}
\caption[promptE_data_MC]{\label{figure:promptE_data_MC}
$E_{\mathrm{vis}}$\ distribution of the prompt events from candidates identified as neutrons produced by muon-induced spallation in the material outside of the KamLAND ID.
}
\end{figure}

\begin{figure}[t]
\begin{center}
\includegraphics[width=\columnwidth]{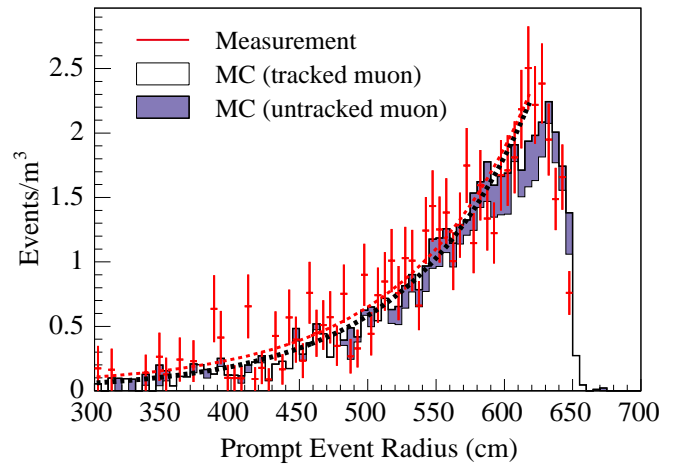}
\end{center}
\caption[promptR_data_MC]{\label{figure:promptR_data_MC}
Radial distribution (from the center of KamLAND) of the prompt event from candidates identified as neutrons produced by muon-induced spallation
in the material outside of the KamLAND ID.
Assuming an exponential distribution,
a fit to the measured data (red dashed lines) yields an attenuation length of \mbox{$70\pm2$\ g$/$cm$^{2}$}.
A similar fit to the Monte Carlo (black dotted lines) yields \mbox{$69\pm2$\ g$/$cm$^{2}$}.
}
\end{figure}

A Monte Carlo simulation based upon  \textsc{geant4} and \textsc{music} (described in Sec.~\ref{section:CosmicRayMuons})
is used to study neutrons produced by muon-induced spallation in the material outside of the KamLAND ID.
Some of these neutrons have sufficient energy to enter the ID where they thermalize and capture.
They can be identified by the coincidence of a prompt signal (for example from \mbox{$n+p$}\ elastic scattering) and a delayed signal from the capture $\gamma$ ray.
This is the same inverse \mbox{$\beta$-decay} reaction signature used for \nuebar\ detection,
\mbox{$\overline{\nu}_{e}+p\rightarrow e^{+}+n$},
where the $e^{+}$\ is the prompt signal, and the $\gamma$ ray from neutron capture is the delayed signal
and therefore a potential background.
These neutrons are also a background for dark-matter experiments that employ nuclear recoils as a detection method.

The primary purpose of this Monte Carlo simulation is to estimate the rate of \textit{untagged fast neutrons},
i.e.\ neutrons produced by muons where the muon is undetected by the KamLAND ID.
A few of these muons are detected by the OD,
either from the Cherenkov radiation produced by the muon itself (\textit{tracked muons}),
or by accompanying electromagnetic and hadronic showers that enter the OD (\textit{untracked muons}).
A prompt signal in coincidence with these tracked and untracked muons followed by a delayed capture $\gamma$\ signal identifies candidates for untagged fast neutrons.

Fast neutrons generated by untracked muons are produced primarily in the rock surrounding the OD,
whereas fast neutrons from tracked muons are primarily produced in the water of the OD.
It is shown in the Monte Carlo that the tracked and untracked muons give distinguishable OD visible energy distributions.
A comparison between Monte Carlo and measurement of the distribution of the number of OD PMTs with signals above threshold for tracked and untracked muons
reveals a deficiency of fast neutrons from untracked muons,
consistent with the underproduction of neutrons by  \textsc{geant4} in concrete reported in Ref.~\cite{Marino2007}.

Figure~\ref{figure:promptE_data_MC} shows the $E_{\mathrm{vis}}$\ distribution of the prompt events from the Monte Carlo simulation
of tracked and untracked muons compared with the measured data.
The measured data and the Monte Carlo simulation correspond to an equal live time exposure of 1368~days.
Figure~\ref{figure:promptR_data_MC} shows the radial distribution of the prompt events relative to the KamLAND center.
Both the measured data and the  \textsc{geant4} simulation exhibit an exponential attenuation of the neutrons as they penetrate farther into the detector,
with the simulation yielding an attenuation length of \mbox{$69\pm2$\ g$/$cm$^{2}$},
consistent with the measurement \mbox{$70\pm2$\ g$/$cm$^{2}$}.

\begin{figure}[t]
\begin{center}
\includegraphics[angle=\DefaultFigureAngle,width=\columnwidth]{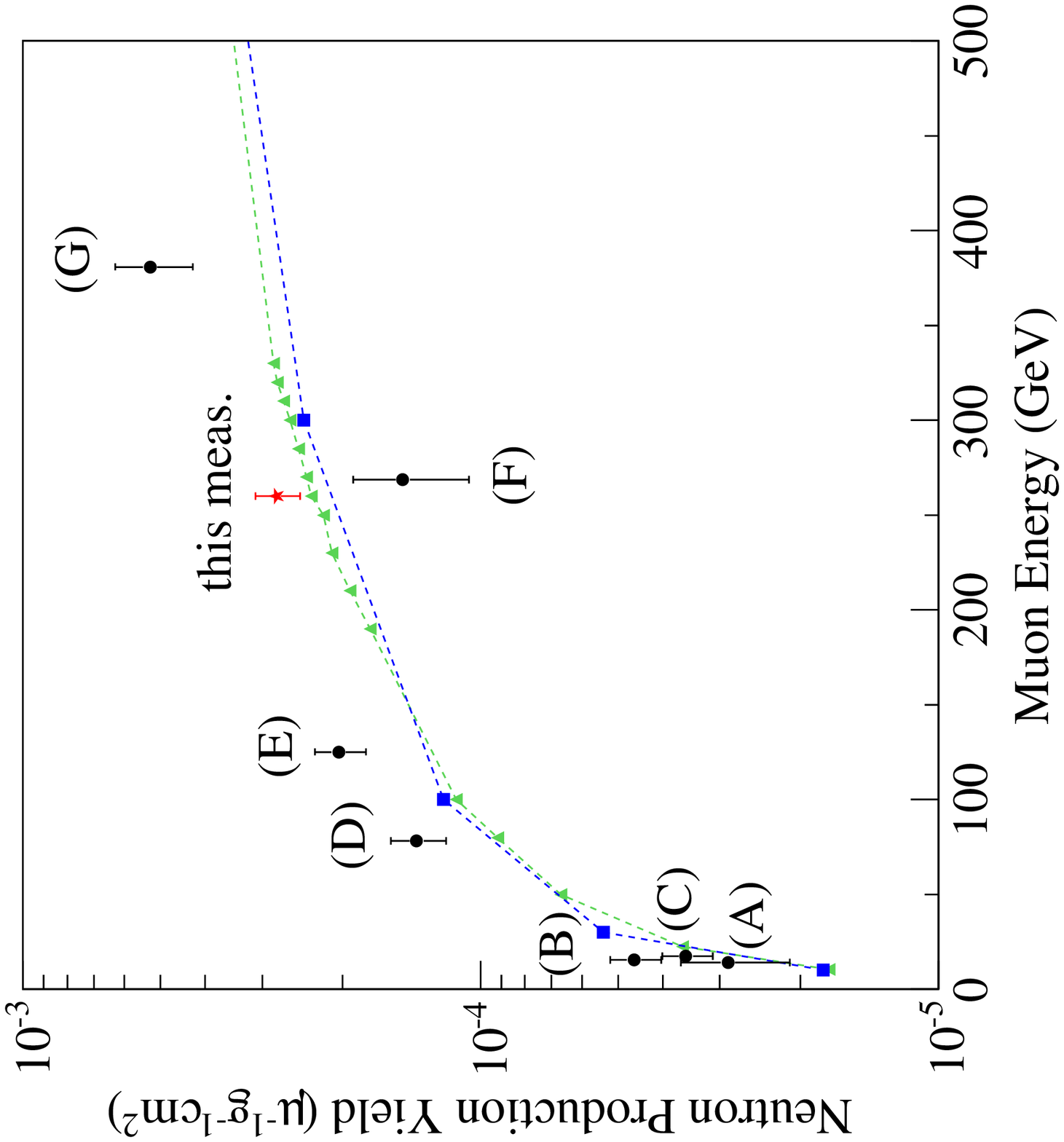}
\end{center}
\caption[ProcessContribution KamLAND]{\label{figure:ProcessContribution-figure2-KamLAND}
Neutron production yield in liquid scintillator as a function of muon energy.
The red star point shows the KamLAND result, \mbox{$Y_{n}=(2.8\pm0.3)\times10^{-4} \mu^{-1}$g$^{-1}$cm$^{2}$}, for \mbox{$260\pm8$\ GeV}.
Other points show the results from experiments at
(A) 20~mwe \cite{Hertenberger1995},
(B) 25~mwe \cite{Bezrukov1973},
(C) 32~mwe \cite{Boehm2000},
(D) 316~mwe \cite{Bezrukov1973},
(E) 570~mwe \cite{Enikeev1987},
(F) 3000~mwe \cite{Aglietta2003},
and (G) 5200~mwe \cite{Aglietta1989}. 
The blue square and green triangle show the  \textsc{geant4} and \textsc{fluka} Monte Carlo predictions,
respectively, from monochromatic muon beams.}
\label{fig:Neutron}
\end{figure}

\subsection{\textsc{fluka}}
\label{subsection:FlukaSimulation}

\begin{table*}
\caption{Simulation of neutron and light isotope production in the KamLAND LS by muon-initiated spallation with \textsc{fluka}:
the isotope production yield by monoenergetic (260~GeV) $\mu^{-}$;
the ratio of the production yields by monoenergetic $\mu^{+}$\ compared to $\mu^{-}$;
the ratio of the production yields by a $\mu^{-}$\ spectrum that matches Fig.~6 (KamLAND curve) in Ref.~\cite{Tang2006} compared with monoenergetic $\mu^{-}$;
the power-law exponent for the production yield \mbox{$Y(E_{\mu})\propto E^{\alpha}_{\mu}$}\ from a fit to the yields from monoenergetic
$\mu^{-}$\ with \mbox{$10\leq E_{\mu}\leq350$\ GeV};
and the primary process for producing the isotope.
The uncertainties are statistical.
}
\begin{ruledtabular}
\begin{tabular}{cccccc}
 & Simulated production yield & \multicolumn{2}{c}{Ratio of Simulated Production Yields for} & & \\ \cline{3-4}
 & $(\times10^{-7} \mu^{-1}$g$^{-1}$cm$^{2})$ & $\mu^{+} / \mu^{-}$ & Spectrum$/$monoenergetic & \raisebox{1.75ex}{Power-law exp.} & \raisebox{1.75ex}{Primary process} \\
\hline
$n$ & $2344\pm4$ & $0.969\pm0.002$ & $0.912\pm0.003$ & $0.779\pm0.001$ & $\pi^{-}+^{1}$H,$^{12}$C \\
$^{11}$C & $460.8\pm1.7$ & $0.971\pm0.005$ & $0.913\pm0.006$ & $0.703\pm0.002$ & $^{12}$C$(\gamma,n)$ \\
$^{7}$Be & $116.8\pm0.9$ & $0.986\pm0.011$ & $0.945\pm0.011$ & $0.684\pm0.004$ & $^{12}$C$(\gamma,n\alpha)$ \\
$^{10}$Be & $44.63\pm0.53$ & $0.960\pm0.018$ & $0.891\pm0.019$ & $0.825\pm0.007$ & $^{12}$C$(n,^{3}$He$)$ \\
$^{12}$B & $30.85\pm0.44$ & $0.970\pm0.021$ & $0.936\pm0.022$ & $0.828\pm0.009$ & $^{12}$C$(n,p)$ \\
$^{8}$Li & $23.42\pm0.39$ & $0.927\pm0.026$ & $0.936\pm0.025$ & $0.821\pm0.010$ & $^{12}$C$(n,p\alpha)$ \\
$^{10}$C & $21.13\pm0.37$ & $0.982\pm0.025$ & $0.915\pm0.027$ & $0.810\pm0.010$ & $^{12}$C$(\pi^{+},np)$ \\
$^{6}$He & $13.40\pm0.29$ & $0.916\pm0.035$ & $0.918\pm0.035$ & $0.818\pm0.013$ & $^{12}$C$(n,2p^{3}$He$)$ \\
$^{8}$B & $6.40\pm0.20$ & $0.996\pm0.045$ & $0.915\pm0.050$ & $0.804\pm0.019$ & $^{12}$C$(\pi^{+},^{2}$H$^{2}$H) \\
$^{9}$Li & $3.51\pm0.15$ & $0.856\pm0.074$ & $0.842\pm0.078$ & $0.801\pm0.026$ & $^{12}$C$(\pi^{-},^{3}$He$)$ \\
$^{9}$C & $1.49\pm0.10$ & $0.850\pm0.114$ & $0.949\pm0.102$ & $0.772\pm0.039$ & $^{12}$C$(\pi^{+},^{3}$H$)$ \\
$^{12}$N & $0.86\pm0.07$ & $0.963\pm0.128$ & $1.006\pm0.120$ & $0.921\pm0.045$ & $^{12}$C$(p,n)$ \\
$^{11}$Be & $0.94\pm0.08$ & $0.842\pm0.145$ & $0.804\pm0.161$ & $0.753\pm0.051$ & $^{12}$C$(n,2p)$ \\
$^{8}$He & $0.35\pm0.05$ & $0.964\pm0.200$ & $0.576\pm0.372$ & $0.926\pm0.078$ & $^{12}$C$(\pi^{-},n3p)$ \\
$^{13}$B & $0.31\pm0.04$ & $1.020\pm0.197$ & $1.062\pm0.176$ & $0.742\pm0.075$ & $^{13}$C$(n,p)$ \\
$^{15}$O & $0.05\pm0.02$ & $1.250\pm0.379$ & $1.635\pm0.234$ & $0.793\pm0.244$ & $^{16}$O$(\gamma,n)$ \\
$^{13}$N & $0.06\pm0.02$ & $1.500\pm0.272$ & $1.190\pm0.401$ & $1.120\pm0.220$ & $^{13}$C$(p,n)$ \\
\end{tabular}
\end{ruledtabular}
\label{table:FLUKA}
\end{table*}

\textsc{fluka} is a mature code that models nuclear and particle physics processes from thermal neutrons to heavy-ion collisions~\cite{Fasso2003,Ferrari2005}.
It has been used previously to model muon-initiated spallation in liquid scintillator~\cite{Wang2001,Kudryavtsev2003,Mei2006,Araujo2005}.
We use \textsc{fluka} version~2006.3b to model neutron and light isotope production from muon-initiated spallation in KamLAND. 
A 40-m-radius by 40-m-high cylinder of KamLAND liquid scintillator is used in the simulation;
the concentric inner cylinder of 20-m radius and 20-m length is used for analysis.

To estimate neutron production yield as a function of muon energy in \textsc{fluka},
monoenergetic beams of $\mu^{-}$\ ranging from 10 to 350~GeV were simulated as in Refs.~\cite{Wang2001,Kudryavtsev2003,Mei2006,Araujo2005}.
Care is taken not to double-count the neutrons involved in reactions like \mbox{$(n,2n)$}.
The results of this simulation are included in Fig.~\ref{fig:Neutron}.
The neutron production yield of this \textsc{fluka} simulation is 10\% lower than previous work~\cite{Wang2001,Kudryavtsev2003,Mei2006,Araujo2005},
but the power-law dependance on muon energy ($E_{\mu}^{\alpha}$, where \mbox{$\alpha=0.77$}) is consistent.
Different scintillator compositions were studied, but they could not explain the deficit.
This deficit is insignificant compared to the discrepancy between these simulations and the data.

The production of light isotopes was studied in the same simulation.
The results,
including the primary production process and power-law exponent,
are summarized in Table~\ref{table:FLUKA}.
For some isotopes the primary production process is much larger than any secondary processes,
as in the case of $^{12}$B.
For other isotopes the primary production process is only slightly larger than the secondary processes, as is the case for $^{9}$Li.
The isotopes produced primarily by $\gamma$ interactions,
$^{11}$C and $^{10}$C,
show the weakest dependence on muon energy.
In comparison,
$^{12}$N and $^{13}$N,
where the primary production mechanism is by $p$ interactions, show the strongest dependence on muon energy.

The use of a monoenergetic $\mu^{-}$ beam overestimates the production of neutrons and light isotopes.
Simulations using a monoenergetic $\mu^{+}$ beam and a beam with the energy spectrum from Ref.~\cite{Tang2006} were also run.
The simulations show that the production yield for $\mu^{+}$\ relative to $\mu^{-}$\ is on average 0.96$\pm$0.01 for the light isotopes.
This reduction is expected, since $\mu^{-}$ may capture, creating spallation products, while $\mu^{+}$ may not.
This ratio, combined with the $\mu^{+}$\ to $\mu^{-}$\ ratio at KamLAND, leads to a correction to the flux of 0.98$\pm$0.06 for light isotopes and 0.981$\pm$0.005 for neutrons.
The reduced production yield due to averaging over the muon spectrum is on average 0.92$\pm$0.02 for the light isotopes,
which is slightly higher than the correction factor suggested by Ref.~\cite{Hagner2000}.
The results of the \textsc{fluka} simulations for KamLAND presented in Table \ref{table:ShimizuSummary} include these two corrections. 

\section{Discussion}

\begin{table*}
\caption{\label{table:ShimizuSummary}
Summary of the neutron and isotope production yields from muon-initiated spallation in KamLAND.
The results of the \textsc{fluka} calculation include corrections for the muon spectrum and the $\mu^{+}$/$\mu^{-}$\ composition of the cosmic-ray muon flux.
}
\begin{ruledtabular}
\begin{tabular}{ccccccc}
 & Lifetime in & Radiation energy & \multicolumn{3}{c}{Yield $(\times10^{-7} \mu^{-1}$g$^{-1}$cm$^{2})$} &Fraction from showering $\mu$\\ \cline{4-6}
 & KamLAND LS & (MeV) & Ref.~\cite{Hagner2000} & \textsc{fluka} calc. & This measurement & This measurement \\
\hline
$n$ & 207.5 $\mu$s & 2.225 (capt. $\gamma$) & --- & $2097\pm13$& $2787\pm311$ & $(64\pm5)\%$ \\
$^{12}$B & 29.1 ms & 13.4 ($\beta^{-}$) & --- & $27.8 \pm 1.9$ & $42.9\pm3.3$ & $(68\pm2)\%$ \\
$^{12}$N & 15.9 ms & 17.3 ($\beta^{+}$) & --- & $0.77 \pm 0.08$ & $1.8\pm0.4$ & $(77\pm14)\%$ \\
$^{8}$Li & 1.21 s & 16.0 ($\beta^{-}\alpha$) & $1.9 \pm 0.8$ & $21.1 \pm 1.4$ & $12.2\pm2.6$ & $(65\pm17)\%$ \\
$^{8}$B & 1.11 s & 18.0 ($\beta^{+}\alpha$) & $3.3 \pm 1.0$ & $5.77 \pm 0.42$ & $8.4\pm2.4$ & $(78\pm23)\%$ \\
$^{9}$C & 182.5 ms & 16.5 ($\beta^{+}$) & $2.3 \pm 0.9$ & $1.35 \pm 0.12$ & $3.0\pm1.2$ & $(91\pm32)\%$ \\
$^{8}$He & 171.7 ms & 10.7 ($\beta^{-}\gamma n$) & & $0.32 \pm 0.05$ & $0.7\pm0.4$ & $(76\pm45)\%$ \\
$^{9}$Li & 257.2 ms & 13.6 ($\beta^{-}\gamma n$) & \raisebox{1.5ex}[1.5ex][0.75ex]{$\left.\begin{array}{c} \\ \end{array}\right\}1.0 \pm 0.3$}
& $3.16 \pm 0.25$ & $2.2\pm0.2$ & $(77\pm6)\%$ \\
$^{11}$C & 29.4 min & 1.98 ($\beta^{+}$) & $421 \pm 68$ & $416 \pm 27$ & $866\pm153$ & $(62\pm10)\%$ \\
$^{10}$C & 27.8 s & 3.65 ($\beta^{+}\gamma$) & $54 \pm 12$ & $19.1 \pm 1.3$ & $16.5\pm1.9$ & $(76\pm6)\%$ \\
$^{11}$Be & 19.9 s & 11.5 ($\beta^{-}$) & $< 1.1$ & $0.84 \pm 0.09$ & $1.1\pm0.2$ & $(74\pm12)\%$ \\
$^{6}$He & 1.16 s & 3.51 ($\beta^{-}$) & $7.5 \pm 1.5$ & $12.08 \pm 0.83$ & --- & --- \\
$^{7}$Be & 76.9 day & 0.478 (EC $\gamma$) & $107 \pm 21$ & $105.3 \pm 6.9$ & --- & --- \\
\end{tabular}
\end{ruledtabular}
\end{table*}

The isotope production yields from muon-initiated spallation in a liquid-scintillator target
were investigated at CERN by earlier experiment~\cite{Hagner2000} using the SPS muon beam with muon energies of 100 and 190~GeV.
Based on those cross-section measurements and the predicted muon energy spectrum at the KamLAND site,
we calculated the isotope production rates by extrapolation,
assuming a power-law of the muon energy $E_{\mu}^{\alpha}$.
The mean muon energy at KamLAND is \mbox{$260\pm8$\ GeV}, so KamLAND provides data to test the extrapolation method at this energy.

The production yields for the isotopes from muon-initiated spallation in KamLAND are provided in Table~\ref{table:ShimizuSummary}. 
On average, the yields from the showering muons ($\sim$15\% of all muons),
whose excess light yield parameter [$\Delta\mathcal{L}$, Eq.~(\ref{equation:ExcessLightYield})]
is greater than $10^{6}$\ p.e.\ \mbox{($\sim$3 GeV)},
constitute \mbox{$(70\pm2)\%$} of the yield from all muons. 
The production yield for $^{11}$C is the largest,
and its measured yield is larger than the \textsc{fluka} calculation by a factor of $\sim$2.
The Borexino Collaboration also reported a similar discrepancy~\cite{Arpesella2008a,Arpesella2008b},
which is consistent with what is observed in KamLAND.
Some measured production yields,
such as $^{8}$Li and $^{10}$C,
deviate significantly from estimates based on the muon beam experiment,
indicating that perhaps estimation by extrapolation is not sufficient.
All isotope yields are consistent within an order of magnitude.

\section{Summary}

We have analyzed KamLAND data to measure production yields of radioactive isotopes and neutrons through muon-initiated spallation in liquid-scintillator.
The neutron production yield is evaluated to be \mbox{$Y_{n}=(2.8\pm 0.3)\times10^{-4} \mu^{-1}$g$^{-1}$cm$^{2}$},
which is higher than the expectation from Monte Carlo simulations based on  \textsc{geant4} and \textsc{fluka}.
Some isotope production yields are found to be inconsistent with extrapolations---based on a power-law dependance with respect to muon energy---of
results from muon beam experiments.

\begin{acknowledgments}
The KamLAND experiment is supported by the COE program under Grant 09CE2003
of the Japanese Ministry of Education, Culture, Sports, Science and Technology;
the World Premier International Research Center Initiative (WPI Initiative), MEXT, Japan;
and under the U.S.\ Department of Energy (DOE) Grants DEFG03-00ER41138 and DE-AC02-05CH11231,
as well as other DOE Grants to individual institutions.
We are grateful to the Kamioka Mining and Smelting Company for supporting the activities in the mine.
We would also like to thank V.\ A.\ Kudryavtsev and M.\ Spurio for helpful comments.
\end{acknowledgments}

\bibliography{MuonSpallation}

\begin{thebibliography}{53}
\expandafter\ifx\csname natexlab\endcsname\relax\def\natexlab#1{#1}\fi
\expandafter\ifx\csname bibnamefont\endcsname\relax
  \def\bibnamefont#1{#1}\fi
\expandafter\ifx\csname bibfnamefont\endcsname\relax
  \def\bibfnamefont#1{#1}\fi
\expandafter\ifx\csname citenamefont\endcsname\relax
  \def\citenamefont#1{#1}\fi
\expandafter\ifx\csname url\endcsname\relax
  \def\url#1{\texttt{#1}}\fi
\expandafter\ifx\csname urlprefix\endcsname\relax\def\urlprefix{URL }\fi
\providecommand{\bibinfo}[2]{#2}
\providecommand{\eprint}[2][]{\url{#2}}

\bibitem[{\citenamefont{Bellini et~al.}()\citenamefont{Bellini, Benziger,
  Bonetti, Avanzini, Caccianiga, Cadonati, Calaprice, Carraro, Chavarria,
  Dalnoki-Veress et~al.}}]{Bellini2008}
\bibinfo{author}{\bibfnamefont{G.}~\bibnamefont{Bellini}},
  \bibinfo{author}{\bibfnamefont{J.}~\bibnamefont{Benziger}},
  \bibinfo{author}{\bibfnamefont{S.}~\bibnamefont{Bonetti}},
  \bibinfo{author}{\bibfnamefont{M.~B.} \bibnamefont{Avanzini}},
  \bibinfo{author}{\bibfnamefont{B.}~\bibnamefont{Caccianiga}},
  \bibinfo{author}{\bibfnamefont{L.}~\bibnamefont{Cadonati}},
  \bibinfo{author}{\bibfnamefont{F.}~\bibnamefont{Calaprice}},
  \bibinfo{author}{\bibfnamefont{C.}~\bibnamefont{Carraro}},
  \bibinfo{author}{\bibfnamefont{A.}~\bibnamefont{Chavarria}},
  \bibinfo{author}{\bibfnamefont{F.}~\bibnamefont{Dalnoki-Veress}},
  \bibnamefont{et~al.} (\bibinfo{collaboration}{Borexino Collaboration}),
  \eprint{astro-ph/0808.2868v1}.

\bibitem[{\citenamefont{Arpesella
  et~al.}(2008{\natexlab{a}})\citenamefont{Arpesella, Bellini, Benziger,
  Bonetti, Caccianiga, Calaprice, Dalnoki-Veress, D'Angelo, de~Kerret, Derbin
  et~al.}}]{Arpesella2008a}
\bibinfo{author}{\bibfnamefont{C.}~\bibnamefont{Arpesella}},
  \bibinfo{author}{\bibfnamefont{G.}~\bibnamefont{Bellini}},
  \bibinfo{author}{\bibfnamefont{J.}~\bibnamefont{Benziger}},
  \bibinfo{author}{\bibfnamefont{S.}~\bibnamefont{Bonetti}},
  \bibinfo{author}{\bibfnamefont{B.}~\bibnamefont{Caccianiga}},
  \bibinfo{author}{\bibfnamefont{F.}~\bibnamefont{Calaprice}},
  \bibinfo{author}{\bibfnamefont{F.}~\bibnamefont{Dalnoki-Veress}},
  \bibinfo{author}{\bibfnamefont{D.}~\bibnamefont{D'Angelo}},
  \bibinfo{author}{\bibfnamefont{H.}~\bibnamefont{de~Kerret}},
  \bibinfo{author}{\bibfnamefont{A.}~\bibnamefont{Derbin}},
  \bibnamefont{et~al.} (\bibinfo{collaboration}{Borexino Collaboration}),
  \bibinfo{journal}{Phys. Lett. B} \textbf{\bibinfo{volume}{658}},
  \bibinfo{pages}{101 } (\bibinfo{year}{2008}{\natexlab{a}}).

\bibitem[{\citenamefont{Arpesella
  et~al.}(2008{\natexlab{b}})\citenamefont{Arpesella, Back, Balata, Bellini,
  Benziger, Bonetti, Brigatti, Caccianiga, Cadonati, Calaprice
  et~al.}}]{Arpesella2008b}
\bibinfo{author}{\bibfnamefont{C.}~\bibnamefont{Arpesella}},
  \bibinfo{author}{\bibfnamefont{H.~O.} \bibnamefont{Back}},
  \bibinfo{author}{\bibfnamefont{M.}~\bibnamefont{Balata}},
  \bibinfo{author}{\bibfnamefont{G.}~\bibnamefont{Bellini}},
  \bibinfo{author}{\bibfnamefont{J.}~\bibnamefont{Benziger}},
  \bibinfo{author}{\bibfnamefont{S.}~\bibnamefont{Bonetti}},
  \bibinfo{author}{\bibfnamefont{A.}~\bibnamefont{Brigatti}},
  \bibinfo{author}{\bibfnamefont{B.}~\bibnamefont{Caccianiga}},
  \bibinfo{author}{\bibfnamefont{L.}~\bibnamefont{Cadonati}},
  \bibinfo{author}{\bibfnamefont{F.}~\bibnamefont{Calaprice}},
  \bibnamefont{et~al.} (\bibinfo{collaboration}{Borexino Collaboration}),
  \bibinfo{journal}{Phys. Rev. Lett.} \textbf{\bibinfo{volume}{101}},
  \bibinfo{pages}{091302} (\bibinfo{year}{2008}{\natexlab{b}}).

\bibitem[{\citenamefont{Hirano et~al.}(2008)\citenamefont{Hirano, Kishimoto,
  Ogawa, Hazama, Umehara, Matsuoka, Ito, and Tsubota}}]{Hirano2008}
\bibinfo{author}{\bibfnamefont{Y.}~\bibnamefont{Hirano}},
  \bibinfo{author}{\bibfnamefont{T.}~\bibnamefont{Kishimoto}},
  \bibinfo{author}{\bibfnamefont{I.}~\bibnamefont{Ogawa}},
  \bibinfo{author}{\bibfnamefont{R.}~\bibnamefont{Hazama}},
  \bibinfo{author}{\bibfnamefont{S.}~\bibnamefont{Umehara}},
  \bibinfo{author}{\bibfnamefont{K.}~\bibnamefont{Matsuoka}},
  \bibinfo{author}{\bibfnamefont{G.}~\bibnamefont{Ito}}, \bibnamefont{and}
  \bibinfo{author}{\bibfnamefont{Y.}~\bibnamefont{Tsubota}},
  \bibinfo{journal}{J. Phys. Conf. Ser.} \textbf{\bibinfo{volume}{120}},
  \bibinfo{eid}{052053} (\bibinfo{year}{2008}).

\bibitem[{\citenamefont{Kishimoto}(2007)}]{Kishimoto2007}
\bibinfo{author}{\bibfnamefont{T.}~\bibnamefont{Kishimoto}},
  \bibinfo{journal}{talk at the International Workshop on "Double Beta Decay
  and Neutrinos", June 11-13, 2007, Osaka, available online at
  http://dbd07.phys.sci.osaka-u.ac.jp/}  (\bibinfo{year}{2007}).

\bibitem[{\citenamefont{Zuber}(2007)}]{Zuber2007}
\bibinfo{author}{\bibfnamefont{K.}~\bibnamefont{Zuber}}, \bibinfo{journal}{Am.
  Inst. Phys. Conf. Proc.} \textbf{\bibinfo{volume}{942}}, \bibinfo{pages}{101}
  (\bibinfo{year}{2007}).

\bibitem[{\citenamefont{Chen}(2006)}]{Chen2006}
\bibinfo{author}{\bibfnamefont{M.~C.} \bibnamefont{Chen}},
  \bibinfo{journal}{Earth, Moon, and Planets} \textbf{\bibinfo{volume}{99}},
  \bibinfo{pages}{221} (\bibinfo{year}{2006}).

\bibitem[{\citenamefont{Grieb et~al.}()\citenamefont{Grieb, Link, Pitt,
  Raghavan, Rountree, and Vogelaar}}]{Grieb2007}
\bibinfo{author}{\bibfnamefont{C.}~\bibnamefont{Grieb}},
  \bibinfo{author}{\bibfnamefont{J.~M.} \bibnamefont{Link}},
  \bibinfo{author}{\bibfnamefont{M.~L.} \bibnamefont{Pitt}},
  \bibinfo{author}{\bibfnamefont{R.~S.} \bibnamefont{Raghavan}},
  \bibinfo{author}{\bibfnamefont{D.}~\bibnamefont{Rountree}}, \bibnamefont{and}
  \bibinfo{author}{\bibfnamefont{R.~B.} \bibnamefont{Vogelaar}},
  \eprint{hep-ex/0705.2769v1}.

\bibitem[{\citenamefont{Undagoitia et~al.}(2008)\citenamefont{Undagoitia, von
  Feilitzsch, G\"{o}ger-Neff, Oberauer, Potzel, Ulrich, Winter, and
  Wurm}}]{Undagoitia2008}
\bibinfo{author}{\bibfnamefont{T.~M.} \bibnamefont{Undagoitia}},
  \bibinfo{author}{\bibfnamefont{F.}~\bibnamefont{von Feilitzsch}},
  \bibinfo{author}{\bibfnamefont{M.}~\bibnamefont{G\"{o}ger-Neff}},
  \bibinfo{author}{\bibfnamefont{L.}~\bibnamefont{Oberauer}},
  \bibinfo{author}{\bibfnamefont{W.}~\bibnamefont{Potzel}},
  \bibinfo{author}{\bibfnamefont{A.}~\bibnamefont{Ulrich}},
  \bibinfo{author}{\bibfnamefont{J.}~\bibnamefont{Winter}}, \bibnamefont{and}
  \bibinfo{author}{\bibfnamefont{M.}~\bibnamefont{Wurm}}, \bibinfo{journal}{J.
  Phys. Conf. Ser.} \textbf{\bibinfo{volume}{120}}, \bibinfo{eid}{052018}
  (\bibinfo{year}{2008}).

\bibitem[{\citenamefont{Hagner et~al.}(2000)\citenamefont{Hagner, von Hentig,
  Heisinger, Oberauer, Sch\"onert, von Feilitzsch, and Nolte}}]{Hagner2000}
\bibinfo{author}{\bibfnamefont{T.}~\bibnamefont{Hagner}},
  \bibinfo{author}{\bibfnamefont{R.}~\bibnamefont{von Hentig}},
  \bibinfo{author}{\bibfnamefont{B.}~\bibnamefont{Heisinger}},
  \bibinfo{author}{\bibfnamefont{L.}~\bibnamefont{Oberauer}},
  \bibinfo{author}{\bibfnamefont{S.}~\bibnamefont{Sch\"onert}},
  \bibinfo{author}{\bibfnamefont{F.}~\bibnamefont{von Feilitzsch}},
  \bibnamefont{and} \bibinfo{author}{\bibfnamefont{E.}~\bibnamefont{Nolte}},
  \bibinfo{journal}{Astropart. Phys.} \textbf{\bibinfo{volume}{14}},
  \bibinfo{pages}{33 } (\bibinfo{year}{2000}).

\bibitem[{\citenamefont{Aglietta et~al.}(1989)}]{Aglietta1989}
\bibinfo{author}{\bibfnamefont{M.}~\bibnamefont{Aglietta}}
  \bibnamefont{et~al.}, \bibinfo{journal}{Nuovo Cimento Soc. Ital. Fis. C}
  \textbf{\bibinfo{volume}{12}}, \bibinfo{pages}{467} (\bibinfo{year}{1989}).

\bibitem[{\citenamefont{Aglietta et~al.}(2003)\citenamefont{Aglietta, Alyea,
  Antonioli, Badino, Bari, Basile, Berezinsky, Bersani, Bertaina, Bertoni
  et~al.}}]{Aglietta2003}
\bibinfo{author}{\bibfnamefont{M.}~\bibnamefont{Aglietta}},
  \bibinfo{author}{\bibfnamefont{E.}~\bibnamefont{Alyea}},
  \bibinfo{author}{\bibfnamefont{P.}~\bibnamefont{Antonioli}},
  \bibinfo{author}{\bibfnamefont{G.}~\bibnamefont{Badino}},
  \bibinfo{author}{\bibfnamefont{G.}~\bibnamefont{Bari}},
  \bibinfo{author}{\bibfnamefont{M.}~\bibnamefont{Basile}},
  \bibinfo{author}{\bibfnamefont{V.}~\bibnamefont{Berezinsky}},
  \bibinfo{author}{\bibfnamefont{F.}~\bibnamefont{Bersani}},
  \bibinfo{author}{\bibfnamefont{M.}~\bibnamefont{Bertaina}},
  \bibinfo{author}{\bibfnamefont{R.}~\bibnamefont{Bertoni}},
  \bibnamefont{et~al.}, \bibinfo{journal}{Phys. Atomic Nuclei}
  \textbf{\bibinfo{volume}{66}}, \bibinfo{pages}{123} (\bibinfo{year}{2003}).

\bibitem[{\citenamefont{Back et~al.}(2006)\citenamefont{Back, Balata, Bellini,
  Benziger, Bonetti, Caccianiga, Calaprice, D'Angelo, de~Bellefon, de~Kerret
  et~al.}}]{Back2006}
\bibinfo{author}{\bibfnamefont{H.}~\bibnamefont{Back}},
  \bibinfo{author}{\bibfnamefont{M.}~\bibnamefont{Balata}},
  \bibinfo{author}{\bibfnamefont{G.}~\bibnamefont{Bellini}},
  \bibinfo{author}{\bibfnamefont{J.}~\bibnamefont{Benziger}},
  \bibinfo{author}{\bibfnamefont{S.}~\bibnamefont{Bonetti}},
  \bibinfo{author}{\bibfnamefont{B.}~\bibnamefont{Caccianiga}},
  \bibinfo{author}{\bibfnamefont{F.}~\bibnamefont{Calaprice}},
  \bibinfo{author}{\bibfnamefont{D.}~\bibnamefont{D'Angelo}},
  \bibinfo{author}{\bibfnamefont{A.}~\bibnamefont{de~Bellefon}},
  \bibinfo{author}{\bibfnamefont{H.}~\bibnamefont{de~Kerret}},
  \bibnamefont{et~al.} (\bibinfo{collaboration}{Borexino Collaboration}),
  \bibinfo{journal}{Phys. Rev. C} \textbf{\bibinfo{volume}{74}},
  \bibinfo{eid}{045805} (\bibinfo{year}{2006}).

\bibitem[{\citenamefont{Antonioli et~al.}(1997)\citenamefont{Antonioli, Ghetti,
  Korolkova, Kudryavtsev, and Sartorelli}}]{Antonioli1997}
\bibinfo{author}{\bibfnamefont{P.}~\bibnamefont{Antonioli}},
  \bibinfo{author}{\bibfnamefont{C.}~\bibnamefont{Ghetti}},
  \bibinfo{author}{\bibfnamefont{E.~V.} \bibnamefont{Korolkova}},
  \bibinfo{author}{\bibfnamefont{V.~A.} \bibnamefont{Kudryavtsev}},
  \bibnamefont{and}
  \bibinfo{author}{\bibfnamefont{G.}~\bibnamefont{Sartorelli}},
  \bibinfo{journal}{Astropart. Phys.} \textbf{\bibinfo{volume}{7}},
  \bibinfo{pages}{357} (\bibinfo{year}{1997}).

\bibitem[{\citenamefont{Fass\`o et~al.}()\citenamefont{Fass\`o, Ferrari,
  Roesler, Sala, Battistoni, Cerutti, Gadioli, Garzelli, Ballarini, Ottolenghi
  et~al.}}]{Fasso2003}
\bibinfo{author}{\bibfnamefont{A.}~\bibnamefont{Fass\`o}},
  \bibinfo{author}{\bibfnamefont{A.}~\bibnamefont{Ferrari}},
  \bibinfo{author}{\bibfnamefont{S.}~\bibnamefont{Roesler}},
  \bibinfo{author}{\bibfnamefont{P.}~\bibnamefont{Sala}},
  \bibinfo{author}{\bibfnamefont{G.}~\bibnamefont{Battistoni}},
  \bibinfo{author}{\bibfnamefont{F.}~\bibnamefont{Cerutti}},
  \bibinfo{author}{\bibfnamefont{E.}~\bibnamefont{Gadioli}},
  \bibinfo{author}{\bibfnamefont{M.}~\bibnamefont{Garzelli}},
  \bibinfo{author}{\bibfnamefont{F.}~\bibnamefont{Ballarini}},
  \bibinfo{author}{\bibfnamefont{A.}~\bibnamefont{Ottolenghi}},
  \bibnamefont{et~al.}, \eprint{hep-ph/0306267}.

\bibitem[{\citenamefont{Ferrari et~al.}(2005)\citenamefont{Ferrari, Sala,
  Fass\'o, and Ranft}}]{Ferrari2005}
\bibinfo{author}{\bibfnamefont{A.}~\bibnamefont{Ferrari}},
  \bibinfo{author}{\bibfnamefont{P.~R.} \bibnamefont{Sala}},
  \bibinfo{author}{\bibfnamefont{A.}~\bibnamefont{Fass\'o}}, \bibnamefont{and}
  \bibinfo{author}{\bibfnamefont{J.}~\bibnamefont{Ranft}},
  \emph{\bibinfo{title}{FLUKA: A multi-particle transport code (program version
  2005)}} (\bibinfo{publisher}{CERN}, \bibinfo{address}{Geneva},
  \bibinfo{year}{2005}).

\bibitem[{\citenamefont{Agostinelli et~al.}(2003)\citenamefont{Agostinelli,
  Allison, Amako, Apostolakis, Araujo, Arce, Asai, Axen, Banerjee, Barrand
  et~al.}}]{Agostinelli2003}
\bibinfo{author}{\bibfnamefont{S.}~\bibnamefont{Agostinelli}},
  \bibinfo{author}{\bibfnamefont{J.}~\bibnamefont{Allison}},
  \bibinfo{author}{\bibfnamefont{K.}~\bibnamefont{Amako}},
  \bibinfo{author}{\bibfnamefont{J.}~\bibnamefont{Apostolakis}},
  \bibinfo{author}{\bibfnamefont{H.}~\bibnamefont{Araujo}},
  \bibinfo{author}{\bibfnamefont{P.}~\bibnamefont{Arce}},
  \bibinfo{author}{\bibfnamefont{M.}~\bibnamefont{Asai}},
  \bibinfo{author}{\bibfnamefont{D.}~\bibnamefont{Axen}},
  \bibinfo{author}{\bibfnamefont{S.}~\bibnamefont{Banerjee}},
  \bibinfo{author}{\bibfnamefont{G.}~\bibnamefont{Barrand}},
  \bibnamefont{et~al.}, \bibinfo{journal}{Nucl. Instr. Meth. A}
  \textbf{\bibinfo{volume}{506}}, \bibinfo{pages}{250 } (\bibinfo{year}{2003}).

\bibitem[{\citenamefont{Allison et~al.}(2006)\citenamefont{Allison, Amako,
  Apostolakis, Araujo, Arce~Dubois, Asai, Barrand, Capra, Chauvie, Chytracek
  et~al.}}]{Allison2006}
\bibinfo{author}{\bibfnamefont{J.}~\bibnamefont{Allison}},
  \bibinfo{author}{\bibfnamefont{K.}~\bibnamefont{Amako}},
  \bibinfo{author}{\bibfnamefont{J.}~\bibnamefont{Apostolakis}},
  \bibinfo{author}{\bibfnamefont{H.}~\bibnamefont{Araujo}},
  \bibinfo{author}{\bibfnamefont{P.}~\bibnamefont{Arce~Dubois}},
  \bibinfo{author}{\bibfnamefont{M.}~\bibnamefont{Asai}},
  \bibinfo{author}{\bibfnamefont{G.}~\bibnamefont{Barrand}},
  \bibinfo{author}{\bibfnamefont{R.}~\bibnamefont{Capra}},
  \bibinfo{author}{\bibfnamefont{S.}~\bibnamefont{Chauvie}},
  \bibinfo{author}{\bibfnamefont{R.}~\bibnamefont{Chytracek}},
  \bibnamefont{et~al.}, \bibinfo{journal}{IEEE Trans. Nucl. Sci.}
  \textbf{\bibinfo{volume}{53}}, \bibinfo{pages}{270} (\bibinfo{year}{2006}).

\bibitem[{\citenamefont{Battino et~al.}(1984)\citenamefont{Battino, Rettich,
  and Tominaga}}]{Battino1984}
\bibinfo{author}{\bibfnamefont{R.}~\bibnamefont{Battino}},
  \bibinfo{author}{\bibfnamefont{T.~R.} \bibnamefont{Rettich}},
  \bibnamefont{and} \bibinfo{author}{\bibfnamefont{T.}~\bibnamefont{Tominaga}},
  \bibinfo{journal}{J.\ Phys.\ Chem.\ Ref.\ Data}
  \textbf{\bibinfo{volume}{13}}, \bibinfo{pages}{563} (\bibinfo{year}{1984}).

\bibitem[{\citenamefont{Hesse et~al.}(1996)\citenamefont{Hesse, Battino,
  Scharlin, and Wilhelm}}]{Hesse1995}
\bibinfo{author}{\bibfnamefont{P.~J.} \bibnamefont{Hesse}},
  \bibinfo{author}{\bibfnamefont{R.}~\bibnamefont{Battino}},
  \bibinfo{author}{\bibfnamefont{P.}~\bibnamefont{Scharlin}}, \bibnamefont{and}
  \bibinfo{author}{\bibfnamefont{E.}~\bibnamefont{Wilhelm}},
  \bibinfo{journal}{J. Chem. Eng. Data} \textbf{\bibinfo{volume}{41}},
  \bibinfo{pages}{195} (\bibinfo{year}{1996}).

\bibitem[{\citenamefont{Kume et~al.}(1983)\citenamefont{Kume, Sawaki, Ito,
  Arisaka, Kajita, Nishimura, and Suzuki}}]{Kume1983}
\bibinfo{author}{\bibfnamefont{H.}~\bibnamefont{Kume}},
  \bibinfo{author}{\bibfnamefont{S.}~\bibnamefont{Sawaki}},
  \bibinfo{author}{\bibfnamefont{M.}~\bibnamefont{Ito}},
  \bibinfo{author}{\bibfnamefont{K.}~\bibnamefont{Arisaka}},
  \bibinfo{author}{\bibfnamefont{T.}~\bibnamefont{Kajita}},
  \bibinfo{author}{\bibfnamefont{A.}~\bibnamefont{Nishimura}},
  \bibnamefont{and} \bibinfo{author}{\bibfnamefont{A.}~\bibnamefont{Suzuki}},
  \bibinfo{journal}{Nucl. Instr. Meth. Phys. Res.}
  \textbf{\bibinfo{volume}{205}}, \bibinfo{pages}{443} (\bibinfo{year}{1983}).

\bibitem[{\citenamefont{Kleinfelder}(Aug. 2003)}]{Kleinfelder2003}
\bibinfo{author}{\bibfnamefont{S.}~\bibnamefont{Kleinfelder}},
  \bibinfo{journal}{IEEE Trans. Nucl. Sci.} \textbf{\bibinfo{volume}{50}},
  \bibinfo{pages}{955} (\bibinfo{year}{Aug. 2003}).

\bibitem[{\citenamefont{McKee et~al.}(2008)\citenamefont{McKee, Busenitz, and
  Ostrovskiy}}]{McKee2008}
\bibinfo{author}{\bibfnamefont{D.~W.} \bibnamefont{McKee}},
  \bibinfo{author}{\bibfnamefont{J.~K.} \bibnamefont{Busenitz}},
  \bibnamefont{and}
  \bibinfo{author}{\bibfnamefont{I.}~\bibnamefont{Ostrovskiy}},
  \bibinfo{journal}{Nucl. Instr. Meth. A} \textbf{\bibinfo{volume}{587}},
  \bibinfo{pages}{272} (\bibinfo{year}{2008}).

\bibitem[{\citenamefont{Berger et~al.}(2009)\citenamefont{Berger, Busenitz,
  Classen, Decowski, Dwyer, Elor, Frank, Freedman, Fujikawa, Galloway
  et~al.}}]{Berger2009}
\bibinfo{author}{\bibfnamefont{B.~E.} \bibnamefont{Berger}},
  \bibinfo{author}{\bibfnamefont{J.}~\bibnamefont{Busenitz}},
  \bibinfo{author}{\bibfnamefont{T.}~\bibnamefont{Classen}},
  \bibinfo{author}{\bibfnamefont{M.~P.} \bibnamefont{Decowski}},
  \bibinfo{author}{\bibfnamefont{D.~A.} \bibnamefont{Dwyer}},
  \bibinfo{author}{\bibfnamefont{G.}~\bibnamefont{Elor}},
  \bibinfo{author}{\bibfnamefont{A.}~\bibnamefont{Frank}},
  \bibinfo{author}{\bibfnamefont{S.~J.} \bibnamefont{Freedman}},
  \bibinfo{author}{\bibfnamefont{B.~K.} \bibnamefont{Fujikawa}},
  \bibinfo{author}{\bibfnamefont{M.}~\bibnamefont{Galloway}},
  \bibnamefont{et~al.} (\bibinfo{collaboration}{KamLAND Collaboration}),
  \bibinfo{journal}{JINST} \textbf{\bibinfo{volume}{4}},
  \bibinfo{pages}{P04017} (\bibinfo{year}{2009}).

\bibitem[{\citenamefont{Birks}(1951)}]{Birks1951}
\bibinfo{author}{\bibfnamefont{J.~B.} \bibnamefont{Birks}},
  \bibinfo{journal}{Proc. Phys. Soc.} \textbf{\bibinfo{volume}{A64}},
  \bibinfo{pages}{874} (\bibinfo{year}{1951}).

\bibitem[{\citenamefont{Birks}(1964)}]{Birks1964}
\bibinfo{author}{\bibfnamefont{J.~B.} \bibnamefont{Birks}},
  \emph{\bibinfo{title}{The Theory and Practice of Scintillation Counting}}
  (\bibinfo{publisher}{Pergamon}, \bibinfo{address}{London},
  \bibinfo{year}{1964}).

\bibitem[{Geo(1997)}]{GeographicalSurvey}
\emph{\bibinfo{title}{Digital Map 50\,m Grid (Elevation)}},
  \bibinfo{organization}{Geographical Survey Institute of Japan}
  (\bibinfo{year}{1997}), \bibinfo{note}{unpublished}.

\bibitem[{\citenamefont{Becherini et~al.}(2006)\citenamefont{Becherini,
  Margiotta, Sioli, and Spurio}}]{Becherini2006}
\bibinfo{author}{\bibfnamefont{Y.}~\bibnamefont{Becherini}},
  \bibinfo{author}{\bibfnamefont{A.}~\bibnamefont{Margiotta}},
  \bibinfo{author}{\bibfnamefont{M.}~\bibnamefont{Sioli}}, \bibnamefont{and}
  \bibinfo{author}{\bibfnamefont{M.}~\bibnamefont{Spurio}},
  \bibinfo{journal}{Astropart. Phys.} \textbf{\bibinfo{volume}{25}},
  \bibinfo{pages}{1 } (\bibinfo{year}{2006}).

\bibitem[{\citenamefont{Honda et~al.}(2007)\citenamefont{Honda, Kajita,
  Kasahara, Midorikawa, and Sanuki}}]{Honda2007}
\bibinfo{author}{\bibfnamefont{M.}~\bibnamefont{Honda}},
  \bibinfo{author}{\bibfnamefont{T.}~\bibnamefont{Kajita}},
  \bibinfo{author}{\bibfnamefont{K.}~\bibnamefont{Kasahara}},
  \bibinfo{author}{\bibfnamefont{S.}~\bibnamefont{Midorikawa}},
  \bibnamefont{and} \bibinfo{author}{\bibfnamefont{T.}~\bibnamefont{Sanuki}},
  \bibinfo{journal}{Phys. Rev. D} \textbf{\bibinfo{volume}{75}},
  \bibinfo{eid}{043006} (\bibinfo{year}{2007}).

\bibitem[{\citenamefont{Casper}(2002)}]{Casper2002}
\bibinfo{author}{\bibfnamefont{D.}~\bibnamefont{Casper}},
  \bibinfo{journal}{Nucl. Phys. B Proc. Suppl.} \textbf{\bibinfo{volume}{112}},
  \bibinfo{pages}{161} (\bibinfo{year}{2002}).

\bibitem[{\citenamefont{Mei and Hime}(2006)}]{Mei2006}
\bibinfo{author}{\bibfnamefont{D.-M.} \bibnamefont{Mei}} \bibnamefont{and}
  \bibinfo{author}{\bibfnamefont{A.}~\bibnamefont{Hime}},
  \bibinfo{journal}{Phys. Rev. D} \textbf{\bibinfo{volume}{73}},
  \bibinfo{eid}{053004} (\bibinfo{year}{2006}).

\bibitem[{\citenamefont{Galbiati and Beacom}(2005)}]{Galbiati2005b}
\bibinfo{author}{\bibfnamefont{C.}~\bibnamefont{Galbiati}} \bibnamefont{and}
  \bibinfo{author}{\bibfnamefont{J.~F.} \bibnamefont{Beacom}},
  \bibinfo{journal}{Phys. Rev. C} \textbf{\bibinfo{volume}{72}},
  \bibinfo{eid}{025807} (\bibinfo{year}{2005}).

\bibitem[{\citenamefont{Tang et~al.}(2006)\citenamefont{Tang, Horton-Smith,
  Kudryavtsev, and Tonazzo}}]{Tang2006}
\bibinfo{author}{\bibfnamefont{A.}~\bibnamefont{Tang}},
  \bibinfo{author}{\bibfnamefont{G.}~\bibnamefont{Horton-Smith}},
  \bibinfo{author}{\bibfnamefont{V.~A.} \bibnamefont{Kudryavtsev}},
  \bibnamefont{and} \bibinfo{author}{\bibfnamefont{A.}~\bibnamefont{Tonazzo}},
  \bibinfo{journal}{Phys. Rev. D} \textbf{\bibinfo{volume}{74}},
  \bibinfo{eid}{053007} (\bibinfo{year}{2006}).

\bibitem[{\citenamefont{{Eidelman} et~al.}(2004)\citenamefont{{Eidelman},
  {Hayes}, {Olive}, {Aguilar-Benitez}, {Amsler}, {Asner}, {Babu}, {Barnett},
  {Beringer}, {Burchat} et~al.}}]{PDG2004}
\bibinfo{author}{\bibfnamefont{S.}~\bibnamefont{{Eidelman}}},
  \bibinfo{author}{\bibfnamefont{K.}~\bibnamefont{{Hayes}}},
  \bibinfo{author}{\bibfnamefont{K.}~\bibnamefont{{Olive}}},
  \bibinfo{author}{\bibfnamefont{M.}~\bibnamefont{{Aguilar-Benitez}}},
  \bibinfo{author}{\bibfnamefont{C.}~\bibnamefont{{Amsler}}},
  \bibinfo{author}{\bibfnamefont{D.}~\bibnamefont{{Asner}}},
  \bibinfo{author}{\bibfnamefont{K.}~\bibnamefont{{Babu}}},
  \bibinfo{author}{\bibfnamefont{R.}~\bibnamefont{{Barnett}}},
  \bibinfo{author}{\bibfnamefont{J.}~\bibnamefont{{Beringer}}},
  \bibinfo{author}{\bibfnamefont{P.}~\bibnamefont{{Burchat}}},
  \bibnamefont{et~al.}, \bibinfo{journal}{{Phys. Lett. B}}
  \textbf{\bibinfo{volume}{592}}, \bibinfo{pages}{1} (\bibinfo{year}{2004}).

\bibitem[{Kam(1977)}]{Kamioka1977}
\bibinfo{type}{Tech. Rep.}, \bibinfo{institution}{Kamioka Mining \& Smelting
  Company} (\bibinfo{year}{1977}), \bibinfo{note}{internal report}.

\bibitem[{\citenamefont{Groom et~al.}(2001)\citenamefont{Groom, Mokhov, and
  Striganov}}]{Groom2001}
\bibinfo{author}{\bibfnamefont{D.~E.} \bibnamefont{Groom}},
  \bibinfo{author}{\bibfnamefont{N.~V.} \bibnamefont{Mokhov}},
  \bibnamefont{and} \bibinfo{author}{\bibfnamefont{S.~I.}
  \bibnamefont{Striganov}}, \bibinfo{journal}{Atomic Data and Nuclear Data
  Tables} \textbf{\bibinfo{volume}{78}}, \bibinfo{pages}{183 }
  (\bibinfo{year}{2001}).

\bibitem[{\citenamefont{Barrett et~al.}(1952)\citenamefont{Barrett, Bollinger,
  Cocconi, Eisenberg, and Greisen}}]{Barrett1952}
\bibinfo{author}{\bibfnamefont{P.~H.} \bibnamefont{Barrett}},
  \bibinfo{author}{\bibfnamefont{L.~M.} \bibnamefont{Bollinger}},
  \bibinfo{author}{\bibfnamefont{G.}~\bibnamefont{Cocconi}},
  \bibinfo{author}{\bibfnamefont{Y.}~\bibnamefont{Eisenberg}},
  \bibnamefont{and} \bibinfo{author}{\bibfnamefont{K.}~\bibnamefont{Greisen}},
  \bibinfo{journal}{Rev. Mod. Phys.} \textbf{\bibinfo{volume}{24}},
  \bibinfo{pages}{133} (\bibinfo{year}{1952}).

\bibitem[{\citenamefont{Mughabghab et~al.}(1981)\citenamefont{Mughabghab,
  Divadeenam, and Holden}}]{Mughabghab1981}
\bibinfo{author}{\bibfnamefont{S.~F.} \bibnamefont{Mughabghab}},
  \bibinfo{author}{\bibfnamefont{M.}~\bibnamefont{Divadeenam}},
  \bibnamefont{and} \bibinfo{author}{\bibfnamefont{N.~E.}
  \bibnamefont{Holden}}, \emph{\bibinfo{title}{Neutron Cross Sections, Volume
  1, Neutron Resonance Parameter and Thermal Cross Sections, Part A
  \mbox{$Z=1-60$}}} (\bibinfo{publisher}{Academic Press}, \bibinfo{address}{New
  York}, \bibinfo{year}{1981}).

\bibitem[{\citenamefont{Baker and Cousins}(1984)}]{Baker1984}
\bibinfo{author}{\bibfnamefont{S.}~\bibnamefont{Baker}} \bibnamefont{and}
  \bibinfo{author}{\bibfnamefont{R.~D.} \bibnamefont{Cousins}},
  \bibinfo{journal}{Nucl. Instr. Meth. Phys. Res.}
  \textbf{\bibinfo{volume}{221}}, \bibinfo{pages}{437} (\bibinfo{year}{1984}).

\bibitem[{\citenamefont{Ajzenberg-Selove}(1990)}]{AjzenbergSelove1990}
\bibinfo{author}{\bibfnamefont{F.}~\bibnamefont{Ajzenberg-Selove}},
  \bibinfo{journal}{Nucl. Phys. A} \textbf{\bibinfo{volume}{506}},
  \bibinfo{pages}{1} (\bibinfo{year}{1990}).

\bibitem[{\citenamefont{Tilley et~al.}(2004)\citenamefont{Tilley, Kelley,
  Godwin, Millener, Purcell, Sheu, and Weller}}]{Tilley2004}
\bibinfo{author}{\bibfnamefont{D.~R.} \bibnamefont{Tilley}},
  \bibinfo{author}{\bibfnamefont{J.~H.} \bibnamefont{Kelley}},
  \bibinfo{author}{\bibfnamefont{J.~L.} \bibnamefont{Godwin}},
  \bibinfo{author}{\bibfnamefont{D.~J.} \bibnamefont{Millener}},
  \bibinfo{author}{\bibfnamefont{J.~E.} \bibnamefont{Purcell}},
  \bibinfo{author}{\bibfnamefont{C.~G.} \bibnamefont{Sheu}}, \bibnamefont{and}
  \bibinfo{author}{\bibfnamefont{H.~R.} \bibnamefont{Weller}},
  \bibinfo{journal}{Nucl. Phys. A} \textbf{\bibinfo{volume}{745}},
  \bibinfo{pages}{155} (\bibinfo{year}{2004}).

\bibitem[{\citenamefont{Winter et~al.}(2006)\citenamefont{Winter, Freedman,
  Rehm, and Schiffer}}]{Winter2006}
\bibinfo{author}{\bibfnamefont{W.~T.} \bibnamefont{Winter}},
  \bibinfo{author}{\bibfnamefont{S.~J.} \bibnamefont{Freedman}},
  \bibinfo{author}{\bibfnamefont{K.~E.} \bibnamefont{Rehm}}, \bibnamefont{and}
  \bibinfo{author}{\bibfnamefont{J.~P.} \bibnamefont{Schiffer}},
  \bibinfo{journal}{Phys. Rev. C} \textbf{\bibinfo{volume}{73}},
  \bibinfo{eid}{025503} (\bibinfo{year}{2006}).

\bibitem[{\citenamefont{Bhattacharya et~al.}(2006)\citenamefont{Bhattacharya,
  Adelberger, and Swanson}}]{Bhattacharya2006}
\bibinfo{author}{\bibfnamefont{M.}~\bibnamefont{Bhattacharya}},
  \bibinfo{author}{\bibfnamefont{E.~G.} \bibnamefont{Adelberger}},
  \bibnamefont{and} \bibinfo{author}{\bibfnamefont{H.~E.}
  \bibnamefont{Swanson}}, \bibinfo{journal}{Phys. Rev. C}
  \textbf{\bibinfo{volume}{73}}, \bibinfo{eid}{055802} (\bibinfo{year}{2006}).

\bibitem[{\citenamefont{Abe et~al.}(2008)\citenamefont{Abe, Ebihara, Enomoto,
  Furuno, Gando, Ichimura, Ikeda, Inoue, Kibe, Kishimoto et~al.}}]{Abe2008}
\bibinfo{author}{\bibfnamefont{S.}~\bibnamefont{Abe}},
  \bibinfo{author}{\bibfnamefont{T.}~\bibnamefont{Ebihara}},
  \bibinfo{author}{\bibfnamefont{S.}~\bibnamefont{Enomoto}},
  \bibinfo{author}{\bibfnamefont{K.}~\bibnamefont{Furuno}},
  \bibinfo{author}{\bibfnamefont{Y.}~\bibnamefont{Gando}},
  \bibinfo{author}{\bibfnamefont{K.}~\bibnamefont{Ichimura}},
  \bibinfo{author}{\bibfnamefont{H.}~\bibnamefont{Ikeda}},
  \bibinfo{author}{\bibfnamefont{K.}~\bibnamefont{Inoue}},
  \bibinfo{author}{\bibfnamefont{Y.}~\bibnamefont{Kibe}},
  \bibinfo{author}{\bibfnamefont{Y.}~\bibnamefont{Kishimoto}},
  \bibnamefont{et~al.} (\bibinfo{collaboration}{KamLAND Collaboration}),
  \bibinfo{journal}{Phys. Rev. Lett.} \textbf{\bibinfo{volume}{100}},
  \bibinfo{eid}{221803} (\bibinfo{year}{2008}).

\bibitem[{\citenamefont{Galbiati et~al.}(2005)\citenamefont{Galbiati, Pocar,
  Franco, Ianni, Cadonati, and Sch\"{o}nert}}]{Galbiati2005a}
\bibinfo{author}{\bibfnamefont{C.}~\bibnamefont{Galbiati}},
  \bibinfo{author}{\bibfnamefont{A.}~\bibnamefont{Pocar}},
  \bibinfo{author}{\bibfnamefont{D.}~\bibnamefont{Franco}},
  \bibinfo{author}{\bibfnamefont{A.}~\bibnamefont{Ianni}},
  \bibinfo{author}{\bibfnamefont{L.}~\bibnamefont{Cadonati}}, \bibnamefont{and}
  \bibinfo{author}{\bibfnamefont{S.}~\bibnamefont{Sch\"{o}nert}},
  \bibinfo{journal}{Phys. Rev. C} \textbf{\bibinfo{volume}{71}},
  \bibinfo{eid}{055805} (\bibinfo{year}{2005}).

\bibitem[{\citenamefont{Ara\'{u}jo et~al.}(2005)\citenamefont{Ara\'{u}jo,
  Kudryavtsev, Spooner, and Sumner}}]{Araujo2005}
\bibinfo{author}{\bibfnamefont{H.}~\bibnamefont{Ara\'{u}jo}},
  \bibinfo{author}{\bibfnamefont{V.}~\bibnamefont{Kudryavtsev}},
  \bibinfo{author}{\bibfnamefont{N.}~\bibnamefont{Spooner}}, \bibnamefont{and}
  \bibinfo{author}{\bibfnamefont{T.}~\bibnamefont{Sumner}},
  \bibinfo{journal}{Nucl. Instr. Meth. A} \textbf{\bibinfo{volume}{545}},
  \bibinfo{pages}{398 } (\bibinfo{year}{2005}).

\bibitem[{\citenamefont{Marino et~al.}(2007)\citenamefont{Marino, Detwiler,
  Henning, Johnson, Schubert, and Wilkerson}}]{Marino2007}
\bibinfo{author}{\bibfnamefont{M.~G.} \bibnamefont{Marino}},
  \bibinfo{author}{\bibfnamefont{J.~A.} \bibnamefont{Detwiler}},
  \bibinfo{author}{\bibfnamefont{R.}~\bibnamefont{Henning}},
  \bibinfo{author}{\bibfnamefont{R.~A.} \bibnamefont{Johnson}},
  \bibinfo{author}{\bibfnamefont{A.~G.} \bibnamefont{Schubert}},
  \bibnamefont{and} \bibinfo{author}{\bibfnamefont{J.~F.}
  \bibnamefont{Wilkerson}}, \bibinfo{journal}{Nucl. Instr. Meth. A}
  \textbf{\bibinfo{volume}{582}}, \bibinfo{pages}{611} (\bibinfo{year}{2007}).

\bibitem[{\citenamefont{Hertenberger et~al.}(1995)\citenamefont{Hertenberger,
  Chen, and Dougherty}}]{Hertenberger1995}
\bibinfo{author}{\bibfnamefont{R.}~\bibnamefont{Hertenberger}},
  \bibinfo{author}{\bibfnamefont{M.}~\bibnamefont{Chen}}, \bibnamefont{and}
  \bibinfo{author}{\bibfnamefont{B.~L.} \bibnamefont{Dougherty}},
  \bibinfo{journal}{Phys. Rev. C} \textbf{\bibinfo{volume}{52}},
  \bibinfo{pages}{3449} (\bibinfo{year}{1995}).

\bibitem[{\citenamefont{Bezrukov et~al.}(1973)\citenamefont{Bezrukov, Beresnev,
  Zatsepin, Ryazhskaya, and Stepanets}}]{Bezrukov1973}
\bibinfo{author}{\bibfnamefont{L.~B.} \bibnamefont{Bezrukov}},
  \bibinfo{author}{\bibfnamefont{V.~I.} \bibnamefont{Beresnev}},
  \bibinfo{author}{\bibfnamefont{G.~T.} \bibnamefont{Zatsepin}},
  \bibinfo{author}{\bibfnamefont{O.~G.} \bibnamefont{Ryazhskaya}},
  \bibnamefont{and} \bibinfo{author}{\bibfnamefont{L.~N.}
  \bibnamefont{Stepanets}}, \bibinfo{journal}{Sov. J. Nucl. Phys.}
  \textbf{\bibinfo{volume}{17}}, \bibinfo{pages}{51} (\bibinfo{year}{1973}).

\bibitem[{\citenamefont{Boehm et~al.}(2000)\citenamefont{Boehm, Busenitz, Cook,
  Gratta, Henrikson, Kornis, Lawrence, Lee, McKinny, Miller
  et~al.}}]{Boehm2000}
\bibinfo{author}{\bibfnamefont{F.}~\bibnamefont{Boehm}},
  \bibinfo{author}{\bibfnamefont{J.}~\bibnamefont{Busenitz}},
  \bibinfo{author}{\bibfnamefont{B.}~\bibnamefont{Cook}},
  \bibinfo{author}{\bibfnamefont{G.}~\bibnamefont{Gratta}},
  \bibinfo{author}{\bibfnamefont{H.}~\bibnamefont{Henrikson}},
  \bibinfo{author}{\bibfnamefont{J.}~\bibnamefont{Kornis}},
  \bibinfo{author}{\bibfnamefont{D.}~\bibnamefont{Lawrence}},
  \bibinfo{author}{\bibfnamefont{K.~B.} \bibnamefont{Lee}},
  \bibinfo{author}{\bibfnamefont{K.}~\bibnamefont{McKinny}},
  \bibinfo{author}{\bibfnamefont{L.}~\bibnamefont{Miller}},
  \bibnamefont{et~al.}, \bibinfo{journal}{Phys. Rev. D}
  \textbf{\bibinfo{volume}{62}}, \bibinfo{pages}{092005}
  (\bibinfo{year}{2000}).

\bibitem[{\citenamefont{Enikeev et~al.}(1987)\citenamefont{Enikeev, Zatsepin,
  Koro\'{l}kova, Kudryavstev, Ma\'{l}gin, Ryazhskaya, and
  Kha\'{l}chukov}}]{Enikeev1987}
\bibinfo{author}{\bibfnamefont{R.~I.} \bibnamefont{Enikeev}},
  \bibinfo{author}{\bibfnamefont{G.~T.} \bibnamefont{Zatsepin}},
  \bibinfo{author}{\bibfnamefont{E.~V.} \bibnamefont{Koro\'{l}kova}},
  \bibinfo{author}{\bibfnamefont{V.~A.} \bibnamefont{Kudryavstev}},
  \bibinfo{author}{\bibfnamefont{A.~S.} \bibnamefont{Ma\'{l}gin}},
  \bibinfo{author}{\bibfnamefont{O.~G.} \bibnamefont{Ryazhskaya}},
  \bibnamefont{and} \bibinfo{author}{\bibfnamefont{F.~F.}
  \bibnamefont{Kha\'{l}chukov}}, \bibinfo{journal}{Sov. J. Nucl. Phys.}
  \textbf{\bibinfo{volume}{46}}, \bibinfo{pages}{883} (\bibinfo{year}{1987}).

\bibitem[{\citenamefont{Wang et~al.}(2001)\citenamefont{Wang, Balic, Gratta,
  Fass\`o, Roesler, and Ferrari}}]{Wang2001}
\bibinfo{author}{\bibfnamefont{Y.-F.} \bibnamefont{Wang}},
  \bibinfo{author}{\bibfnamefont{V.}~\bibnamefont{Balic}},
  \bibinfo{author}{\bibfnamefont{G.}~\bibnamefont{Gratta}},
  \bibinfo{author}{\bibfnamefont{A.}~\bibnamefont{Fass\`o}},
  \bibinfo{author}{\bibfnamefont{S.}~\bibnamefont{Roesler}}, \bibnamefont{and}
  \bibinfo{author}{\bibfnamefont{A.}~\bibnamefont{Ferrari}},
  \bibinfo{journal}{Phys. Rev. D} \textbf{\bibinfo{volume}{64}},
  \bibinfo{pages}{013012} (\bibinfo{year}{2001}).

\bibitem[{\citenamefont{Kudryavtsev et~al.}(2003)\citenamefont{Kudryavtsev,
  Spooner, and McMillan}}]{Kudryavtsev2003}
\bibinfo{author}{\bibfnamefont{V.~A.} \bibnamefont{Kudryavtsev}},
  \bibinfo{author}{\bibfnamefont{N.~J.~C.} \bibnamefont{Spooner}},
  \bibnamefont{and} \bibinfo{author}{\bibfnamefont{J.~E.}
  \bibnamefont{McMillan}}, \bibinfo{journal}{Nucl. Instr. Meth. A}
  \textbf{\bibinfo{volume}{505}}, \bibinfo{pages}{688} (\bibinfo{year}{2003}).

\end{thebibliography}

\end{document}